\documentclass[a4paper,11pt]{article}

\pdfoutput=1

\usepackage[usenames,dvipsnames]{color}
\usepackage{jcappub}
\usepackage{epsfig}
\usepackage{amssymb}
\usepackage{amsmath}
\usepackage{graphicx}
\usepackage{multirow}
\usepackage{verbatim}
\usepackage{xspace}
\usepackage{placeins}
\usepackage{tabularx}

\definecolor{darkblue}{rgb}{0,0,0.5}
\definecolor{darkgreen}{rgb}{0.1,0,0.3}
\definecolor{darkred}{rgb}{0.6,0,0}

\defcitealias{GS98}{GS98}
\defcitealias{AGSS}{AGSS09}
\defcitealias{AGS05}{AGS05}

\newcommand{\nc}{\newcommand}

\nc{\ba}{\begin{eqnarray}}
\nc{\ea}{\end{eqnarray}}

\newcommand\s{\sigma}

\nc{\ga}{\gamma}
\nc{\om}{\omega}

\nc{\x}{{\bf x }}
\nc{\mx}{m_\chi}
\nc{\mnuc}{m_{\rm nuc}}
\nc{\kk}{{\bf k }}
\nc{\f}{{\bf f }}
\nc{\e}{{\bf e }}
\nc{\T}{ \theta (s_i (t)- \s) }
\nc{\TT}{ \theta (s_i (t_{ r \, i } )- \s) }
\nc{\br}{   (s_i (t)- \s)  }
\nc{\gta}{\gamma \rightarrow a}
\nc{\Dag}{\Delta_{a \gamma}}
\nc{\Dosc}{\Delta_{osc}}
\nc{\Dpl}{\Delta_{pl}}
\nc{\Da}{\Delta_a}
\nc{\gag}{g_{a \gamma}}
\nc{\wpl}{\omega_{pl}}
\nc{\hr}{$h_{res}$}
\nc{\ud}{\,\mathrm{d}}
\nc{\dr}{\delta h_{res}}
\nc{\igev}{GeV$^{-1}$}
\nc{\ssi}{\sigma_{\mathrm{SI}}}
\nc{\ssd}{\sigma_{\mathrm{SD}}}
\nc{\tq}{\tilde \q}
\nc{\qmin}{q_{\mathrm{min}}}
\nc{\qmax}{q_{\mathrm{max}}}
\nc{\dmin}{\delta_{\mathrm{min}}}
\nc{\dmax}{\delta_{\mathrm{max}}}
\nc{\DS}{\textsf{DarkStec}\xspace}
\nc{\GS}{\textsf{GARSTEC}\xspace}
\nc{\dstars}{\textsf{DarkStars}\xspace}
\nc{\vrel}{v_{\rm rel}}
\nc{\qref}{q_{\rm ref}}
\nc{\sv}{\langle \sigma v \rangle_{\rm ann}}
\nc{\er}{E_\mathrm{R}}
\begin{document}

\title{Updated constraints on velocity and momentum-dependent asymmetric dark matter}

\author[1,2]{Aaron C. Vincent,}
 \emailAdd{aaron.vincent@durham.ac.uk}
 \affiliation[1]{Institute for Particle Physics Phenomenology (IPPP), Department of Physics, Durham University,   Durham DH1 3LE, UK}
\author[2]{Pat Scott} 
 \emailAdd{p.scott@imperial.ac.uk}
 \affiliation[2]{Department of Physics, Imperial College London, Blackett Laboratory, Prince Consort Road, London SW7 2AZ, UK}
\author[3]{and Aldo Serenelli}
 \emailAdd{aldos@ice.csic.es}
 \affiliation[3]{Institut de Ci\`encies de l'Espai (ICE-CSIC/IEEC), Campus UAB, Carrer de Can Magrans s/n, 08193 Cerdanyola del VallŽs, Spain}

\abstract{We present updated constraints on dark matter models with momentum-dependent or velocity-dependent interactions with nuclei, based on direct detection and solar physics.  We improve our previous treatment of energy transport in the solar interior by dark matter scattering, leading to significant changes in fits to many observables.  Based on solar physics alone, DM with a spin-independent $q^{4}$ coupling provides the best fit to data, and a statistically satisfactory solution to the solar abundance problem.  Once direct detection limits are accounted for however, the best solution is spin-dependent $v^2$ scattering with a reference cross-section of 10$^{-35}$\,cm$^2$ (at a reference velocity of $v_0=220$\,km\,s$^{-1}$), and a dark matter mass of about 5\,GeV.}

\subheader{IPPP/16/45}

\maketitle

\section{Introduction}
Despite tremendous success in predicting neutrino fluxes and describing the bulk structure of the Sun, the Standard Solar Model (SSM) still fails to reproduce key observables relating to helioseismology. After the downward revision of solar photospheric abundances over 10 years ago \cite{APForbidO,CtoO,AspIV,AspVI,AGS05,ScottVII,Melendez08,Scott09Ni,AGSS,AGSS_NaCa,AGSS_FePeak,AGSS_heavy} it has become clear that the predicted sound speed profile, position of the base of the convection zone and surface helium abundance are all several standard deviations away from the values obtained by direct helioseismological inversion \cite{Bahcall:2004yr, Basu:2004zg, Bahcall06, Yang07, Basu08, Serenelli:2009yc}. Despite over a decade of effort, no adequate solution has been found to reconcile solar modelling with these precision observables \cite{TurckChieze:1993dw,Guzik:2010ck,Bahcall05,Serenelli:2016nms}, prompting a search for solutions beyond the Standard Model. 

As the Sun travels through the dark matter halo of the Milky Way, it inevitably interacts with the dark matter population\footnote{For a review of dark matter, see e.g. \cite{Lisanti:2016jxe} and references therein.}. If a DM particle scatters with a nucleus in the Sun to a velocity less than the local escape velocity, it becomes gravitationally bound. From there, it will quickly settle into an equilibrium orbit near the solar centre, governed by the thermodynamics of the weakly-interacting DM gas in the steep gravitational potential of the Sun.  The effects of dark matter capture in stars has been studied in depth since the 1980s \cite{Krauss86, Gaisser86, Griest87, Gandhi:1993ce, Bottino:1994xp, Bergstrom98b,2002PhRvL..88o1303L,2002MNRAS.331..361L, Barger02, Desai04, Desai08, IceCube09, IceCube09_KK, IC40DM, SuperK11, IC22Methods, Silverwood12, IC79, Guo2013, Catena15b, Rott13, Bernal:2012qh, SalatiSilk89, BouquetSalati89a, Moskalenko07, Bertone07, Spolyar08, Fairbairn08, Scott08a, Iocco08a, Iocco08b, Scott09, Casanellas09, Ripamonti10, Zackrisson10a, Zackrisson10b, Scott11,Lopes:2012af}. In the absence of self-annihilation (i.e.\  asymmetric dark matter, ADM \cite{Petraki13, Zurek14}), a large population of dark matter can accumulate.  This can act as a heat conductor, acquiring kinetic energy from the hot core, and releasing it via interactions with nuclei in the cooler outer regions.  This can lead to a slightly shallower temperature gradient with height in the star. Despite the small population of DM (at most $\sim$1 particle per $10^{10}$ baryons), these small adjustments in the thermal gradient can have measurable effects on our own Sun.  These include the solar structure itself -- including the sound speed $c_s(r)$ and convective zone radius $r_{CZ}$ -- and on neutrino fluxes from fusion processes, due to their strong dependence on the core temperature.

Precision solar models that include capture and heat transport from standard spin-dependent or spin-independent ADM can be built to satisfy the solar radius, age and luminosity  \cite{Taoso10,Cumberbatch:2010hh,Lopes:2012}.  Whilst these models can bring some helioseismological observables into better agreement with data than the SSM, this comes at the cost of an elastic scattering cross-section that is several orders of magnitude higher than allowed by direct detection experiments. These models also lead to drastic underproduction of solar neutrinos. 

However, the kinetic regime probed by solar capture is very different from earth-based direct detection experiments: the Sun preferentially captures slower-moving particles from the DM halo, as these are more likely to scatter to sub-escape velocities, whereas direct detection relies on large enough velocities to produce observable recoils in a target material. This observation is partly what led to efforts to examine DM models with less trivial interactions with the Standard Model \cite{VincentScott2013, Lopes:2014, Lopes14, Vincent2014, Vincent:2015gqa}. Such interactions are generic in particle physics: scattering mediated by a massive particle generally gives rise to non-relativistic elastic scattering cross section that is proportional to some positive power of the momentum transfer $q$ or relative velocity $v$ (see e.g.\ \cite{Fan2010}). Conversely, new long-range forces can yield cross sections proportional to negative powers of these quantities (e.g. \cite{Panci:2014gga}). 

Recently, we found \cite{Vincent2014,Vincent:2015gqa} that a light asymmetric particle with a momentum-dependent interaction with quarks could be captured in large enough quantities in the Sun and conduct heat in such a way that helioseismic observables could be brought into excellent agreement with data. Although this led to a reduction in the predicted neutrino fluxes, the significant improvement in sound speed and convective zone depth was sufficient to produce a marked improvement over the SSM of 6 standard deviations and a potential solution to the solar composition problem. 
New searches by direct detection experiments at low scattering threshold have since ruled out the best fit from these studies. CRESST-II~\cite{Angloher:2016jsl} analysed the specific model highlighted in \cite{Vincent2014,Vincent:2015gqa} excluding it to high significance. The even more sensitive CDMSlite analysis \cite{Agnese:2015nto} released by the SuperCDMS collaboration around the same period confirms this result. 

We have furthermore revised the formalism of \cite{VincentScott2013}, and found that the conductive luminosity had been over (under) estimated in models with a cross section proportional to a positive (negative) power of q or $v$. We have corrected this error, finding that the best fit point in each model has moved in the parameter space. The best fits that we find are as good as those we found in \cite{Vincent2014,Vincent:2015gqa}, though the required cross sections are higher, in some cases by as much as an order of magnitude. Given the strength of the direct detection bounds that we derive in this work, this is of little consequence as the previously-favoured parameter space is now entirely ruled out.

Our goals in this work are therefore 1) to update our solar simulations, correcting the error in luminosity, and 2) to confront these results with constraints from direct detection, to determine whether such models can indeed lead to a solution to the solar composition problem. We find that the parameter space that remains, after taking into account the recent CDMSlite results, is highly restricted. Although the models that obey these bounds do not fit the solar data as well as some that are excluded by direct detection, they are nonetheless capable of producing very significant improvements over the SSM (see also \cite{Lopes14} for similar results).

This paper is structured as follows. In Sec.\ \ref{sec:solarDM} we outline the class of models that we are considering, and present the relevant equations for capture by the Sun and heat transport within it. In Sec.\ \ref{sec:dd} we briefly discuss constraints from direct detection on light DM particles able to affect solar structure. We present the results of our simulations in Sec.\ \ref{sec:results}, and conclude in Sec.\ \ref{sec:conclusion}.

\section{Dark matter in the Sun}
\label{sec:solarDM}
The method of obtaining the capture rate and subsequent energy transport due to momentum or velocity-dependent DM in the Sun is presented in detail in Refs.~\cite{VincentScott2013,Vincent:2015gqa}. Here, we show the main results, along with a crucial correction to the transported luminosity formula, Eq.\ \ref{LTEtransport}. Reflecting the nature of most concrete models so far proposed in the theory literature, we focus on models with isoscalar (identical proton and neutron) DM-nucleon couplings with the Standard Model.  We parameterise the resulting differential cross-sections as 
\begin{subequations}
\label{qdepvdep}
\begin{minipage}{0.4\linewidth}
  \begin{equation}
    \label{qdep}
    \sigma = \sigma_0 \left(\frac{q}{q_0}\right)^{2n}
  \end{equation}
\end{minipage}
\begin{minipage}{0.18\linewidth}
  \flushright and
\end{minipage}
\begin{minipage}{0.4\linewidth}
  \begin{equation}
    \label{vdep}
    \sigma = \sigma_0 \left(\frac{\vrel}{v_0}\right)^{2n}.
  \end{equation}
\end{minipage}
\end{subequations}\vspace{0.5mm}\\
Each of these can be spin-independent (SI), coupling coherently to all nucleons; or spin-dependent (SD), coupling only to the spin of unpaired nucleons. This ``form factor'' approach \cite{Feldstein:2009tr}, encompasses both non-relativistic effective operators that can come from effective point-like interactions \cite{Fan:2010gt,Kumar:2013iva,Fitzpatrick13}, as well as certain classes of long-range forces (e.g.  \cite{Panci:2014gga}). The cross section for scattering with a nucleus is then 
\begin{equation}
\sigma_{N,i} = \frac{\mnuc^2 (\mx + m_\mathrm{p})^2}{m_\mathrm{p}^2(\mx + \mnuc)^2}\left[\sigma_{\rm SI} A_i^2 + \sigma_{\rm SD}\frac{4(J_i +1)}{3J_i}|\langle S_{p,i}\rangle +\langle S_{n,i}\rangle |^2 \right].
\label{DMnucleus}
\end{equation}

\subsection{Capture in the Sun}
\label{sec:cap}
In the absence of annihilation or evaporation, the population of DM in the Sun is simply given by $\dot N_\chi(t) = C_\odot(t)$:
\begin{equation}
C_\odot(t) = 4\pi \int_0^{R_\odot} r^2 \int_0^\infty \frac{f_\odot(u)}{u} w \Omega(w) \ud u\ud r.
\label{caprate}
\end{equation}
Here, $f_\odot(u)$ is the DM speed distribution in the Sun's frame, and $\Omega(w)$ encodes the kinematics of scattering below the local escape velocity $v_{esc}(r)$. $w(r) \equiv \sqrt{u^2 + v_{esc}^2}$ is the local DM velocity inside the star's gravitational potential. Expressing the DM-to-nucleon mass ratio as $\mu_i \equiv m_\chi/m_{N_i}$, this quantity is:
\begin{equation}
 \Omega(w) = \frac{2}{m_\chi w} \sum_i \sigma_{N,i} n_i(r,t) \frac{\mu_{i,+}^2}{\mu_i}\Theta\left(\frac{\mu_i v_{\rm esc}^2}{\mu_{i,-}^2} - u^2 \right) GFFI(\mu,u,v_{esc}),
 \label{eq:omega}
\end{equation} 
where $GFFI$ is the generalized form factor integral, and $n_i(r)$ is the number density of each nuclear species in the Sun. In the case of a constant cross section, the $GFFI$ is simply an integral over the nuclear form factor $F_i(E_\mathrm{R})$:
\begin{equation}
GFFI_{n=0} = \int_{\mx u^2/2}^{\mx w^2 \mu_i/2\mu_{i,+}^2} |F_i(\er)|^2 \ud \er,
\end{equation}
where the integral limits are set by the kinematics required to downscatter the DM velocity enough for it to be captured.

For velocity-dependent scattering, the required modification of the capture rate is straightforward: the integrand in Eq.\ \ref{caprate} is simply multiplied by an overall factor of $[w(r)/v_0]^{2n}$.

When the cross section is momentum-dependent, the factors of $(q/q_0)^{2n}$ must be included inside the $GFFI$. For scattering with hydrogen (assuming a point-like particle, i.e. $F_1(E_\mathrm{R}) = 1$): 
\begin{equation}
GFFI_{n\ne0,\mathrm{H}} = \left(\frac{p}{q_0}\right)^{2n}\frac{m_\chi w^2}{2\mu^n}
\begin{cases}
 \frac{1}{1+n}\left[\left(\frac{\mu}{\mu_+^2}\right)^{n+1} - \left(\frac{u^2}{w^2}\right)^{n+1}\right],         &(n \neq -1) \\
 \ln \left(\frac{\mu }{\mu_+^2}\frac{w^2}{u^2}\right),        &(n = -1) \\ 
\end{cases}
\label{ffHresult}
\end{equation}
where $p = m_\chi w$. For heavier nuclei, a parameterisation of the form factor must be chosen. Significant work over the past few years has gone into computing accurate form factors that model the nuclear response function using effective field theory; these yield interactions close to those in Eqs.\ \ref{qdepvdep} and Ref.\ \cite{Catena:2015uha} has computed these response functions for capture in the Sun.  The standard Helm \cite{Helm:1956zz} parameterisation is %
\begin{equation}
|F_i(\er)|^2 = \exp{\left(-\frac{\er}{E_i}\right)},
\label{heavyformfactor}
\end{equation}
where $E_i$ is a constant quantity for each nuclear species $i$, given by
\begin{equation}
E_i = \frac{5.8407 \times 10^{-2}}{m_{N,i} (0.91 m_{N,i}^{1/3} + 0.3)^2} \mathrm{GeV}.
\end{equation}
For spin-independent capture and transport, we use the following 15 elements: H, He, C, N, O, Ne, Mg, Na, Al, Si, S, Ar, Ca, Fe and Ni. Though heavier elements are suppressed in abundance, the $A^2$ coherence factor in Eq.~\ref{DMnucleus} can enhance the capture rate off these elements by as much as four orders of magnitude.

We have furthermore checked in Ref.\ \cite{Vincent:2015gqa} that for most models, the deviations with respect to Helm are minimal, as low-momentum scattering tends to dominate the capture rate. There is one exception: when computing spin-dependent capture rates, we neglect elements heavier than hydrogen. These do not benefit from the $A^2$ coherent enhancement present for spin-independent interactions, and their very low abundance further suppresses their contribution. However, in the $q^4$ SD case, which can be matched to an $\mathcal{O}_6 \equiv (\vec S_\chi \cdot \vec q/m_N)(\vec S_N \cdot \vec q/m_N) $ coupling, the total capture rate in an evolved solar model is actually dominated by scattering with nitrogen.  This does not turn out to be very helpful, as we will find that the cross-section required to produce an effect in the Sun with this operator is ruled out by five orders of magnitude.  We thus use the Helm  form factor for every case, and the generalised form factor integral for momentum-dependent interactions becomes:
\begin{equation}
GFFI_{n\ne0,i\ne\mathrm{H}} = \left(\frac{p}{q_0}\right)^{2n}\frac{E_i}{(B \mu)^n}\left[ \Gamma\left(1+n,B\frac{u^2}{w^2}\right) - \Gamma\left(1+n,B\frac{\mu}{\mu_+^2}\right) \right],
\label{ffZresult}
\end{equation}
where $B\equiv \frac12 \mx w^2 / E_i$, and $\Gamma(m,x)$ is the (upper) incomplete gamma function. 

\begin{figure}[h]
\includegraphics[width=0.5\textwidth]{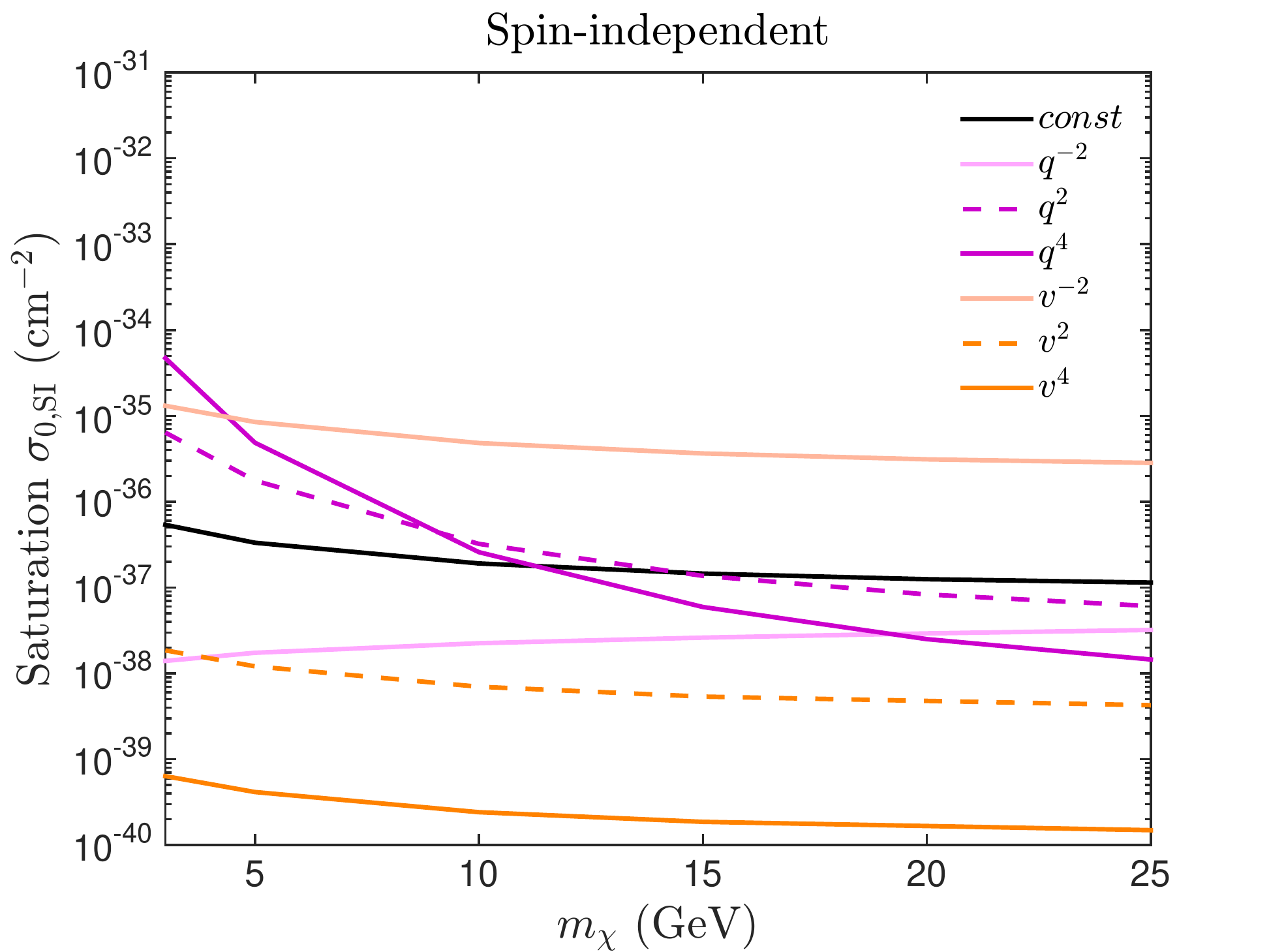} \includegraphics[width=0.5\textwidth]{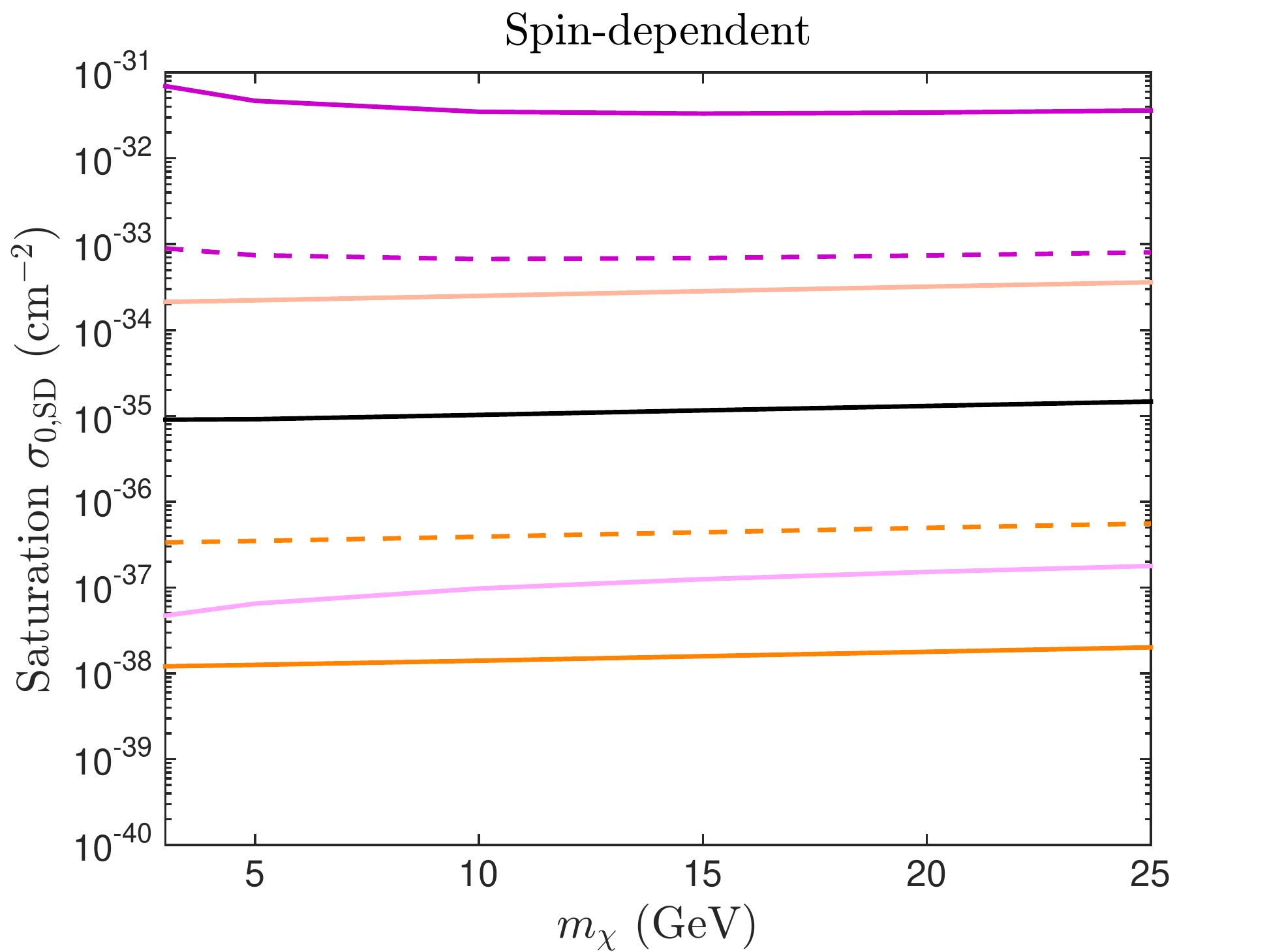} 
\caption{Cross section at which the capture rate computed in \DS saturates the geometric limit \ref{eq:capcutoff}. Left: spin-independent interactions, coupling to every element; right: spin-dependent interactions with hydrogen only. Despite their small abundances, heavier elements contribute significantly to spin-independent capture, thanks in large part to the coherent $ \sigma_N \propto A^2$ enhancement.}
\label{fig:saturation}
\end{figure}

The Sun cannot capture more dark matter than the so-called geometric limit, in which all of the DM that intercepts the solar disk is captured. The total capture rate is therefore the smallest of Eq.\ \ref{caprate} and
\begin{eqnarray}
  C_\mathrm{max}(t) &=& \pi R_\odot^2(t) \int_0^\infty \frac{f_\odot(u)}{u}w^2(u,R_\odot) \ud u \label{eq:capcutoff} \\
  &=& \frac{1}{3}\pi\frac{\rho_\chi}{m_\chi}R_\odot^2(t)\left(e^{-\frac{3}{2}\frac{u_\odot^2}{u_0^2}}\sqrt\frac{6}{\pi}u_0 + \frac{6G_{\rm N}M_\odot + R_\odot(u_0^2 + 3 u_\odot^2)}{R_\odot u_\odot}\mathrm{Erf}{\left[\sqrt{\frac{3}{2}}\frac{u_\odot}{u_0}\right]} \right), \nonumber
\end{eqnarray}
where $u_0=270$\,km\,s$^{-1}$ is the dispersion of the Maxwell-Boltzmann DM velocity distribution, $u_\odot=220$\,km\,s$^{-1}$ is the velocity of the Sun relative to the DM halo, and $\rho_\chi= 0.38$ GeV cm$^{-3}$. is the local density of dark matter.  We show the cross section that saturates this bound in Fig.~\ref{fig:saturation}, for each of the models we consider in this paper. Strictly speaking, the transition from Eq.\ \ref{caprate} to \ref{eq:capcutoff} should be smooth, rather than a sharp cutoff as we impose here. To accurately model this transition, the full optical depth must be including in Eq.\ \ref{caprate}. This leads to a suppression of the capture rate as $\sigma_0$ approaches the saturation cross section, with the ultimate effect of slightly suppressing the effect of DM close to this threshold. As this computation is relatively involved, we address it in an upcoming work \cite{BusoniEvap}.

\begin{figure}[p]
\includegraphics[width=0.48\textwidth]{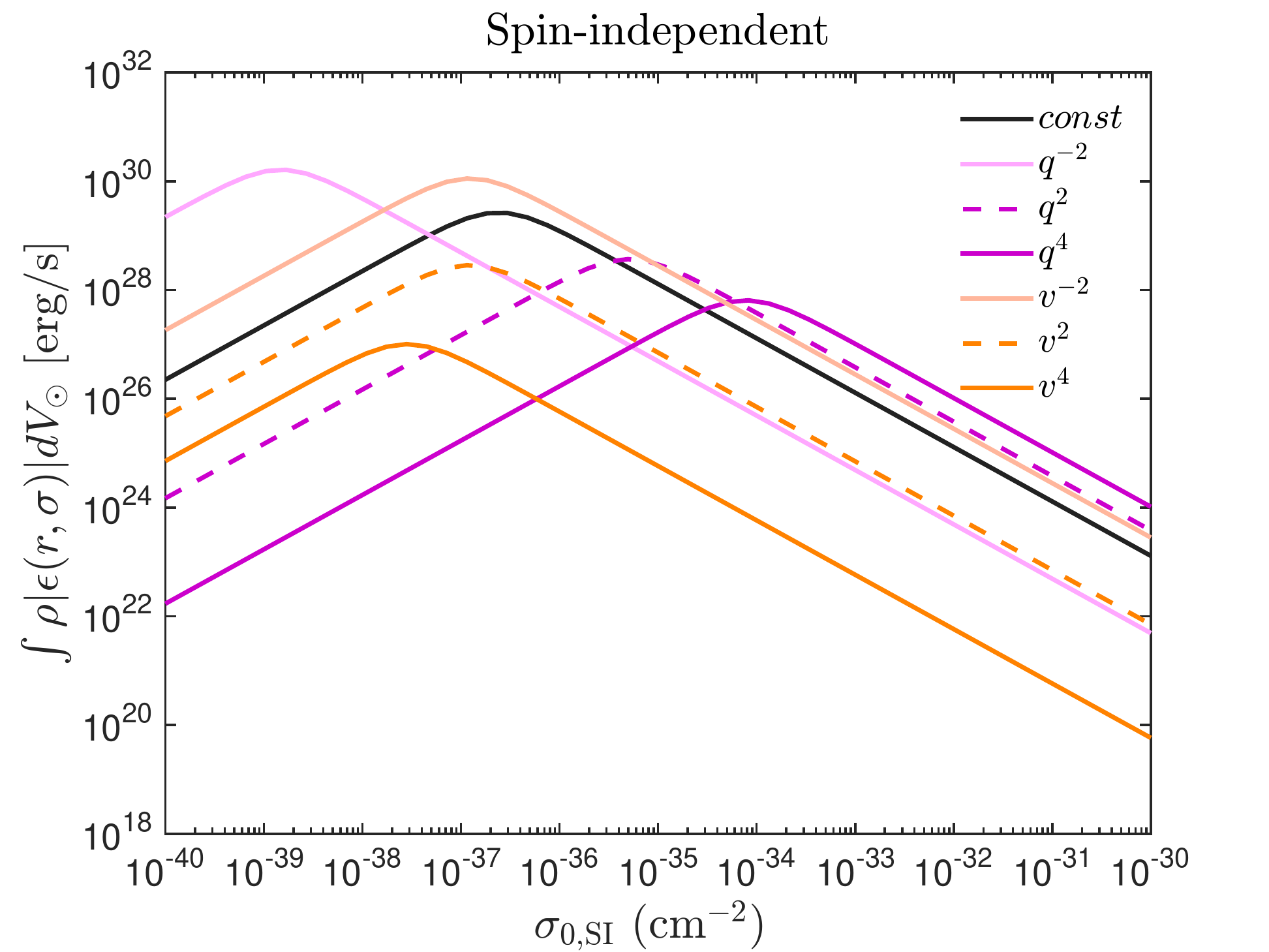}\hspace{0.04\textwidth}\includegraphics[width=0.48\textwidth]{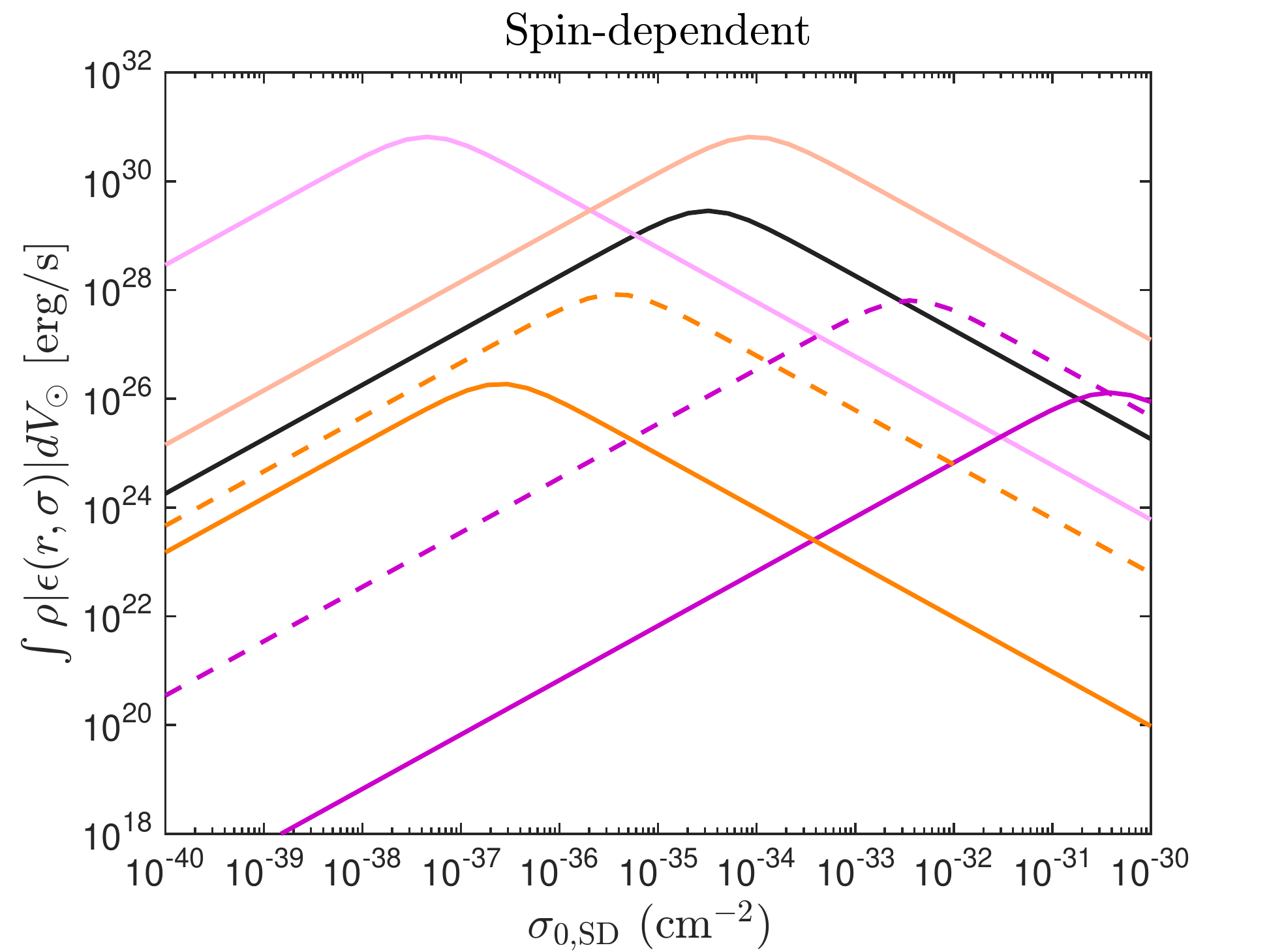}
\caption{The transition from local thermal equilibrium (LTE) to the non-local, isothermal regime of energy transport by spin-independent (\textit{left}) and spin-dependent DM scattering (\textit{right}).   Left of the peaks is the non-local regime, rightwards is the LTE regime.  The total energy transport is plotted for a fixed DM mass ($\mx = 10$\,GeV) and number ratio of DM to baryons ($n_\chi/n_{\rm b} = 10^{-15}$). }
\label{fig:knudtras}
\end{figure}

\begin{figure}[p]
\includegraphics[width=0.48\textwidth]{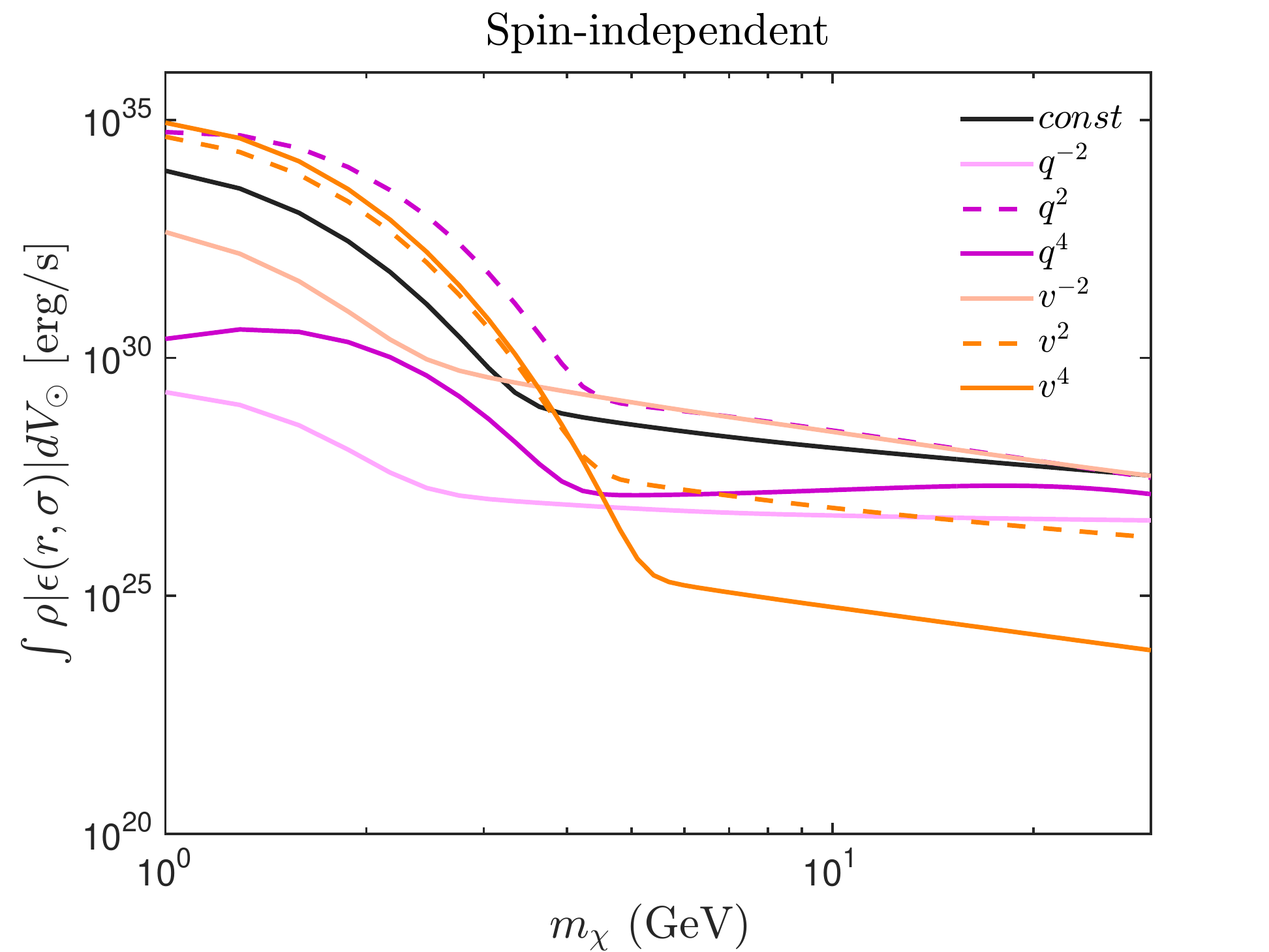}\hspace{0.04\textwidth}\includegraphics[width=0.48\textwidth]{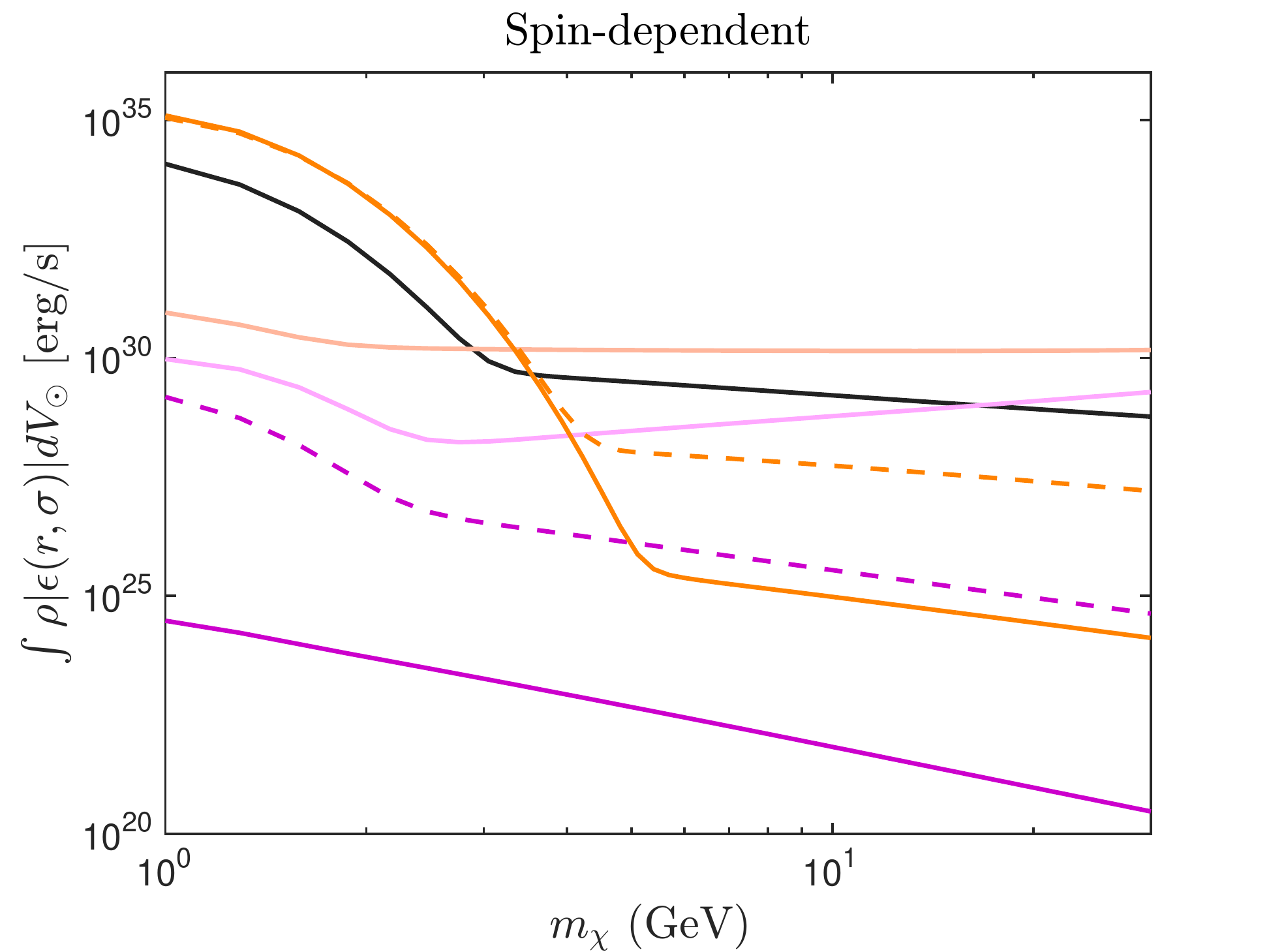}
\caption{Mass dependence of energy transport by spin-independent (\textit{left}) and spin-dependent DM scattering (\textit{right}), for a fixed scattering cross-section ($\sigma_0 = 10^{-35}$\,cm$^{2}$) and number ratio of DM to baryons ($n_\chi/n_{\rm b} = 10^{-15}$).  Peaks are where DM is relatively closely matched in mass to H and/or He.}
\label{fig:knudtras2}
\end{figure}

\begin{figure}[p]
\includegraphics[width=0.48\textwidth]{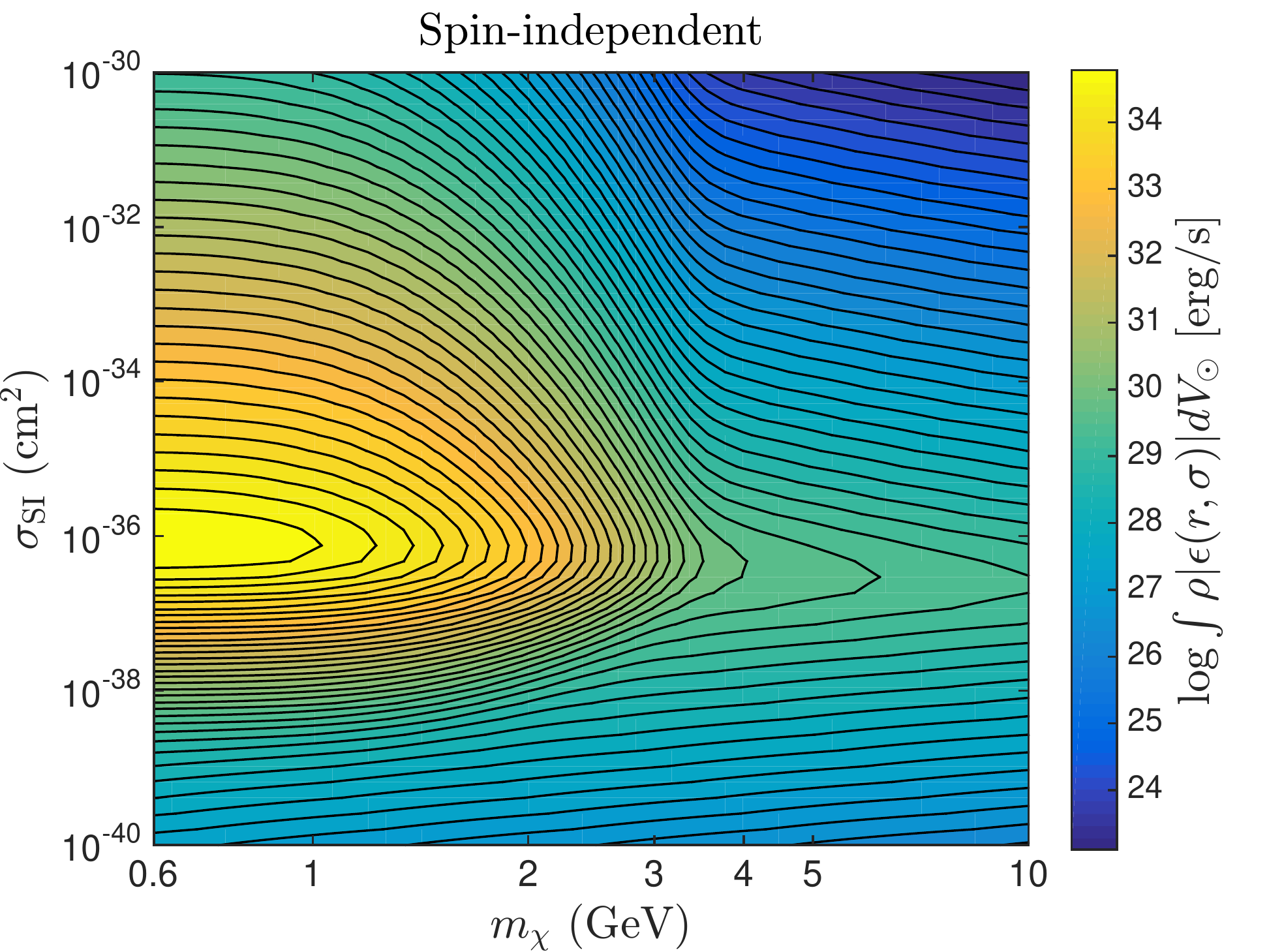}\hspace{0.04\textwidth}\includegraphics[width=0.48\textwidth]{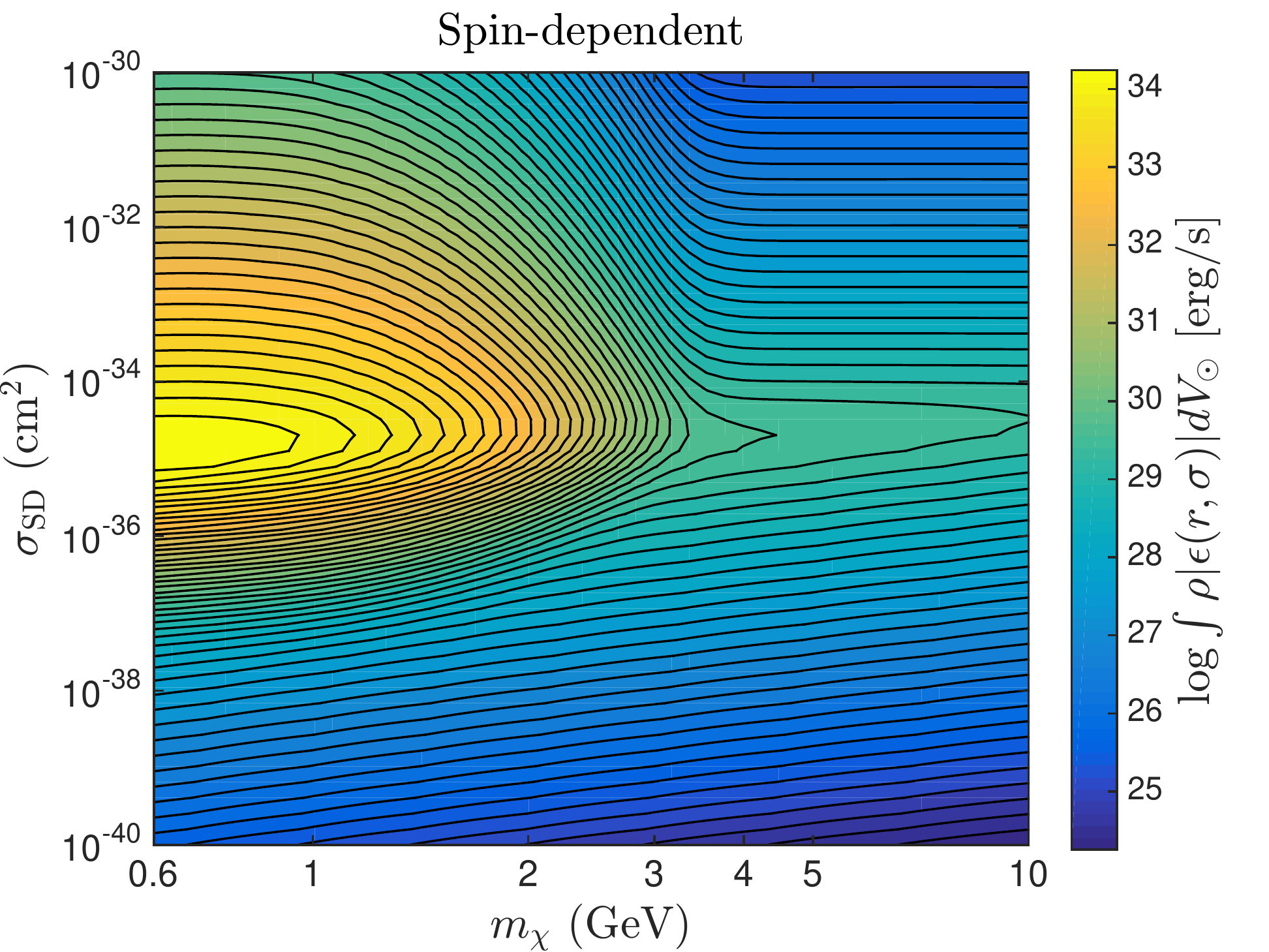}
\caption{Joint dependence on cross-section and mass of energy transport by $q-$ and $v$-independent interactions, assuming a fixed number ratio of DM to baryons ($n_\chi/n_{\rm b} = 10^{-15}$).}
\label{fig:knudtras3}
\end{figure}

\subsection{Conductive energy transport by dark matter}
\label{sec:transport}
In the local thermal equilibrium (LTE) regime, where the mean DM interscattering distance $l_\chi$ is much smaller than the scale of the DM distribution $r_\chi$, the equilibrium distribution of DM particles in the gravitational potential $\phi(r)$ of the Sun is given by \cite{GouldRaffelt90a,Scott09,VincentScott2013}
\begin{equation}
\label{LTEdens}
    n_{\chi,{\rm LTE}}(r) = n_{\chi,\mathrm{LTE}}(0)\left[\frac{T(r)}{T(0)}\right]^{3/2} \exp\left[-\int^r_0 \ud r'\,\frac{k_{\rm B}\alpha(r')\frac{\ud T(r')}{\ud r'} + 
      m_\chi\frac{\ud \phi(r')}{\ud r'}}{k_{\rm B}T(r')}\right],
\end{equation}
where $r = 0$ represents the centre of the Sun. The conductive luminosity is:
\begin{equation}
\label{LTEtransport}
    L_{\chi,{\rm LTE}}(r)= 4\pi r^2 \kappa(r)n_{\chi,{\rm LTE}}(r)l_\chi(r) \left[\frac{k_\mathrm{B}T(r)}{m_\chi}\right]^{1/2}k_\mathrm{B}\frac{\ud T(r)}{\ud r},
\end{equation}
where we have removed the erroneous factor $\zeta^{2n}$ which was present in \cite{VincentScott2013,Vincent:2015gqa}. Below, we will find that this has the effect of displacing the best fit regions in parameter space, to higher (lower) values of $\sigma_0$ for positive (negative) values of $n$. $v_T(r) \equiv \sqrt{2 k_{\rm B} T(r)/\mx}$ is related to the typical thermal velocity \cite{GouldRaffelt90a}; $v_0$ and $q_0$ are respectively the reference velocity and momentum defined in Eq.\ \ref{qdepvdep}. The effect of our correction to Eq.~\ref{LTEtransport} is to suppress transport by models with positive powers of $q$ and $v$, and enhance it for negative powers. The rate of energy transported per unit mass of stellar material is:
\begin{equation}
\label{epsLTE}
    \epsilon_{\chi,{\rm LTE}}(r) = \frac{1}{4\pi r^2 \rho(r)}\frac{\ud L_{\chi,{\rm LTE}}(r)}{\ud r}.
\end{equation}
This quantity is usually expressed in units of erg g$^{-1}$ s$^{-1}$. The molecular diffusion coefficient $\alpha(r,\mu)$ and conduction coefficient $\kappa(r,\mu)$ are computed by perturbatively solving a Boltzmann collision equation in the diffuse gas limit. This method was introduced by Gould and Raffelt \cite{GouldRaffelt90a} and generalised in Ref.~\cite{VincentScott2013}; as before, we use the tabulated coefficients given in the latter reference. 

In the non-LTE, ``Knudsen'' regime $K \equiv l_\chi/r_\chi \gg 1$, the DM distribution becomes isothermal:
\begin{equation}
\label{isodens}
n_{\chi,{\rm iso}}(r,t) = N(t)\pi^{-3/2}r_\chi^{-3}\exp(-r^2/r_\chi^2),
\end{equation}
Gould and Raffelt \cite{GouldRaffelt90a} found that the transition to the LTE regime could be well described by the interpolating functions \cite{Bottino02,Scott09}:
\begin{equation}
\mathfrak{h}(r) = 1 + \left(\frac{r - r_\chi}{r_\chi}\right)^3 \hspace{2cm}\mathfrak f(K) = \left[1+ \left(\frac{K}{K_0}\right)^{1/\tau}\right]^{-1}, \label{fK}
\end{equation}
such that
\begin{equation}
\label{nchi}
n_\chi(r) = \mathfrak f(K) n_{\chi,{\rm LTE}} + \left[1 - \mathfrak f(K) \right] n_{\chi,{\rm iso}},
\end{equation}
\noindent and
\begin{equation}
\label{ltransport}
L_{\chi,\rm total}(r,t) = \mathfrak f(K) \mathfrak h(r,t)L_{\chi,{\rm LTE}}(r,t).
\end{equation}

Energy transport has the opposite behaviour as a function of the interaction cross section in the LTE and Knudsen regimes. In the LTE regime (large $\sigma_0$), a higher scattering rate suppresses the typical distance travelled, confining DM and suppressing energy transport. Conversely, in the Knudsen (small $\sigma_0$) regime, the inter-scattering distance is already large, and energy transfer is instead dominated by the collisional efficiency: luminosity thus increases with increasing cross section. Heat conduction is therefore maximised at the boundary between these regimes, when inter-scattering distances are large, but interaction rates are still large enough to allow efficient energy transfer. We show this behaviour in Fig.\ \ref{fig:knudtras}. It follows that the largest effect of momentum or velocity-dependent dark matter on solar observables will occur when the cross-section is closely matched to the Knudsen peak (as long as a sufficient amount of DM can be captured with this cross section).

In Fig.\ \ref{fig:knudtras2} we show the equivalent for the impact of DM mass, illustrating the strong enhancement of energy transport at low masses due to mass-matching between DM and hydrogen or helium.  This corresponds to transitions in $\alpha$ and $\kappa$ at $m_\chi\sim m_N$ for different interactions \cite{VincentScott2013}, and can also be seen in the fact that transport is maximised at lower masses for SD scattering, reflecting the fact that H plays a role due to its spin, but He does not.  The combination of this effect with the transition from LTE to non-local transport can be seen in Fig.\ \ref{fig:knudtras3}.  Whilst this figure shows contours for momentum and velocity-independent scattering only, other interactions lead to a very similar pattern, just scaled or shifted in the $x$, $y$ and/or $z$ directions.

Finally, as the DM mass is lowered below ~5 GeV, it may pick up sufficient speed from thermal collisions to escape the sun. This can lead to significant loss through evaporation. However, the exact amount of evaporation depends sensitively on the mass and cross section. Since this dependence is not not known, we also explore this region of the parameter space. In an upcoming work \cite{BusoniEvap} we aim to compute these precise evaporation masses, for every model explored here. 

\section{Direct detection bounds}
\label{sec:dd}
Recently, the SuperCDMS collaboration has presented their lowest-threshold analysis ever, \cite{Agnese:2015nto}. This 70\,kg\,day run was sensitive to recoil energies above 56\,eVee (or around 300\,eV nuclear recoil energies), allowing them to set exclusion bounds on DM masses extending below 2 GeV. Though CRESST-II \cite{Angloher:2016jsl} have presented analyses of specific models in this work, we use CDMSlite data because it is both more sensitive and easier to recast into new limits. Since CDMS uses a germanium crystal target, it presents the further advantage of being sensitive to spin-dependent interactions: 7.6\% of natural Ge is in the form of $^{73}$Ge ($J = 9/2$), allowing spin-dependent bounds to be derived with the same data set. In contrast, the CaWO$_4$ target used by CRESST-II presents no appreciable target mass with nonzero spin.

\begin{figure}[h]
\includegraphics[width=0.5\textwidth]{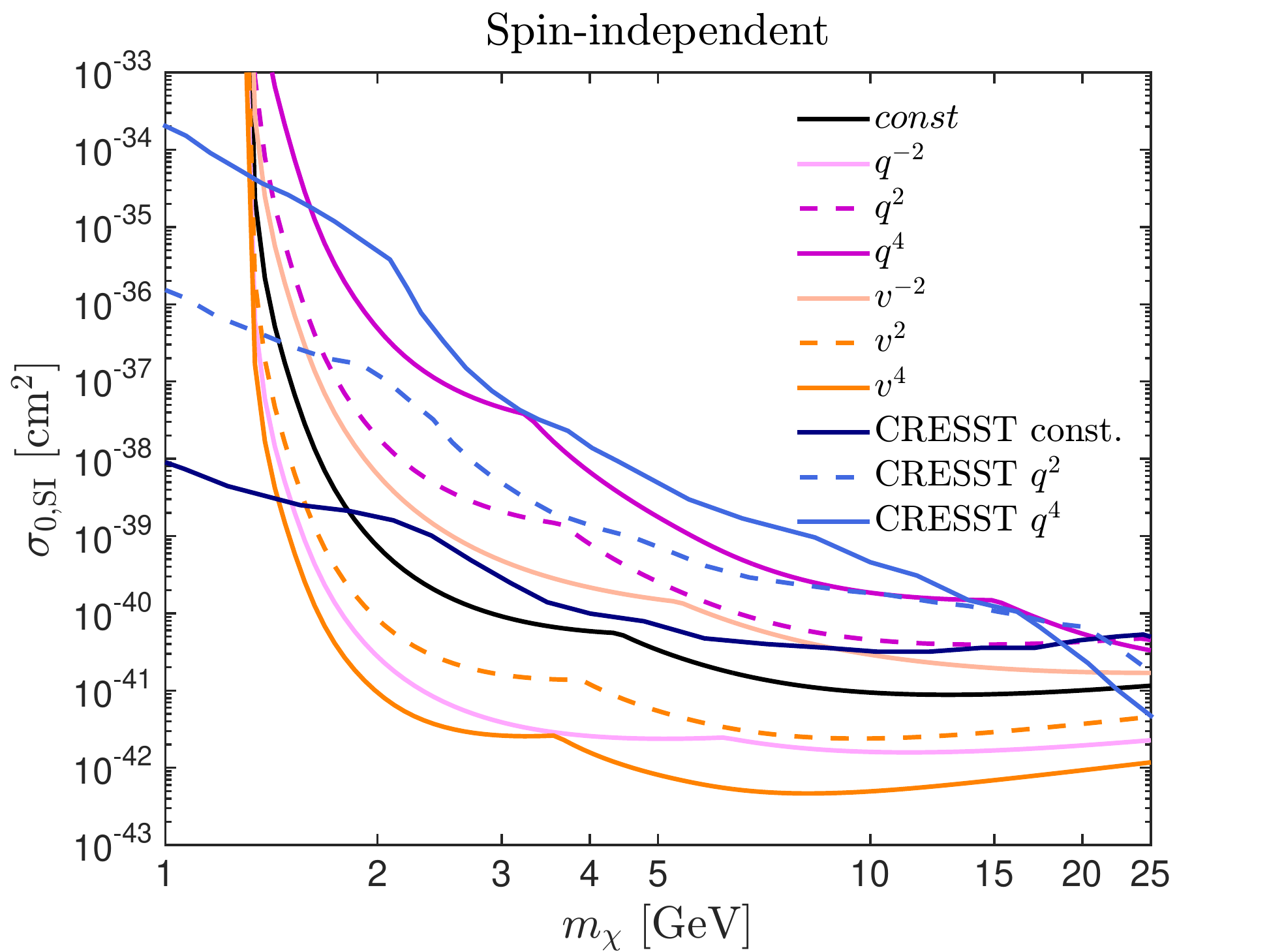} \includegraphics[width=0.5\textwidth]{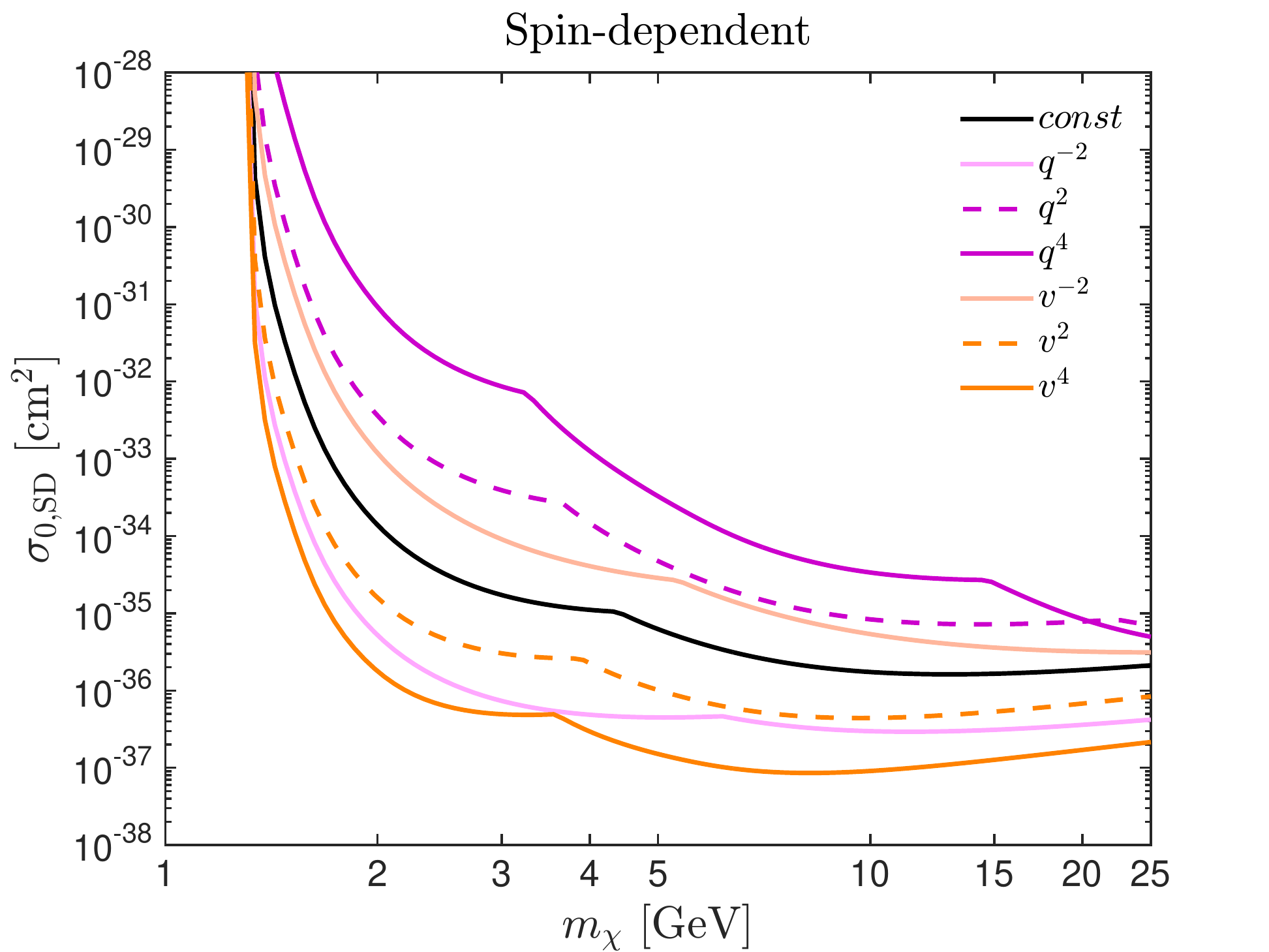} 
\caption{CDMSlite upper limits on $\sigma_0$ for the models that we consider, from the event rates and instrument details given in Ref.\ \cite{Agnese:2015nto}. To compare with our Solar results, we use the same values of $v_0 = 220$ km/s and $q_0$ = 40 MeV. We also show results obtained by the CRESSST-II low-threshold analysis taken from \cite{Angloher:2016jsl}, which are slightly less constraining than CDMSlite for DM masses above 1.4 GeV.}
\label{fig:cdmslite}
\end{figure}

Assuming different SI and SD couplings, but equal couplings to protons and neutrons for each type of interaction (i.e. isoscalar couplings), the differential nuclear recoil rate is  
\begin{align}
\frac{dR}{dE_\mathrm{R}} = \frac{\rho_\chi}{2 m_\chi\mu_\mathrm{p}} \int^\infty_{v_{\rm min}} \frac{f(v)}{v} \bigg(&\sigma_{SI,\mathrm{p}}(q,v)A^2F_{SI}^2(E_\mathrm{R}) + \nonumber\\
&\sigma_{SD,\mathrm{p}}(q,v)\frac{4(J+1)}{3J}\left[\langle S_\mathrm{n} \rangle + \langle{S_\mathrm{p}}\rangle\right]^2F_{SD}^2(E_\mathrm{R})\bigg) dv,
\end{align}
where $\mu_\mathrm{p}\sim\mu_\mathrm{n}$ is the DM-nucleon reduced mass and $v_{\rm min} \equiv \sqrt{m_NE_R/2\mu_N^2}$ is the minimum DM velocity required to produce a recoil with energy $E_R$. In this expression, $\mu_N$ is the DM-nucleus reduced mass. 
Using the binned data, efficiency curve and ionisation yield presented in Ref.\ \cite{Agnese:2015nto} and a simple Poisson analysis, along with the same halo parameters as in our solar analysis, we obtain the exclusion curves presented in Fig.\ \ref{fig:cdmslite}. For SI limits (left panel) we use the same form factor as for solar capture.  For the spin-dependent limits (right panel), we assume scattering only on 7.6\% of the target, and use the spin expectations and isoscalar structure function $S_{00}$ from Ref.\ \cite{Klos:2013rwa}, such that
\begin{equation}
\frac{4(J+1)}{3J}\left[\langle S_\mathrm{n} \rangle + \langle{S_\mathrm{p}}\rangle\right]^2F_{SD}^2(E_\mathrm{R}) = \frac{16\pi}{3(2J+1)}S_{00}(E_\mathrm{R}).
\end{equation}
Although we show limits for isoscalar couplings only, it is worth noting that most of the unpaired spin contribution from $^{73}$Ge comes from neutrons, so limits on ADM that couples mainly to protons would be significantly weaker.  The SI limits in Fig.\ \ref{fig:cdmslite} are slightly more constraining than the limits presented by CRESST-II \cite{Angloher:2016jsl}, for the spin-independent $q^2$ and $q^{4}$ cases.
\FloatBarrier

\section{Solar bounds}
\label{sec:results}
To produce our updated constraints, we employ the \DS code, developed for and described in Ref.~\cite{Vincent:2015gqa}. \DS is a combination of the legendary \textsf{GARSTEC} solar evolution code \cite{weiss:2008, Serenelli11}, and the lightweight capture and transport routines from \textsf{DarkStars} \cite{Scott08a, Scott09, Scott09b}. These were modified to include the full capture and transport technology described above, for momentum and velocity dependent dark matter. To obtain our results, we perform a scan over a grid of masses $m_\chi = \{3, 5, 10, 15, 20, 25\}$ GeV and one cross section per decade from $\sigma_{0} = 10^{-40}$ to $10^{-30}$ cm$^2$, totalling over 900 simulations, or approximately 2.5 CPU years. 

A full description of the observables that we use is given in Ref.\ \cite{Vincent:2015gqa}. Here we summarise the salient points. Solar neutrino fluxes and small frequency separations are highly sensitive to the impacts of DM on the core of the Sun, and typically provide the strongest constraints on DM models.  The depth of the convection zone is sensitive to impacts of DM on the temperature gradient closer to the surface, and generally depends more subtly on the overall diffusivity of DM in the Sun. The surface helium abundance and radial profile of the sound speed profile probe the entire radiative region, but contribute comparatively little additional constraining power over other observables.

 Plots of the neutrino fluxes, depth of the convection zone, sound-speed and small-separation likelihoods vs $m_\chi$ and $\sigma_0$ are presented in Appendix \ref{app:results}.  We also provide the full results of all our simulations, including each of these observables, their likelihoods and the combined likelihood presented in Sec.\ \ref{sec:combined}, as machine-readable supplementary material. 

\begin{figure}
\centering
\includegraphics[width=0.8\textwidth]{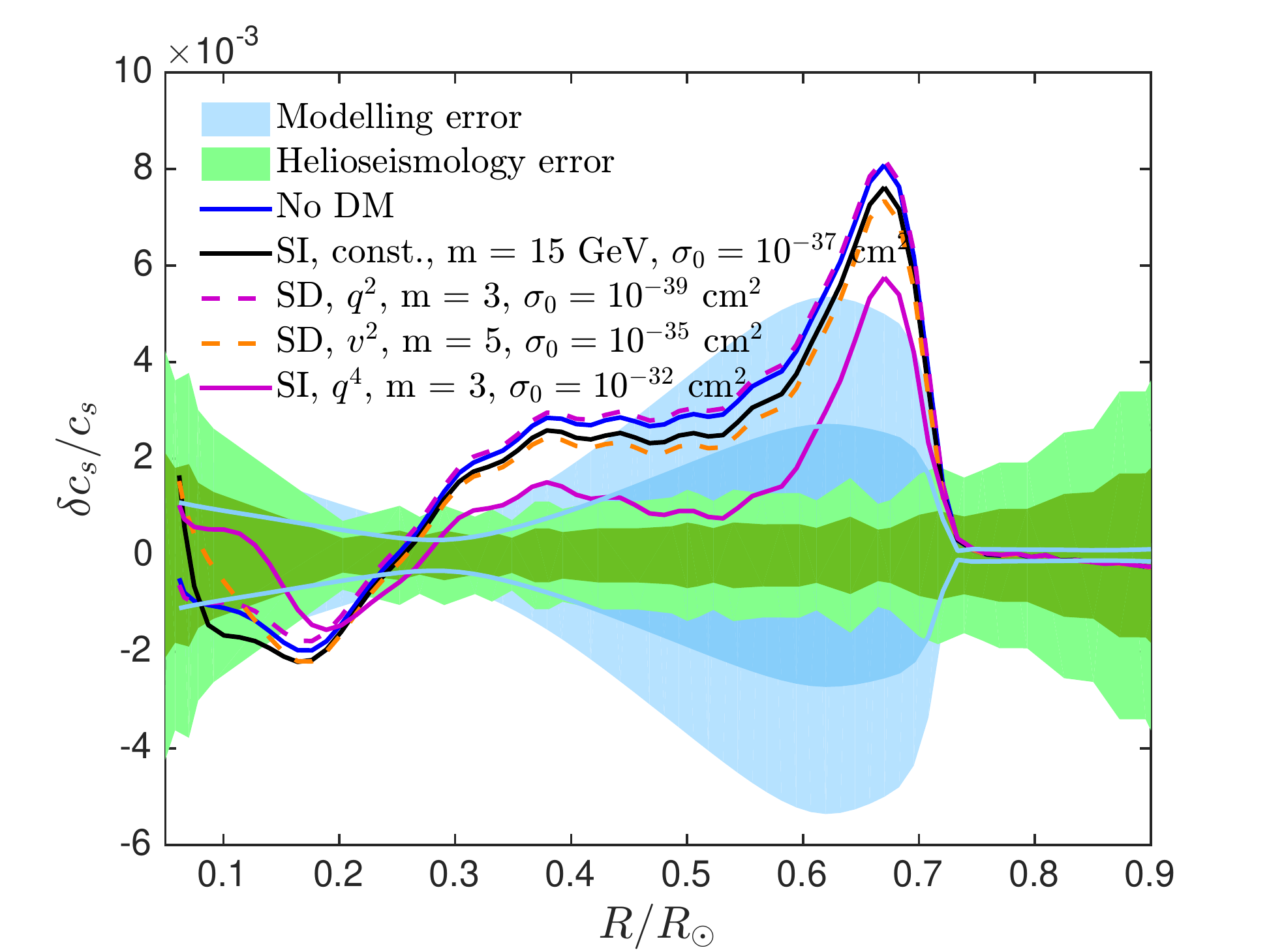}
\caption{Deviation of radial sound speed profile from values inferred from combined SoHO and BiSON helioseismology data \cite{basu:2009} using our SSM. We show the best fit for a constant cross section in black, and the two best fits that are also allowed by CDMSlite data, the spin-dependent $v^2$ and $q^2$ models (indicated by daggers in Table \ref{best_fit_tab}). We finally show the best overall fit, a spin-independent $q^4$ DM model. However, this model is ruled out by direct detection data by many orders of magnitude. The dark and light blue bands represent the 1 and 2$\sigma$ errors on modelling, estimated from error propagation of  uncertainties in SSM inputs; the green bands represent the 1 and 2$\sigma$ errors on helioseismological inversions.}
\label{fig:cs}
\end{figure}

\begin{figure}
\includegraphics[width=.5\textwidth]{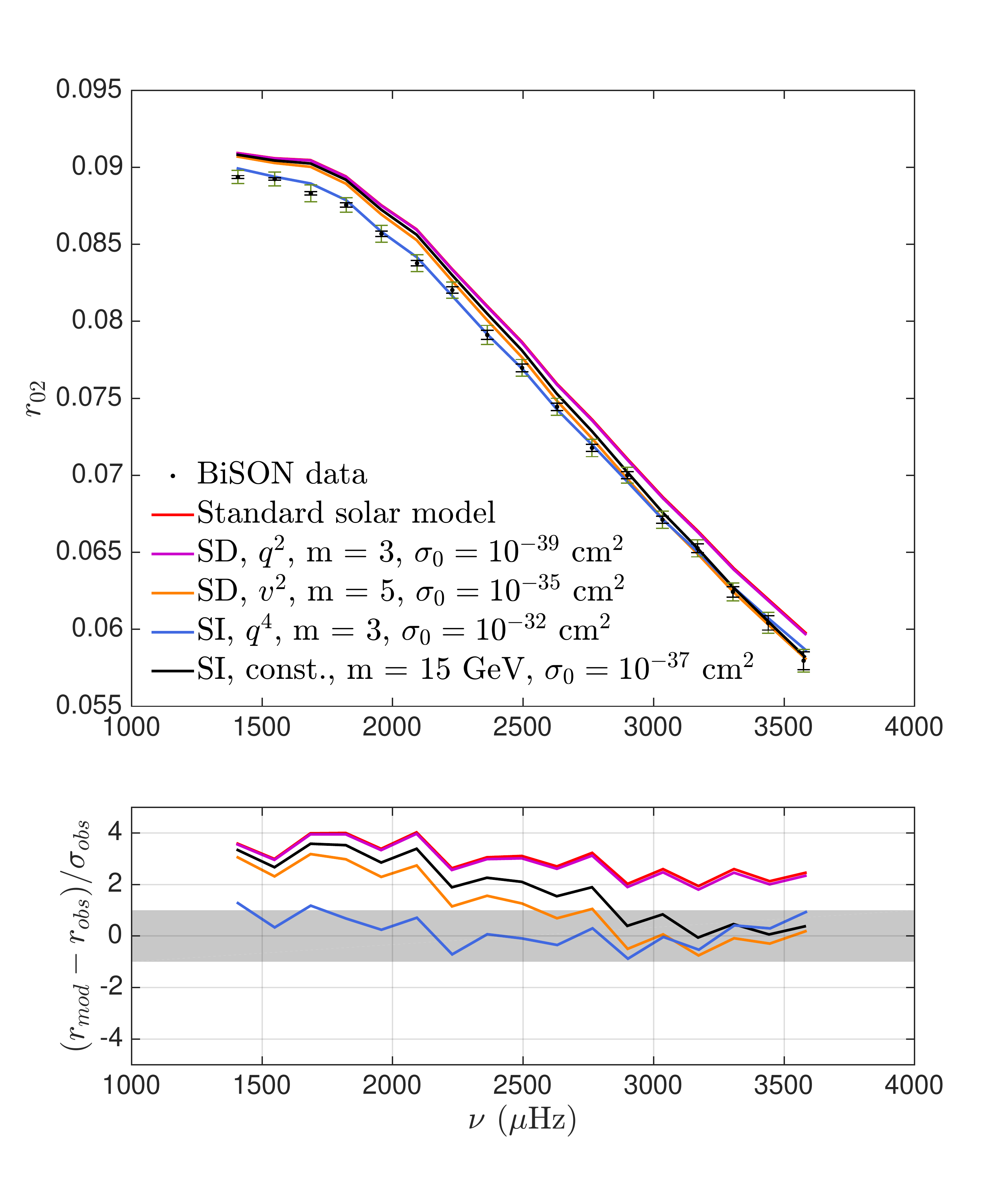} \includegraphics[width=.5\textwidth]{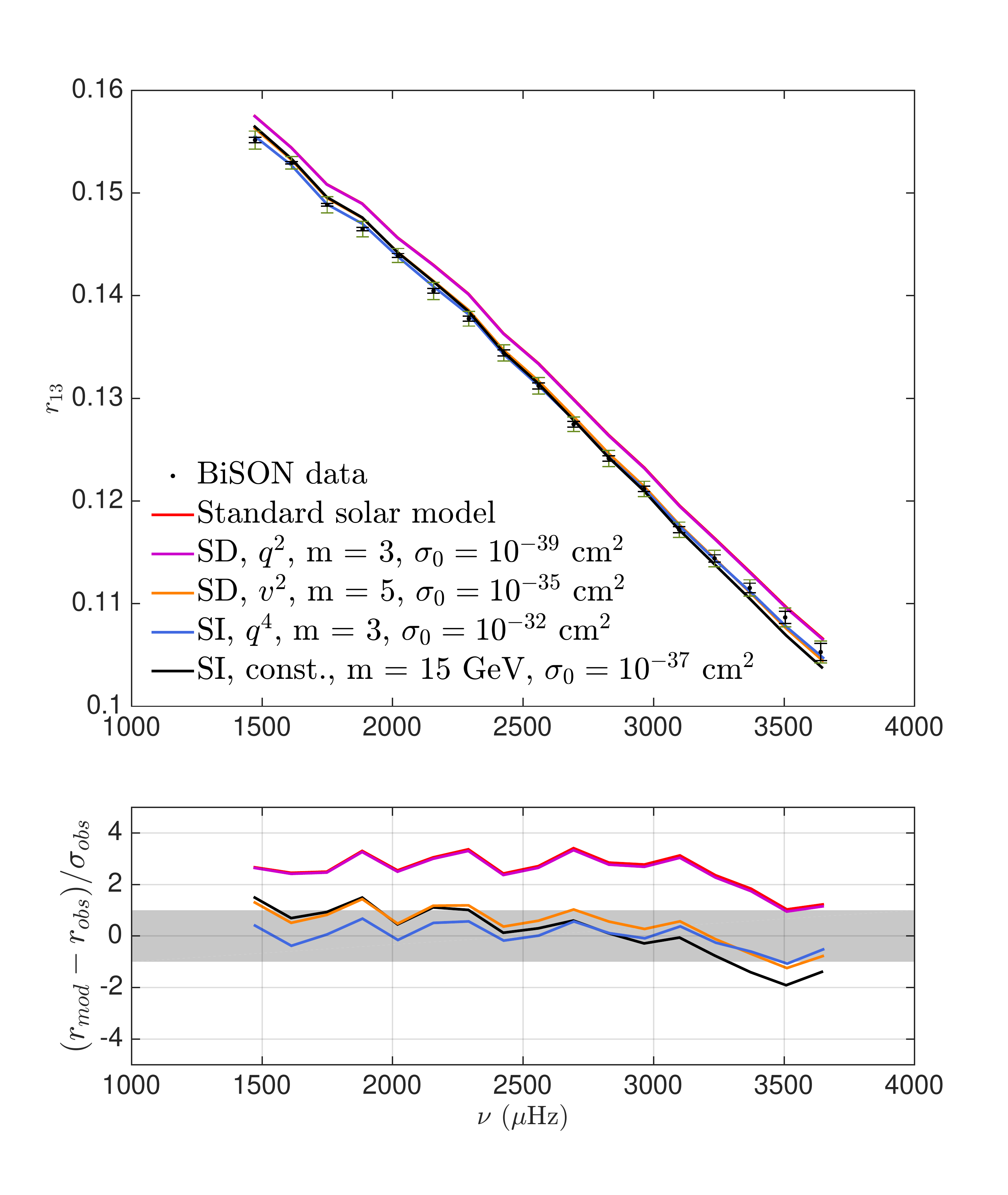}
\caption{Small frequency separations $r_{02}$ (\textit{left}) and $r_{13}$ (\textit{right}) as defined in Eq.\ \ref{eq:rdef}, for various dark matter models. DM models correspond to the best-fit parameters for two couplings returning the best overall $p$-values (see Table \ref{best_fit_tab}), without being excluded by CDMSlite, along with the best constant case (black), and the overall best fit case (pink).  The latter two are however excluded by direct detection. Predictions are compared with helioseismological observations from the BiSON experiment \cite{Basu:2006vh}. Inner black error bars correspond to observational error, whereas outer (green) bars also include modelling error. Below each figure we show the residuals with respect to BiSON data, in units of the total error. Though the $v^2$ model with $\mx = 5$ GeV yields substantial improvement in $r_{13}$ for all frequencies, it does less well at low frequencies ($\lesssim 2500$ Hz) with respect to the measured $r_{02}$. }
\label{fig:rij}
\end{figure}

Our reference SSM uses the AGSS09 photospheric abundances \cite{AGSS}.
\begin{description}
\item[Solar neutrino fluxes:] Minute changes in the core temperature can have substantial effects on the solar neutrino fluxes, due to their steep temperature-dependence. The main fusion process in the Sun, $p+p \rightarrow  {^2\mathrm{H}} + e^+ + \nu_e$, is quite insensitive to changes due to DM energy transport, as the luminosity condition enforces a fairly constant reaction rate. However, the subdominant flux of neutrinos from the decay of $^8$B  produced in the $pp$ and $pep$ chains is an especially good probe of the core temperature, as are the two lines at $E_\nu = 862$ keV and 384 keV from the decay of $^7$Be. While these are measured at 3\% and 5\% accuracy, respectively, the uncertainties in their production rates are closer to 14\% and 7\%. We add these errors in quadrature when assessing the goodness of fit. As the SSM is in very good agreement with the measured neutrino fluxes, this serves as a constraint on the total effect DM can have in the solar core. The changes in neutrino fluxes in different models are shown in Figs.\ \ref{SIboronfluxes}-\ref{SDBerylliumfluxes}. When uncertainties are compared on the same footing, our constant, SD results are in agreement with \cite{Taoso10}. 
\item[Depth of the convection zone:] The radius at which hydrostatic equilibrium breaks down, radiative pressure is no longer balanced by gravity, and convective energy transport begins, depends crucially on the radial temperature and pressure gradients. The height of the convection zone can be inferred from helioseismological measurements, $R_{\mathrm{CZ},\odot} = 0.713 \pm 0.001\,R_\odot$, while the SSM predicts a much higher value: $R_{\rm CZ,SSM} = 0.722 \pm 0.004\,R_{\odot}$.  As DM deposits energy at larger radii, the absolute temperature gradient increases slightly, thus lowering the location of the base of the convection zone. This is shown in Figs~\ref{SIrc} and \ref{SDrc}.
\item[Surface helium abundance:] As the initial mass fractions of hydrogen, helium and metals cannot be directly measured, they (along with the mixing length parameter $\alpha_{MLT}$) are allowed to vary as \DS attempts to find a solution for the input parameters. The present-day surface helium abundance is thus predicted by these parameters, when combined with the physics entering into the SSM. The measured value is $Y_s = 0.2485 \pm 0.0034$, while the SSM yields $Y_s =0.2356 \pm 0.0035$. The surface helium abundance essentially reflects the initial helium abundance, which is strongly constrained by the requirement that the model must reproduce the observed solar luminosity.  Any change in initial helium abundance must come with a corresponding change in the initial hydrogen fraction, which modifies the resulting solar luminosity unless the core temperature is also drastically altered.  In extreme cases, DM can have enough of an effect on the core temperature to change the resulting surface helium abundance expressed by a model -- so we do include this observable in our fits -- but typically, the effects on the core temperature are not drastic enough for this observable to be significantly affected.
\item[Sound speed:] The radial sound speed profile can be inferred by inverting the oscillation frequencies of the different angular modes projected onto the solar surface. Fig.\ \ref{fig:cs} shows the sound speed profile obtained with our SSM (blue solid line), when compared with the errors from helioseismological measurement and inversions (green regions) and modelling (blue regions). The change in the temperature, pressure and density due to the redistribution of heat by DM can be quite large -- as long as the  cross section is large enough to saturate the capture rate and hit the Knudsen transition. Fig.\ \ref{fig:cs} also shows the effect of three models which are allowed by DD constraints, along with one which is not (SI, $q^4$, solid magenta line), but which illustrates the strong effect DM can have. We also show in Figs.\ \ref{SIcschsq} and \ref{SDcschsq} the goodness of fit of each of our models to the sound speed profile from helioseismology observations with SoHO and BiSON \cite{basu:2009}, using a chi-squared with 5 equally-spaced points between $R = 0.1 R_\odot $ and $0.67 R_\odot$, where helioseismic errors are minimal. Given the large modelling errors, however, and the fact that $c_s$ is highly correlated with the more precise frequency separation ratios, we do not include the sound speed in our total $\chi^2$. 
\item[Frequency separation ratios:] It is possible to remove a large fraction of the systematic errors that enter sound speed inversions near the core by directly using specific combination of frequencies. Two most commonly used quantities are the so-called small frequency separation ratios \cite{roxburgh:2003}: 
\begin{equation}
r_{02}(n) = \frac{d_{02}(n)}{\Delta_1(n)}, \, \, \, \, r_{13}(n) = \frac{d_{13}(n)}{\Delta_0(n+1)},
\label{eq:rdef}
\end{equation}
where $\Delta_l(n) \equiv \nu_{n,l} - \nu_{n-1,l}$ and
\begin{equation}
d_{l,l+2}(n) \equiv \nu_{n,l} - \nu_{n-1,l+2} \simeq -(4l + 6) \frac{\Delta_l(n)}{4 \pi^2 \nu_{n,l}}\int_0^{R_\odot} \frac{dc_s}{dr}\frac{dr}{r}.
\label{eq:ddef}
\end{equation}
These have the additional advantage of being insensitive to details in the structure of the external layers of the Sun, which are not properly modeled in SSMs. The $1/r$ dependence ensures sensitivity to the properties of the solar core, and the modeling and observational uncertainties on $r_{02}$ and $r_{13}$ are much smaller than for Solar neutrinos. These frequency separation ratios are shown in Fig.\ \ref{fig:rij} for the SSM (red), and several models including energy transport by ADM. 
\end{description}

%
\subsection{Combined constraints}
\label{sec:combined}
We define a combined chi-squared, including all of the quantities listed above except the sound speed profile:
\begin{eqnarray}
\chi^2 &=& \frac{(1 - \phi^\nu_{\rm B,obs}/\phi^\nu_{\rm B})^2}{\sigma_{\rm B}^2} + \frac{(1 - \phi^\nu_{\rm Be,obs}/\phi^\nu_{\rm Be})^2}{\sigma_{\rm Be}^2} + \frac{(r_{\rm CZ}  - r_{\rm CZ,obs} )^2}{\sigma_{\rm CZ} ^2} + \frac{(Y_{\rm S} - Y_{\rm S,obs} )^2}{\sigma_{Y_{\rm S}}^2} \nonumber \\
&&+ \chi^2_{r_{02}} + \chi^2_{r_{13}} ,
 \label{eq:fullchisq}
\end{eqnarray}
where $\phi^\nu_{\rm B}$ and $\phi^\nu_{\rm Be}$ correspond to the boron-8 and beryllium-7 neutrino fluxes, $r_{\rm CZ}$ is the depth of the convection zone, $Y_S$ is the surface helium abundance, and $r_{ij}$ are the small frequency separation ratios. The sound speed profiles are strongly correlated with the small separations, and less precise, so we do not use them. The subscript ``obs'' indicates the observed value. The errors $\sigma_i$ include both modelling and observational error, added in quadrature. 

\begin{figure}[p]
\begin{tabular}{c@{\hspace{0.04\textwidth}}c}
\multicolumn{2}{c}{\includegraphics[height = 0.32\textwidth]{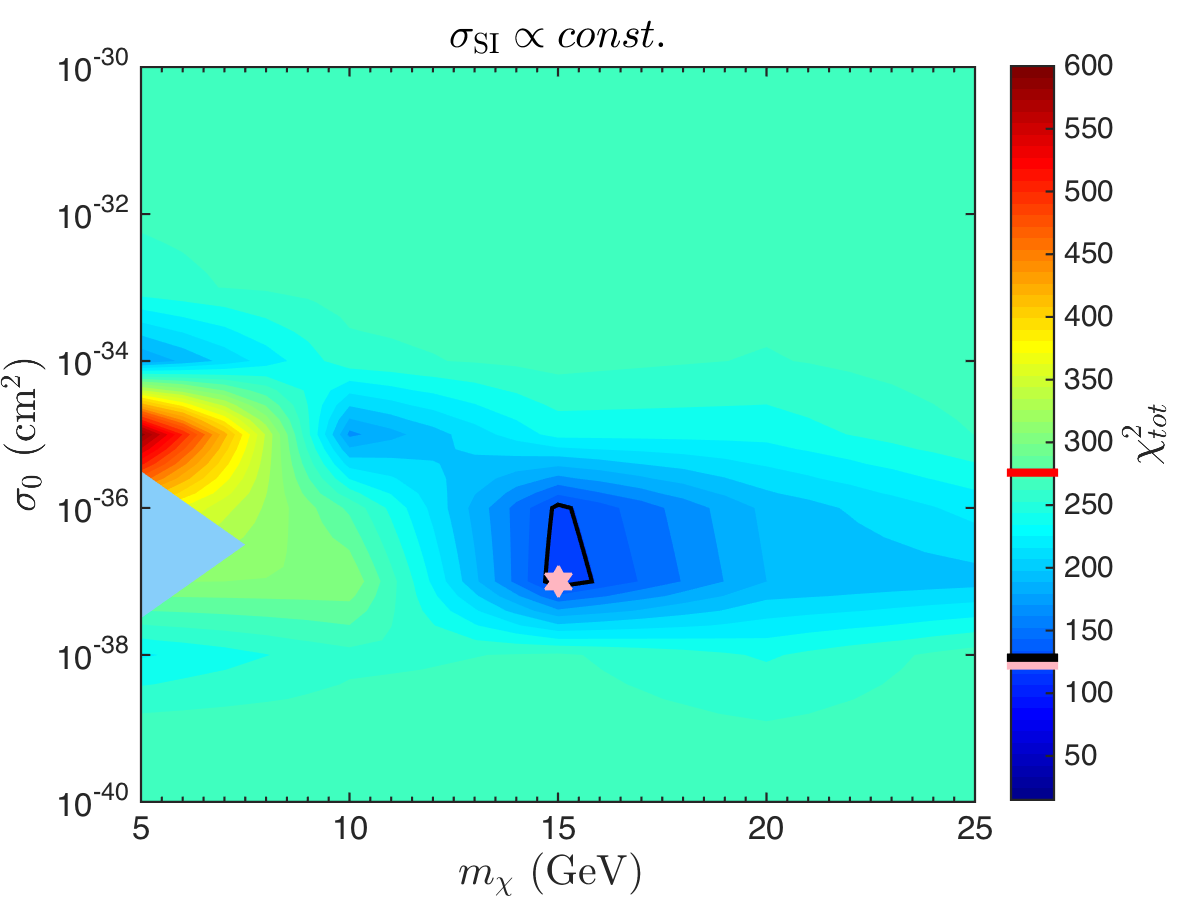}} \\
\includegraphics[height = 0.32\textwidth]{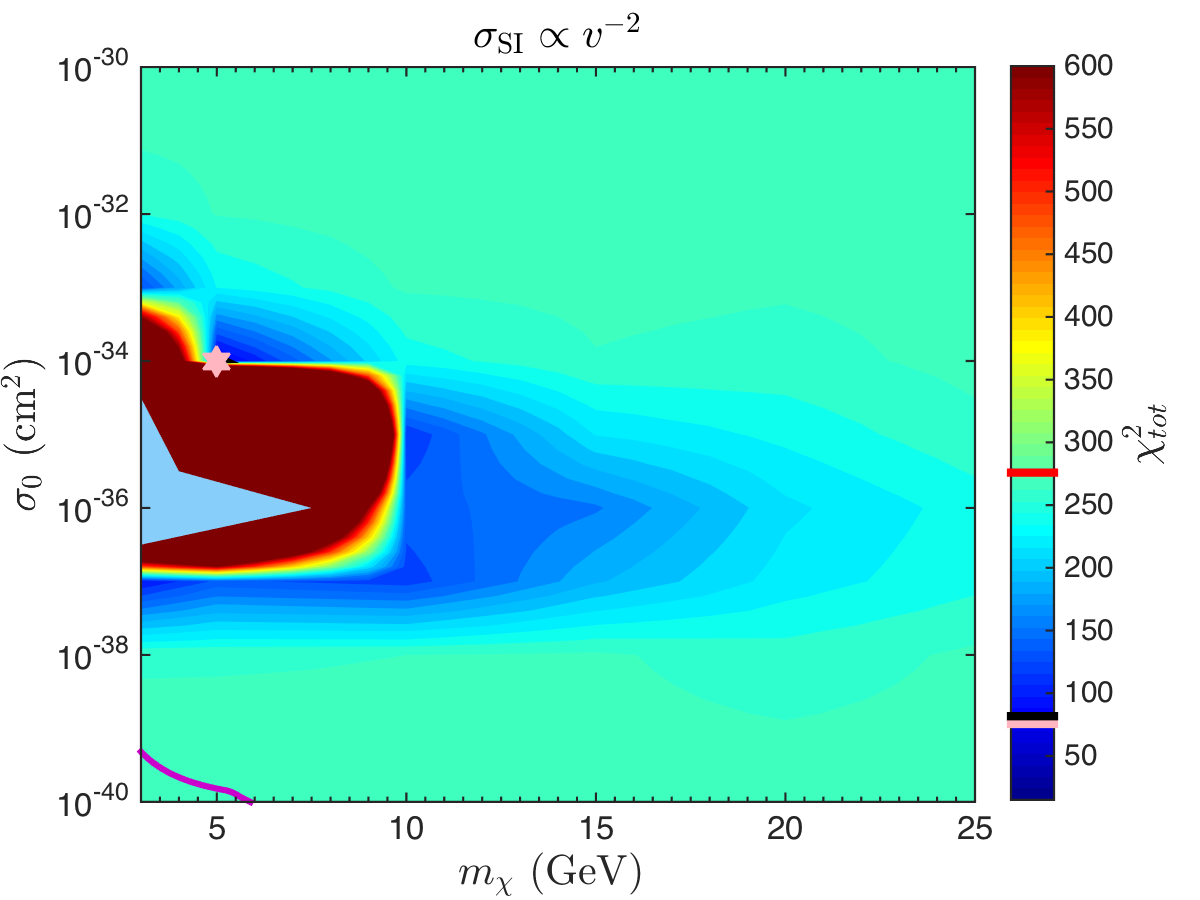} & \includegraphics[height = 0.32\textwidth]{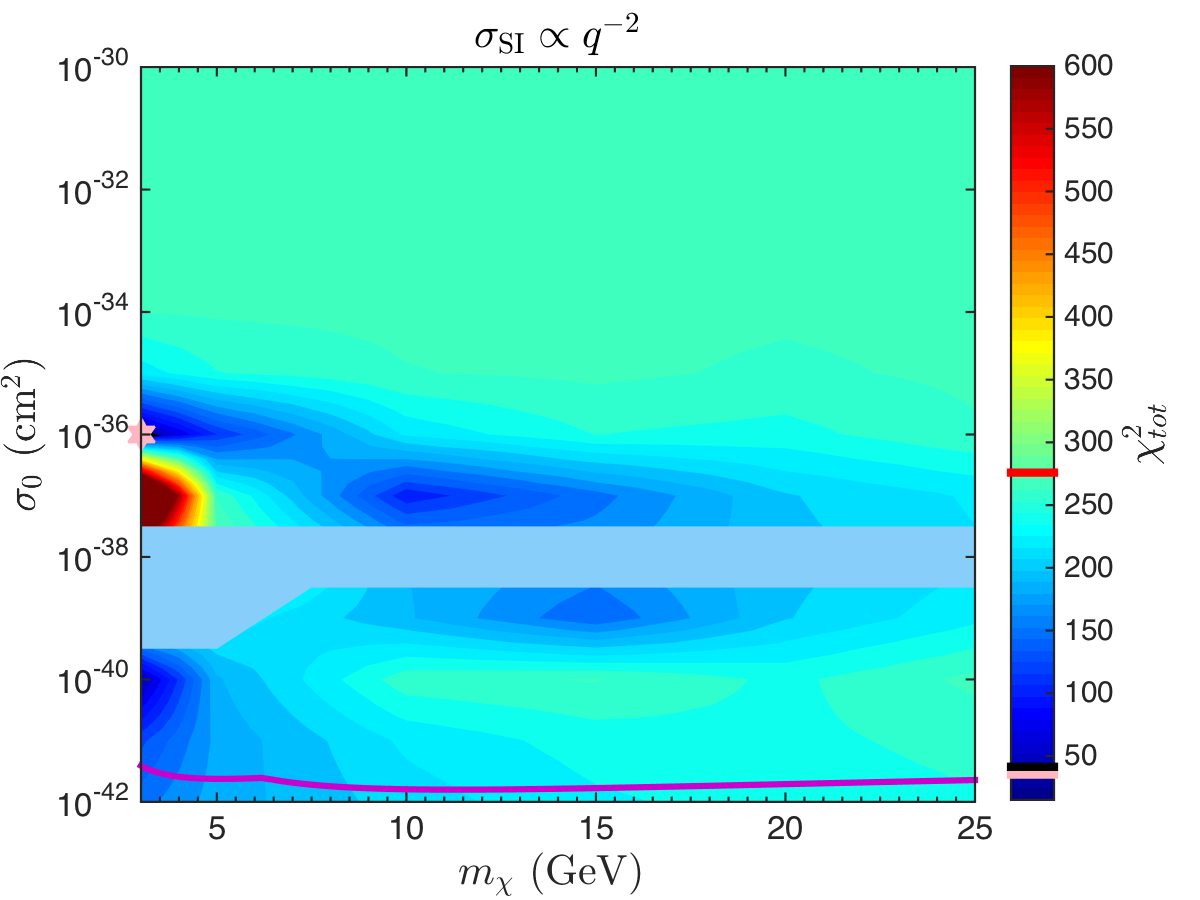} \\
\includegraphics[height = 0.32\textwidth]{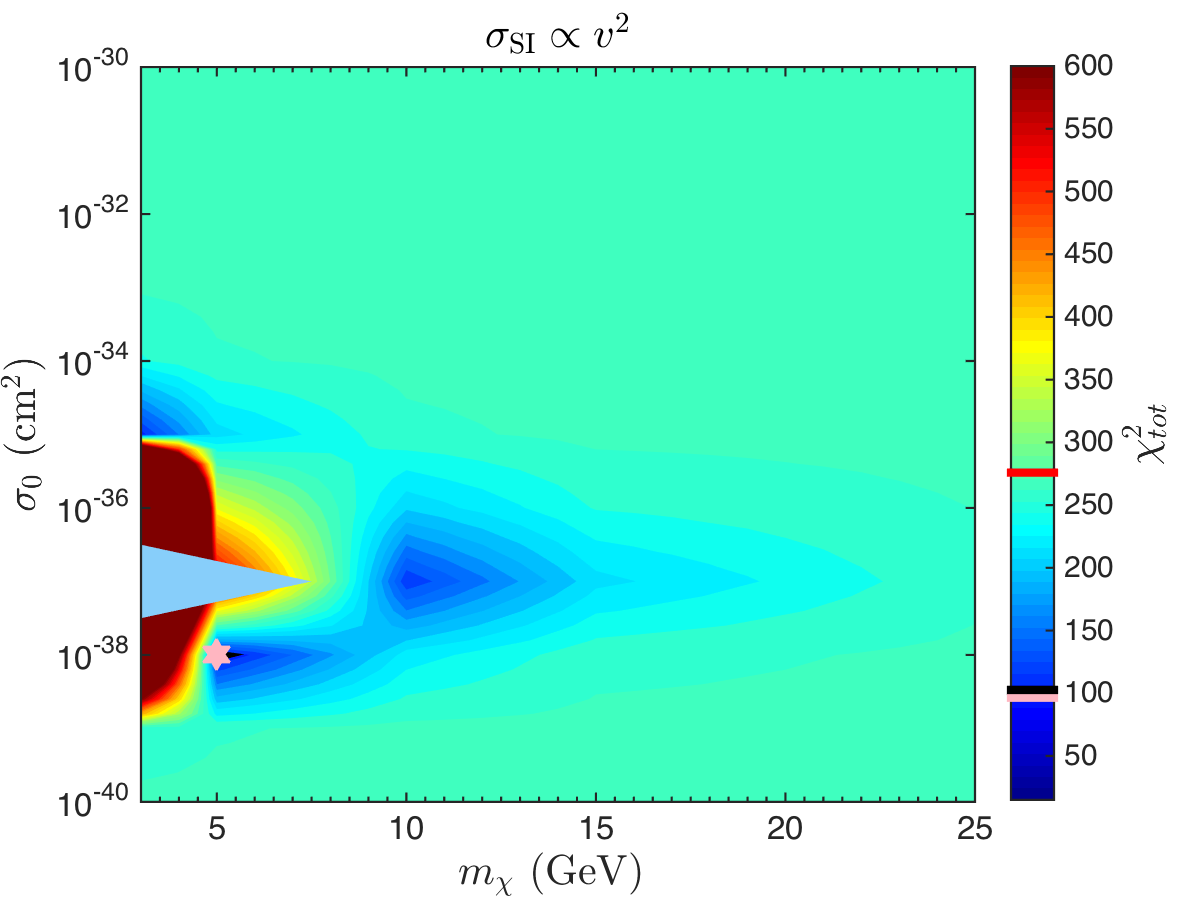} & \includegraphics[height = 0.32\textwidth]{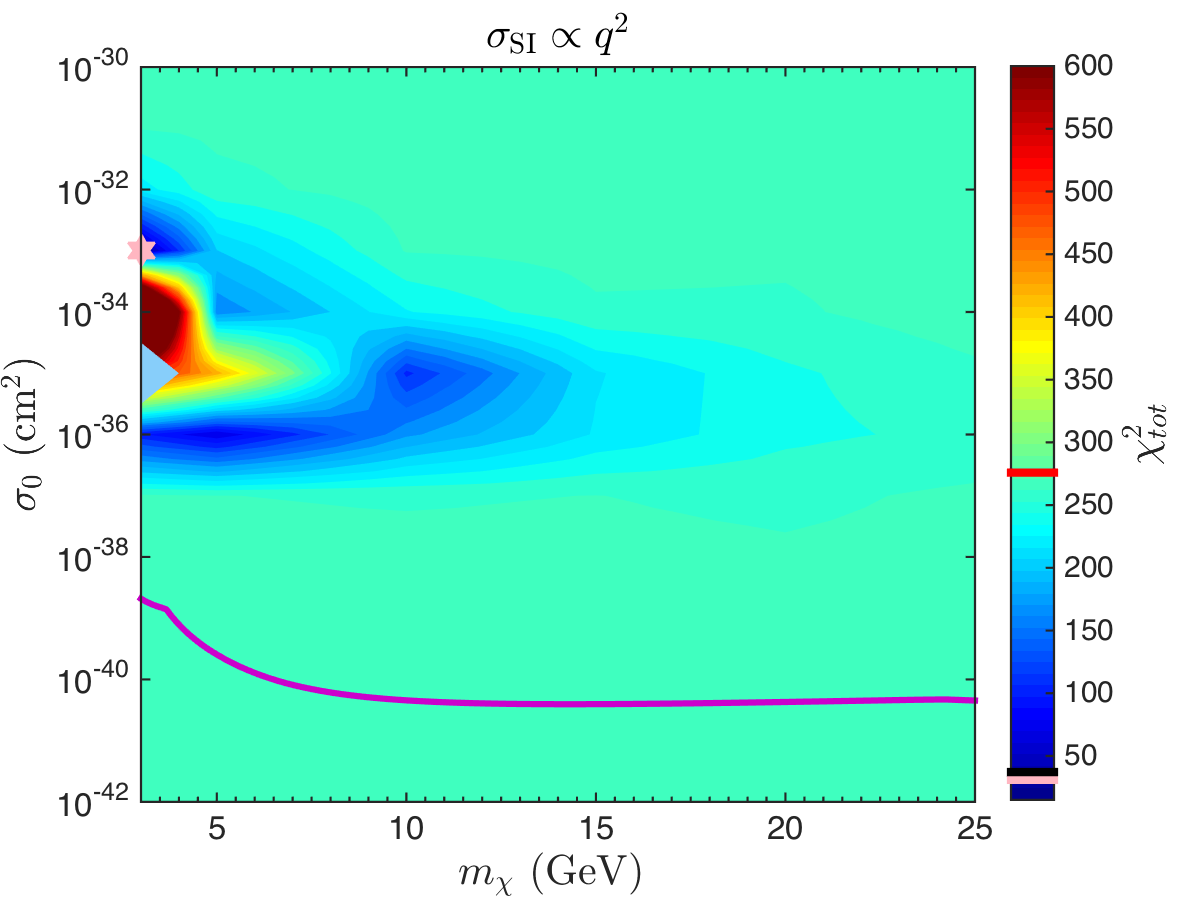} \\
\includegraphics[height = 0.32\textwidth]{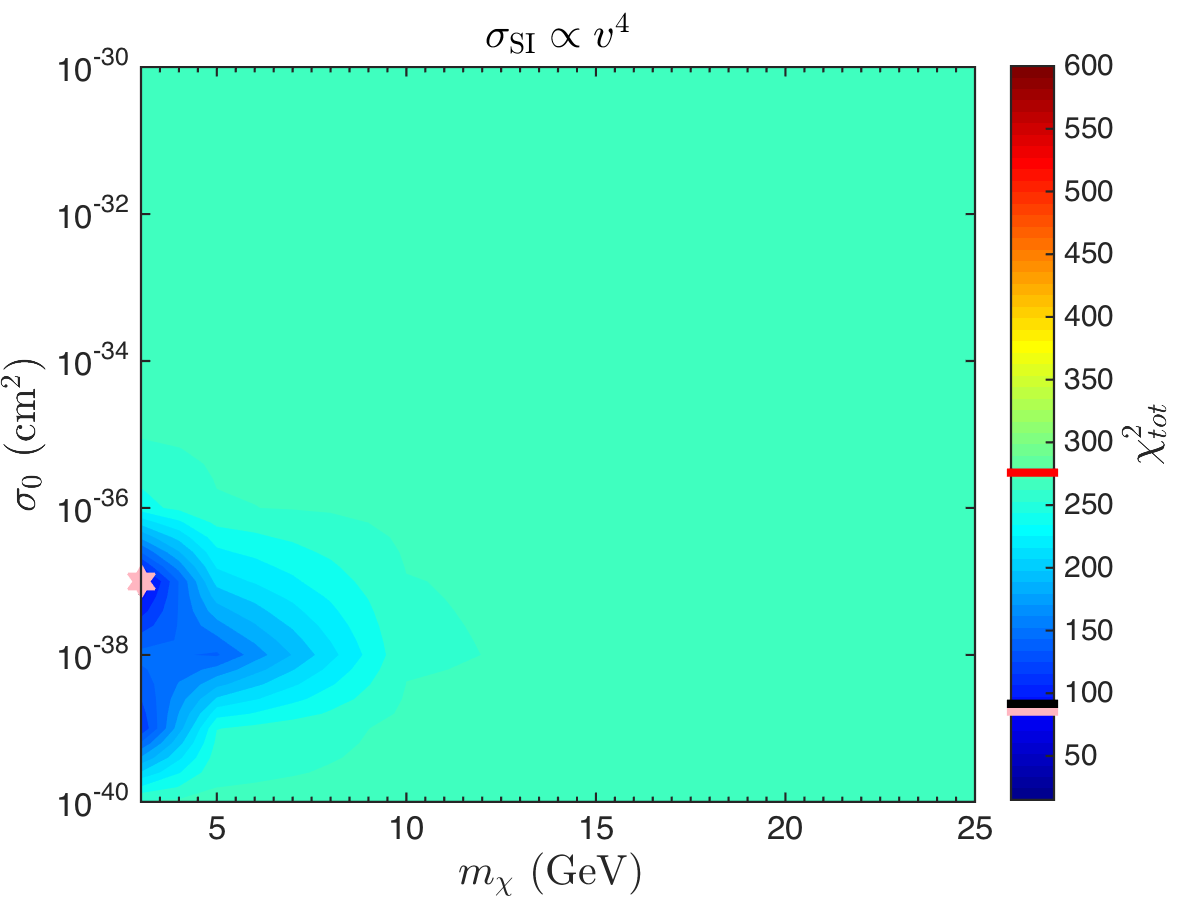} & \includegraphics[height = 0.32\textwidth]{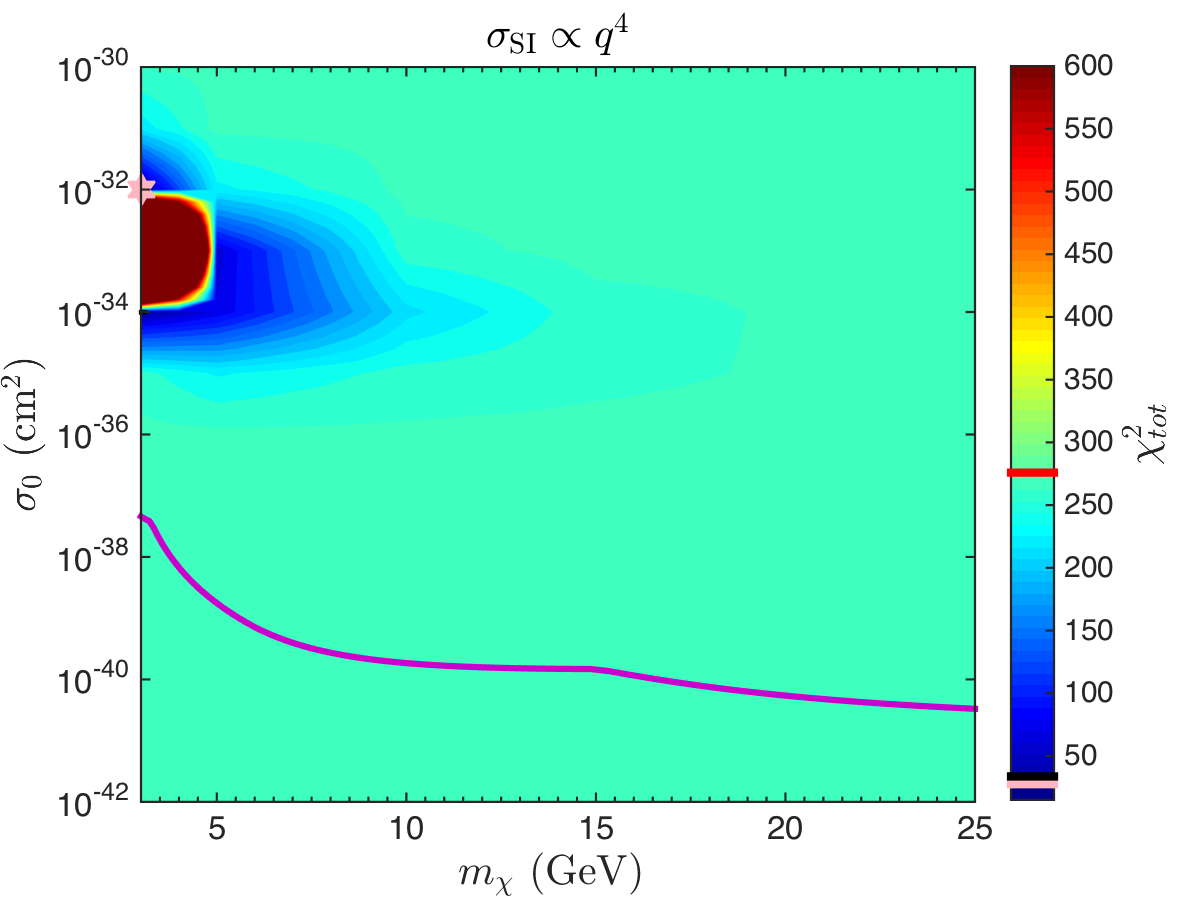} \\
\end{tabular}
\caption{Composite likelihood (Eq.\ \ref{eq:fullchisq}) including $^8$B and $^7$Be neutrino flux measurements, surface helium abundance $Y_{\rm S}$, depth of the convection zone and small frequency separations. Cross-sections are normalised such that $\sigma = \sigma_0(v/v_0)^{2n}$ or $\sigma = \sigma_0(q/q_0)^{2n}$, with $v_0 = 220$\,km\,s$^{-1}$ and $q_0 = 40$\,MeV. Best fits are shown as pink stars and pink lines on colour bars. 3$\sigma$ deviations from the best fits are marked in black, and red lines show the $\chi^2$ value of the SSM. Magenta lines correspond to the CDMSlite 90\% upper limits shown in Fig.\ \ref{fig:cdmslite}. Panels in which the CDMS line is not seen are completely ruled out. Simulations carried out in the masked regions did not converge, due to the significant heat conduction by the DM particles, leading in extreme cases to density inversions in the core.}
\label{SIchisq}
\end{figure}

\begin{figure}[p]
\begin{tabular}{c@{\hspace{0.04\textwidth}}c}
\multicolumn{2}{c}{\includegraphics[height = 0.32\textwidth]{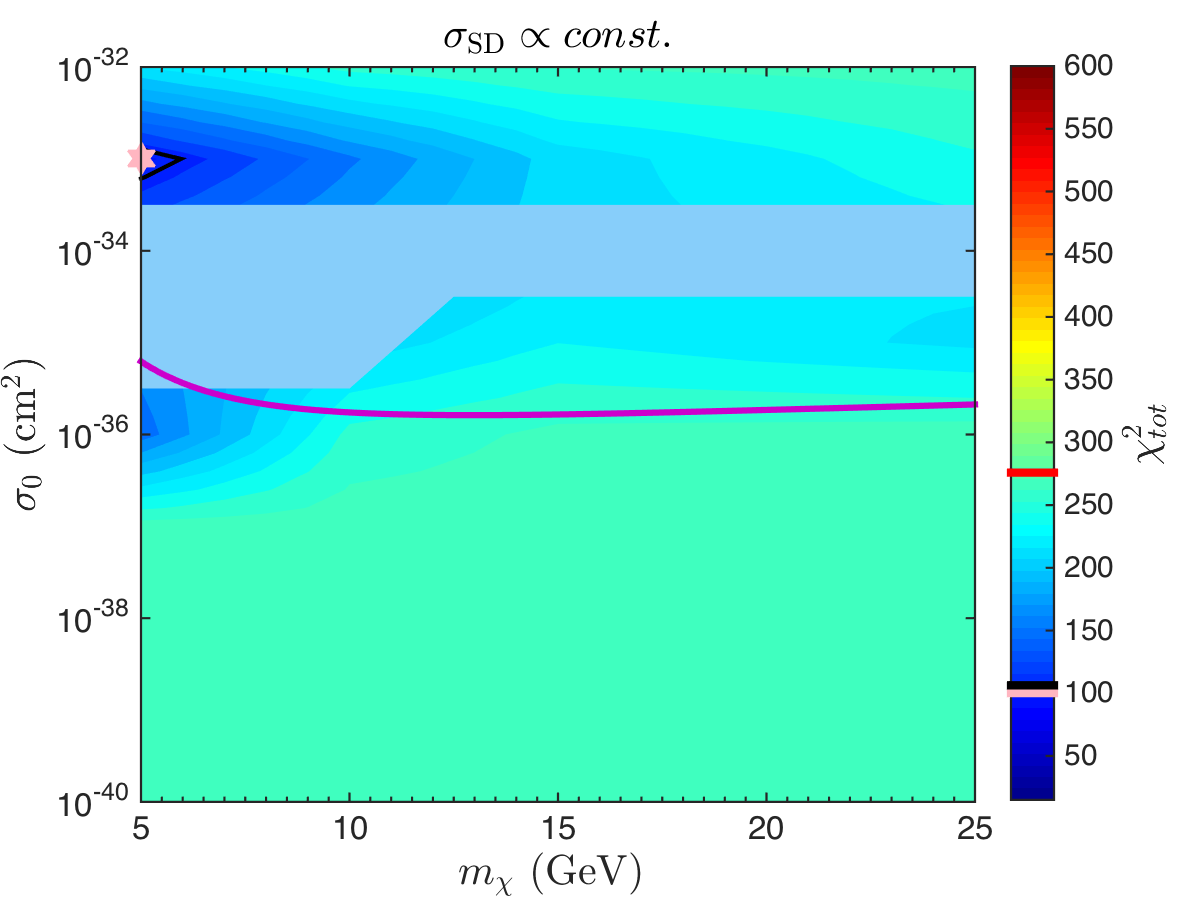}} \\
\includegraphics[height = 0.32\textwidth]{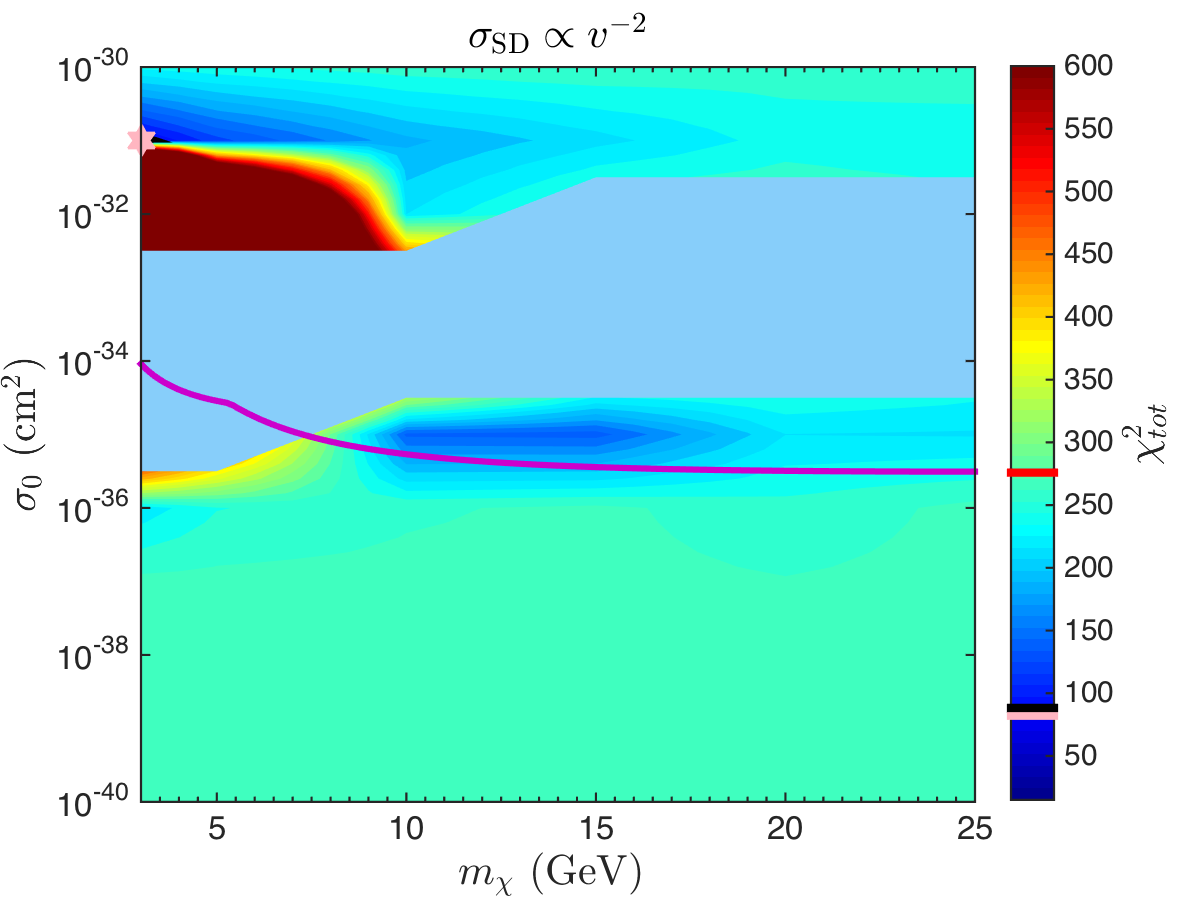} & \includegraphics[height = 0.32\textwidth]{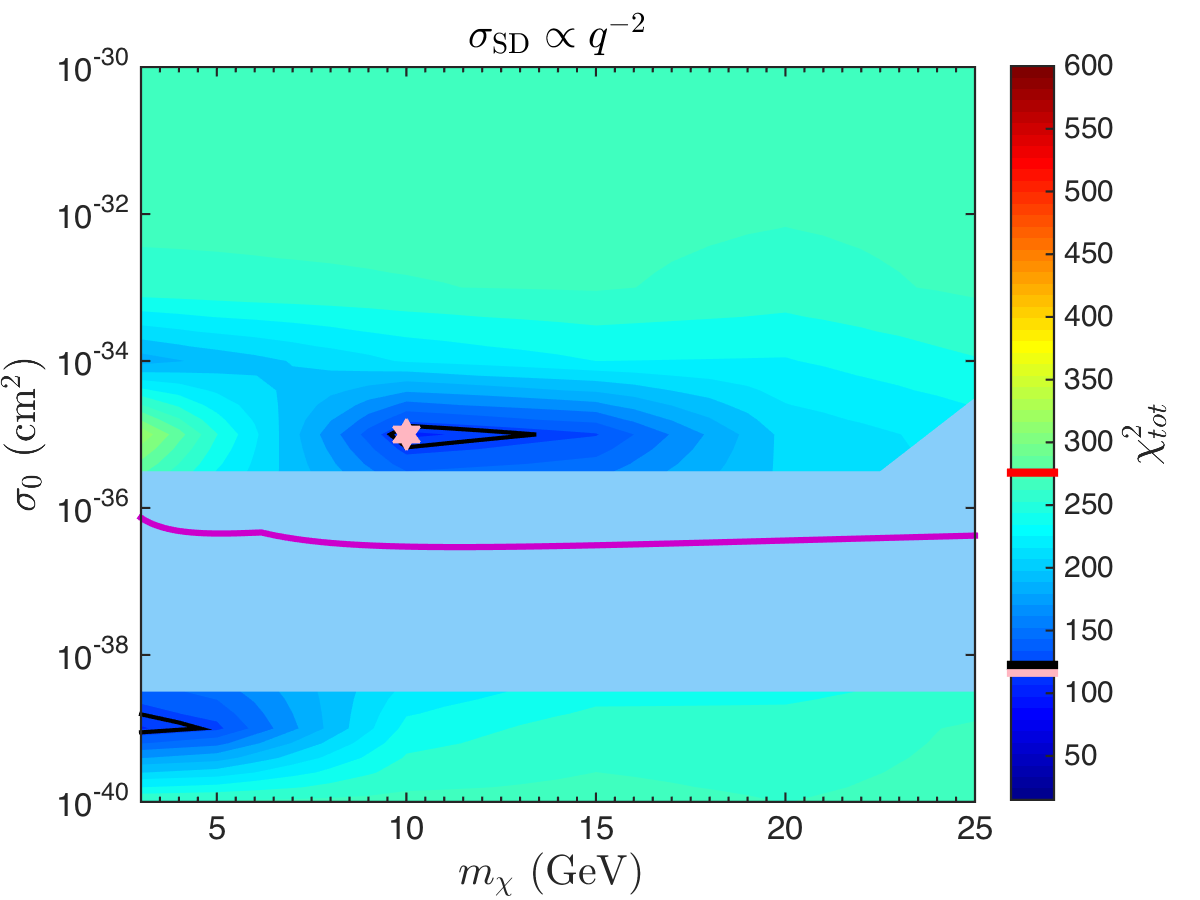} \\
\includegraphics[height = 0.32\textwidth]{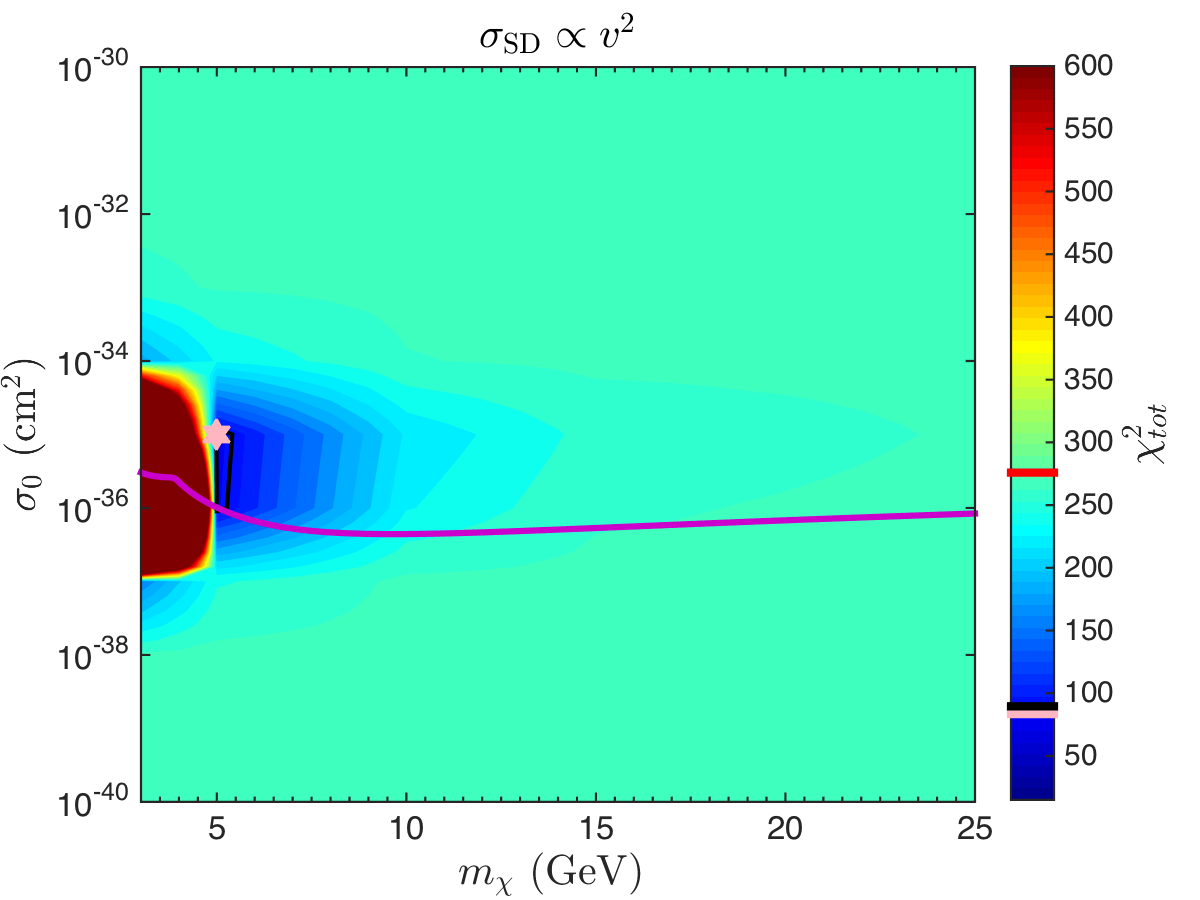} & \includegraphics[height = 0.32\textwidth]{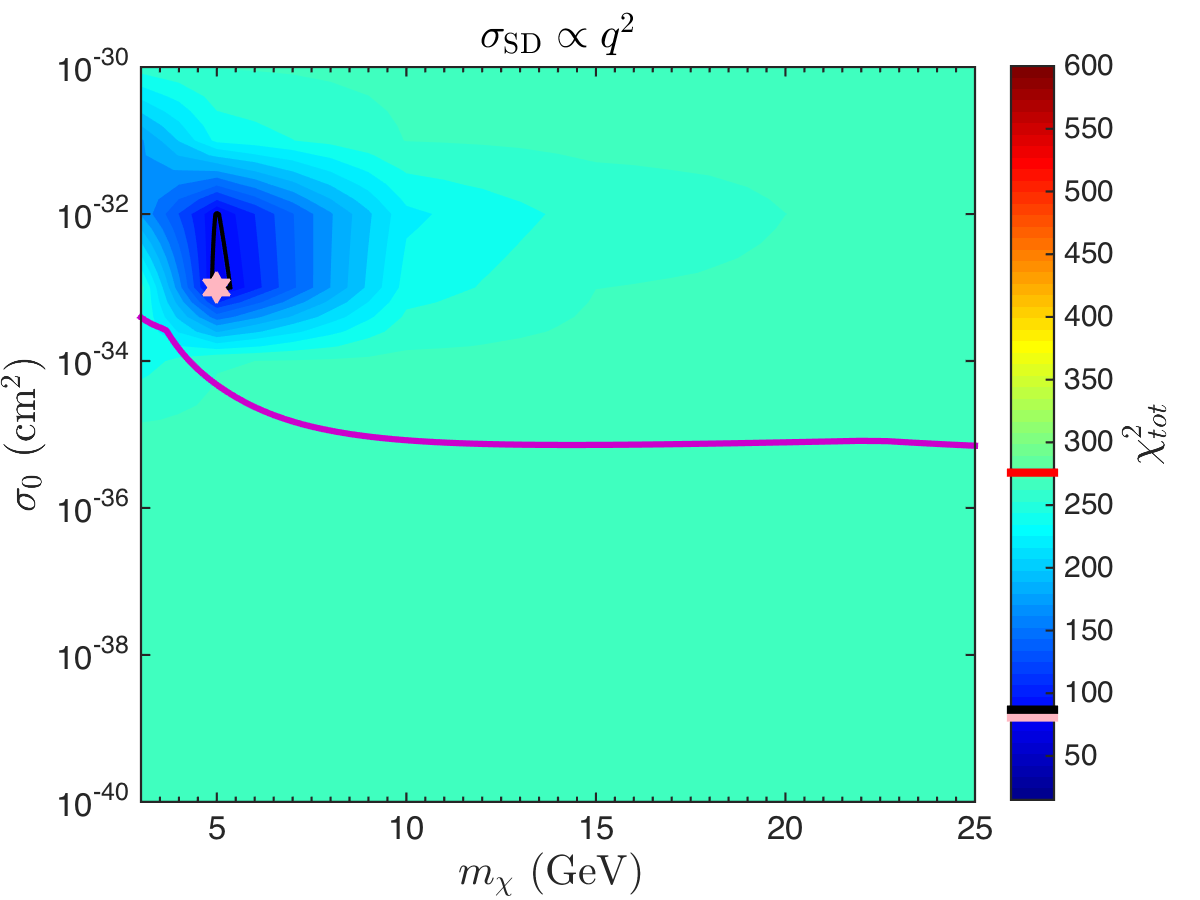} \\
\includegraphics[height = 0.32\textwidth]{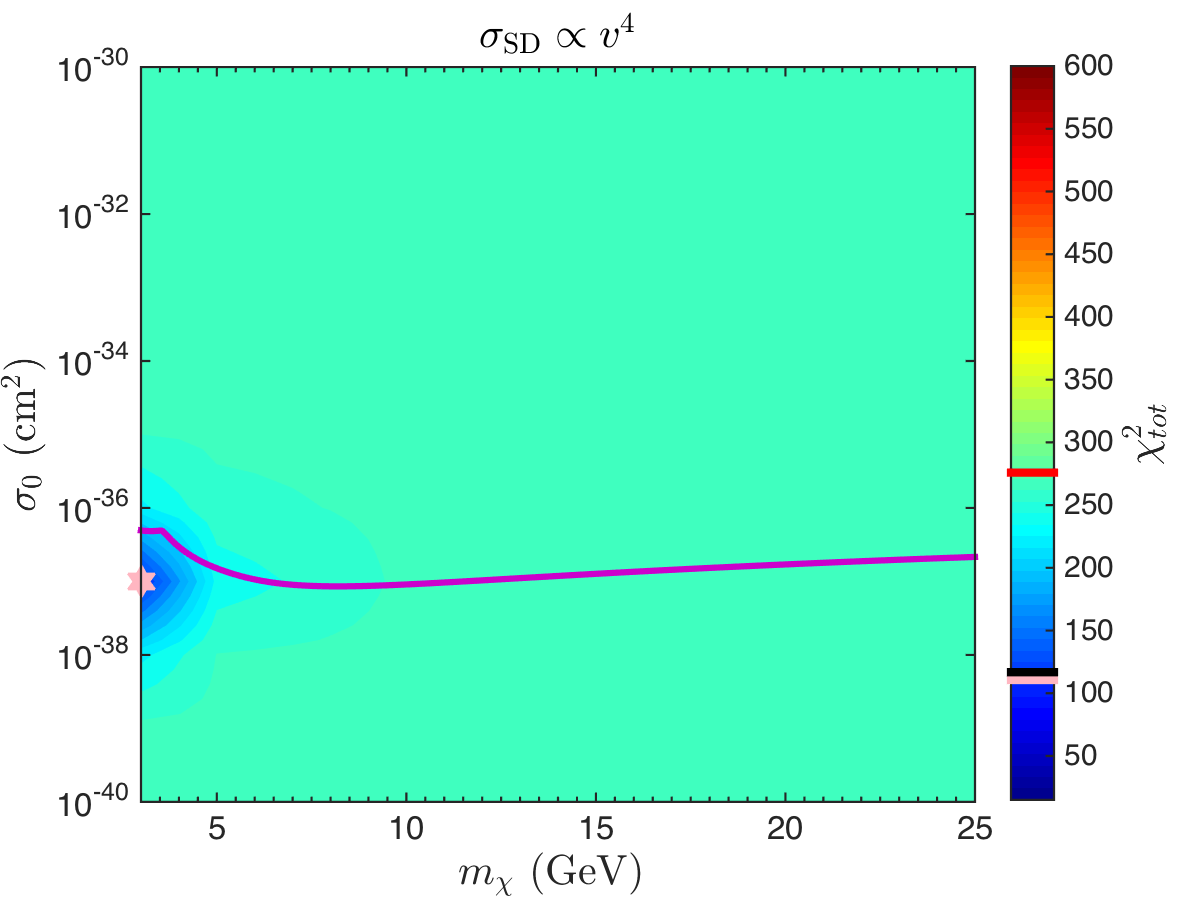} & \includegraphics[height = 0.32\textwidth]{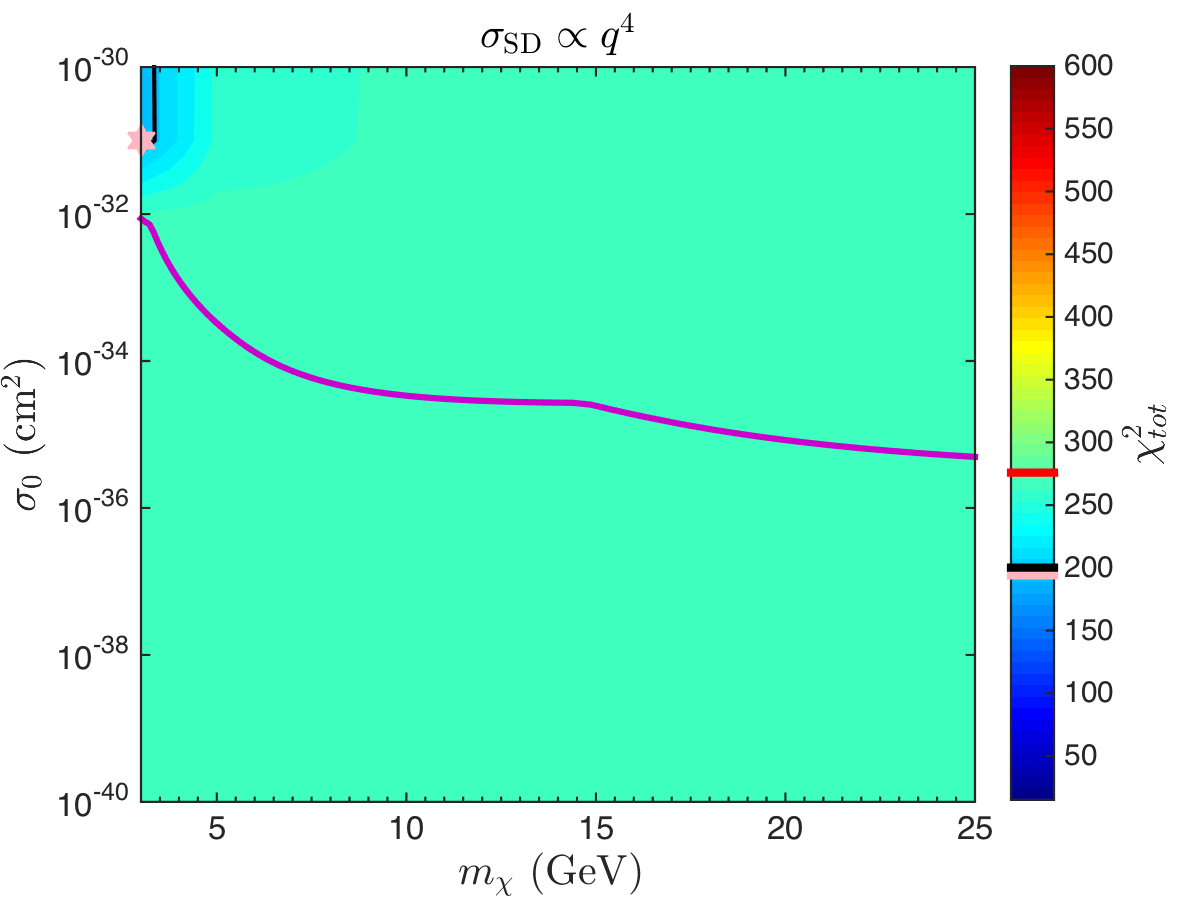} \\
\end{tabular}
\caption{As per Fig.\ \ref{SIchisq}, but for spin-dependent couplings.  }
\label{SDchisq}
\end{figure}

 The results of these combined constraints are given in Fig.\ \ref{SIchisq} for spin-independent couplings, and Fig.\ \ref{SDchisq} in the spin-dependent case. The thick magenta lines in these figures are the constraints that we obtained above from the latest CDMSlite data. There is very little parameter space in which DM is unconstrained by direct detection, yet has a strong enough effect on the Sun to change the fit. 

The best-fit points for each model are given in Table \ref{best_fit_tab}. In all cases, the SI direct detection limits are too strong to allow any of our best-fit points. In only two spin-dependent cases ($v^2$ and $v^4$) are the best-fit points below the CDMSLite limits.  In one other case (SD $q^{-2}$), there is a region that is almost as good as the best fit; we list this point instead. These allowed models are labelled with a dagger ($\dagger$) in Table \ref{best_fit_tab}.

Despite the low p-values, each of the allowed models still represents a remarkable improvement over the SSM. This improvement is driven by the much better fits to the small frequency separations, assisted by a drop in the base of the convection zone. The former observation tells us that the main improvement comes from better modelling of the region near the solar core. However, the increased tension with measured neutrino fluxes suggests that the decrease in core temperature predicted by models of ADM in the Sun may be too strong. A full DM solution to the solar abundance problem must therefore strike a delicate balance between softer thermal gradients and a lower core temperature.

\begin{table}
\caption{Standard Solar Model (SSM) and best fit (b.f.) values for each of the models we consider, along with observable quantities. Neutrino fluxes are from Ref.~\cite{Abe:2010hy} while inferred helioseismological quantities are from Ref.~\cite{Serenelli11} and references therein. The DM mass and cross-sections are in GeV and cm$^2$, respectively. The $^8$B neutrino flux $\phi^\nu_{\mathrm{B}}$ is in units of $10^{-6}$\,cm$^{-2}$\,s$^{-1}$ and the $^7$Be neutrino flux $\phi^\nu_{\mathrm{Be}}$ is expressed in $10^{-9}$ cm$^{-2}$ s$^{-1}$. The full $\chi^2$ is defined in Eq.\ \ref{eq:fullchisq} and includes the neutrino fluxes, surface helium abundance $Y_S$, depth of the convection zone and small frequency separations. A dagger ($\dagger$) denotes models that are not ruled out by CDMSlite.}
\label{best_fit_tab}
\vspace{1mm}
\begin{tabular*}{\textwidth}{| c |  @{\extracolsep{\fill}} l | r c c c c r r |}
\hline
\multicolumn{2}{|c|}{Model} & $(m_\chi, \sigma_0)_{b.f.}$ & $\phi^\nu_{\mathrm{B}}$ & $\phi^\nu_{\mathrm{Be}}$ & $R_{CZ}/R_\odot$ & $Y_s$ & $\chi^2$ & $p$ \phantom{100} \\ \hline \hline
\multicolumn{2}{|c|}{SSM} 		    & -- \phantom{100} & 4.95  &4.71   & 0.722  & 0.2356 & 275.9 & $ < 10^{-10}$ \\ \hline 
\multirow{7}{*}{$\sigma_{\rm SI}  $} &$ const.$ & (15,$10^{-37}$) & 3.48 & 4.37 & 0.721 & 0.2348 &  122.2& $ < 10^{-10}$ \\
							&$ q^{-2}$ & (3,$10^{-36}$) & 3.9 & 4.38 & 0.719 & 0.2336 & 35.17 & 0.508 \\
							&$ q^{2}$ & (3,$10^{-33}$) & 3.93 & 4.38 & 0.719 & 0.2334 & 31.02 & 0.704 \\
							&$ q^{4}$ & (3,$10^{-32}$) & 3.92 & 4.36 & 0.718 & 0.2331 & 27.52 & 0.844 \\
							&$ v^{-2}$ & (5,$10^{-34}$) & 3.82 & 4.41 & 0.72 & 0.2345 & 75.54 & 1.26 $\times 10^{-4}$\\
							&$ v^{2}$ & (5,$10^{-38}$) & 3.47 & 4.28 & 0.72 & 0.234 & 96.48 & 1.99$\times 10^{-7}$ \\
							&$ v^{4}$ & (3,$10^{-37}$) & 4.31 & 4.51 & 0.72 & 0.2343 & 85.38 & 6.84$\times 10^{-6}$ \\ \hline
\multirow{7}{*}{$\sigma_{\rm SD}  $} &$ const.$ & (5,$10^{-33}$) & 3.36 & 4.27 & 0.72 & 0.2341 & 100.2 & 5.8$\times 10^{-8}$ \\
							& $ q^{-2}$	 &$\dagger$ (3,$10^{-39}$)  &	3.22 & 4.16 & 0.718 & 0.2333 & 119.48 &  $ < 10^{-10}$ \\
							&$ q^{2}$ & (5,$10^{-33}$) & 3.85 & 4.42 & 0.721 & 0.2346 & 80.7 & 2.82$\times 10^{-5}$ \\
							&$ q^{4}$ & (3,$10^{-31}$) & 4.69 & 4.64 & 0.721 & 0.2352 & 194 & $ < 10^{-10}$ \\						
							&$ v^{-2}$ & (3,$10^{-31}$) & 4.11 & 4.48 & 0.72 & 0.2346 & 82.15 & 1.83$\times 10^{-5}$ \\
							&$ v^{2}$ &$\dagger$ (5,$10^{-35}$) & 3.88 & 4.43 & 0.721 & 0.2346 & 83.44 & 1.24$\times 10^{-5}$ \\
							&$ v^{4}$ & $\dagger$ (3,$10^{-37}$) & 4.39 & 4.54 & 0.72 & 0.2346 & 110.6 & 1.63$\times 10^{-9}$ \\
   \hline \hline
\multicolumn{2}{|c|}{Obs.} 			& -- & 5.00& 4.82 & 0.713   & 0.2485    & -- & -- \\ \hline
\multicolumn{2}{|c|}{Obs. error} 		& -- &  3 \% & 5\%  & 0.001 &  0.0034 & -- & -- \\ 
\multicolumn{2}{|c|}{Model error} 	& -- &  14 \% & 7\%  &  0.004 & 0.0035 & -- & -- \\ \hline
\hline
\end{tabular*}
\end{table}

We end this section by noting that we also provide as supplementary material online a complete table of the boron and beryllium neutrino fluxes, surface helium, radius of the convection zone and small frequency separations for all of our models, in addition to the partial and total chi-squared values defined in Eq.\ \ref{eq:fullchisq}.  This table can be used for quick lookup and interpolation, e.g.\ for global fits of DM models.
\FloatBarrier
\section{Conclusions}
\label{sec:conclusion}
We have presented new limits on momentum and velocity-dependent dark matter from direct detection, and an updated study of their effects on state of the art simulations of the Sun. A correction of the LTE conduction formalism (Eq.\ \ref{LTEtransport}), has shifted the best fits in the parameter space, and the very strong limits from the low-threshold analysis of CDMSlite have ruled out all but a small part of the parameter space of models that can improve the SSM. Our solver could not find solutions for certain values in the parameter space, ostensibly because they led to changes in the solar structure that were too large with respect to the SSM (which it was originally designed to compute). This is not to say that these models are ruled out, however. We note that in some cases (notably spin-dependent $v^{-2}$), the non-convergence region lies on top of the region that is probed by direct detection. A better exploration of this parameter space could be very interesting indeed. 

Of the models that we have found that provide a good fit, only some can be obtained  from the current paradigm of simplified models of dark matter. For instance, a spin-dependent $v^4$ cross section does not occur in such setups. A fermion DM candidate interacting via a vector mediator can give rise to an SD, $v^2$ cross section, however. This requires a purely vector coupling to the DM, and an axial coupling at the quark level. However, this setup also gives a similar contribution proportional to $q^2$ (see e.g.\ \cite{Dent:2015zpa}), which would be three orders of magnitude larger using our normalisations of $q_0$ and $v_0$, and thus excluded by experiment. This is not to say that this model cannot be obtained, only that it is not straightforward using current tools. A $q^{-2}$ cross section, on the other hand, is closer to a long-range force, which is again not covered in the simplified model approach. A model that can yield a similar type of behaviour  would be an electromagnetic dipole or anapole interaction. These models actually couple with mixed powers of $v$ and $q$, and thus require explicit recalculation of the molecular diffusion and conduction coefficients $\alpha$ and $\kappa$, and is the subject of a dedicated study \cite{Ben}. 

Finally, we note that the effects of evaporation have not been included here, as the full evaporation formalism must be generalised to the form factor case from scratch. Because the relevant DM masses are so close to the evaporation limit for constant cross-sections ($m_{evap} \sim 4$ GeV), this may have a an important impact on the remaining parameter space. Full evaporation rates for momentum- and velocity-dependent interactions will be computed in detail in an upcoming publication \cite{BusoniEvap}. 

It is clear from the fits we present here that the addition of heat transport by dark matter in the Sun can improve the SSM significantly.  The improvement is well beyond what one would naively expect from adding just two degrees of freedom.  Even if DM is not the solution, the fact that the true solution can be accurately mimicked by DM would seem to indicate something fundamental about the physical processes behind the solution.  Given the types of DM interactions that work, it is clear that if DM \textit{is} to explain the solar composition problem, then a strong model-building effort is required to connect our results with a more robust model of dark matter that can be embedded in a UV-complete theory.

\section*{Acknowledgements}
We thank Ben Geytenbeek for pointing out the crucial correction to the transported luminosity. ACV thanks Elias Lopez Asamar, David Cerde\~no and Ji-Haeng Huh for valuable help with the CDMSlite bounds. We are grateful to the SOM group at IFIC (Universitat de Val\`encia -- CSIC), for allowing calculations to be performed on SOM2  funded by PROMETEO/2009/116, PROMETEOII/2014/050 and FPA2011-29678.  PS is supported by STFC (ST/K00414X/1 and ST/N000838/1) and AS acknowledges support from grants 2014SGR-1458 (Generalitat de Catalunya), ESP2014-56003-R and ESP2015-66134-R (MINECO).

\appendix
\section{Contour plots of solar observables}
\label{app:results}
In this appendix we show contour plots for individual quantities impacted by the presence of dark matter in the Sun. In order, we show the boron-8 and beryllium-7 neutrino fluxes, the depth of the convection zone and the goodness of fit ($\chi^2$) computed for the sound speed profile and for the small frequency separations $r_{02}$ and $r_{13}$. 

\begin{figure}[p]
\begin{tabular}{c@{\hspace{0.04\textwidth}}c}
\multicolumn{2}{c}{\includegraphics[height = 0.32\textwidth]{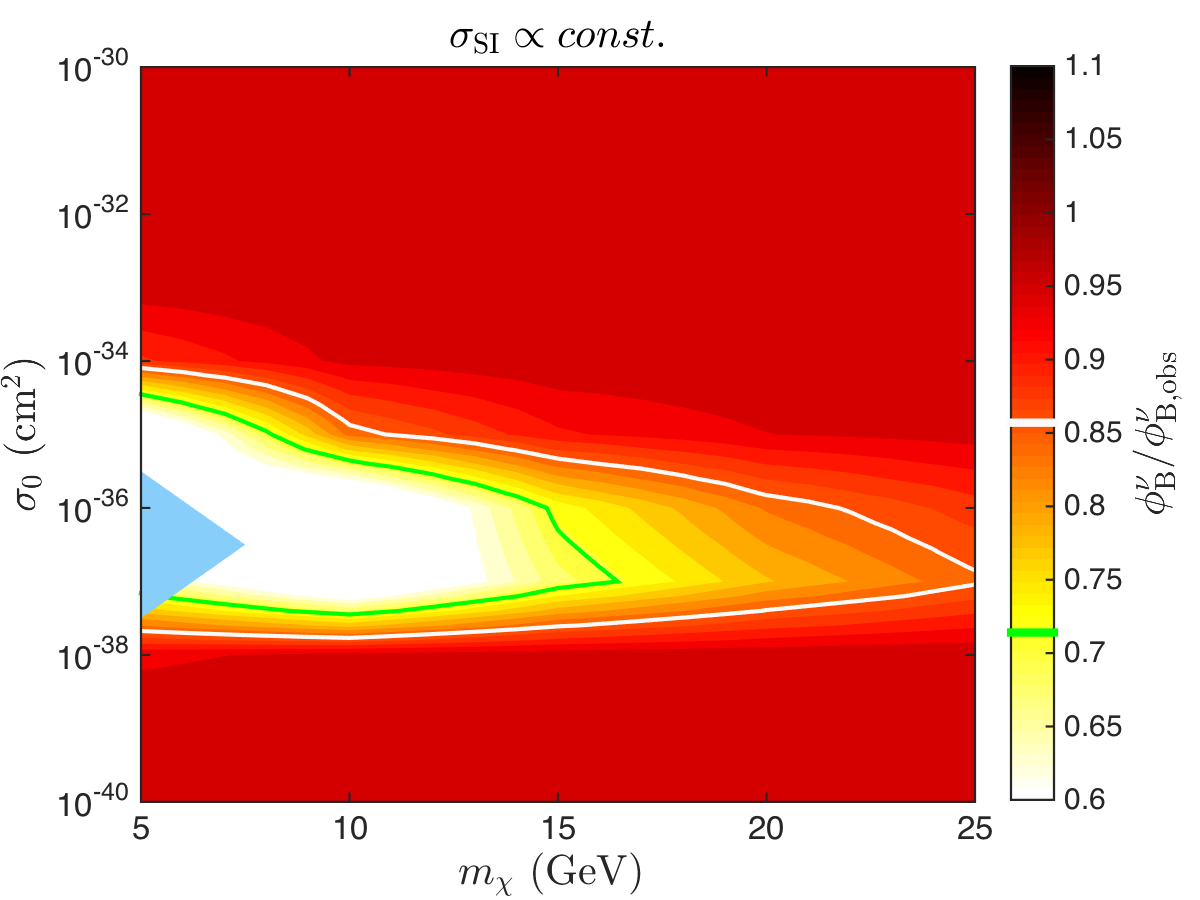}} \\
\includegraphics[height = 0.32\textwidth]{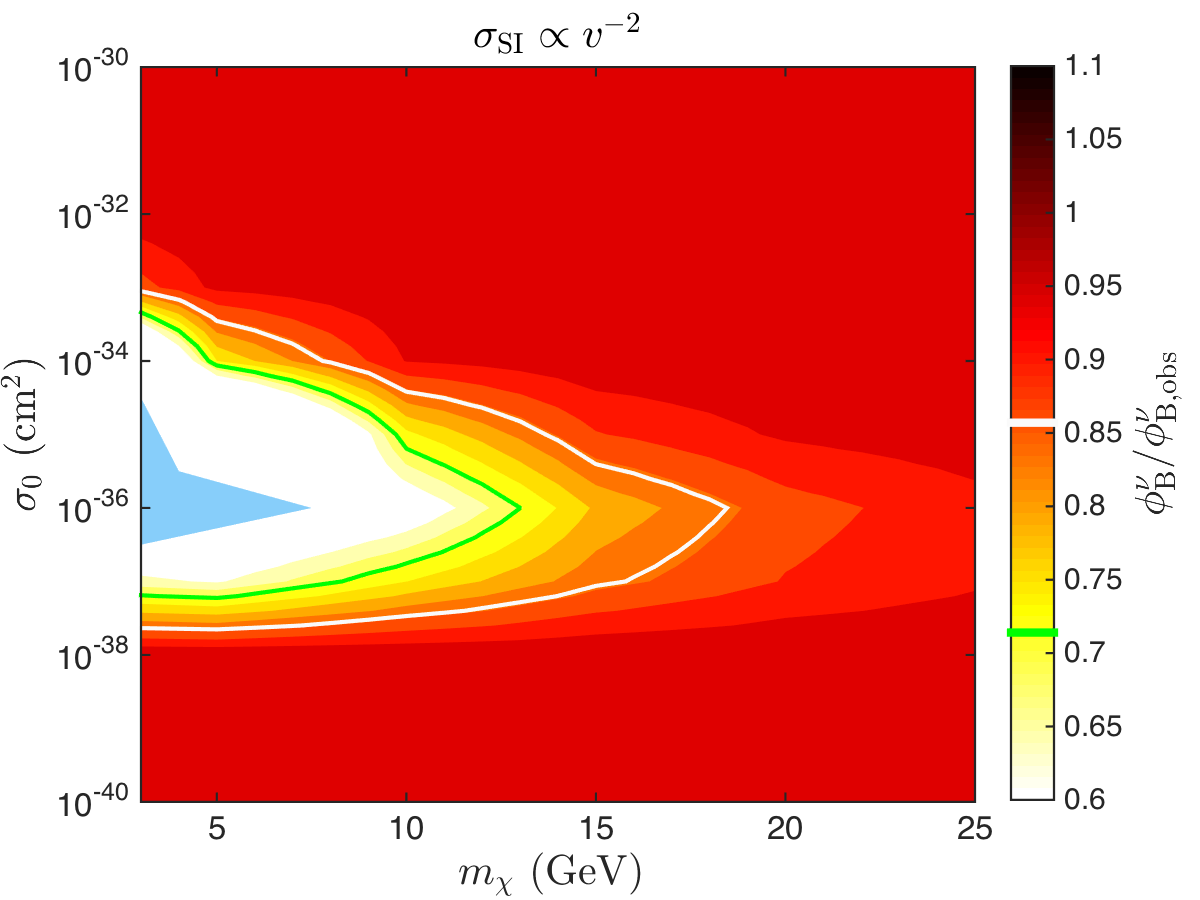} & \includegraphics[height = 0.32\textwidth]{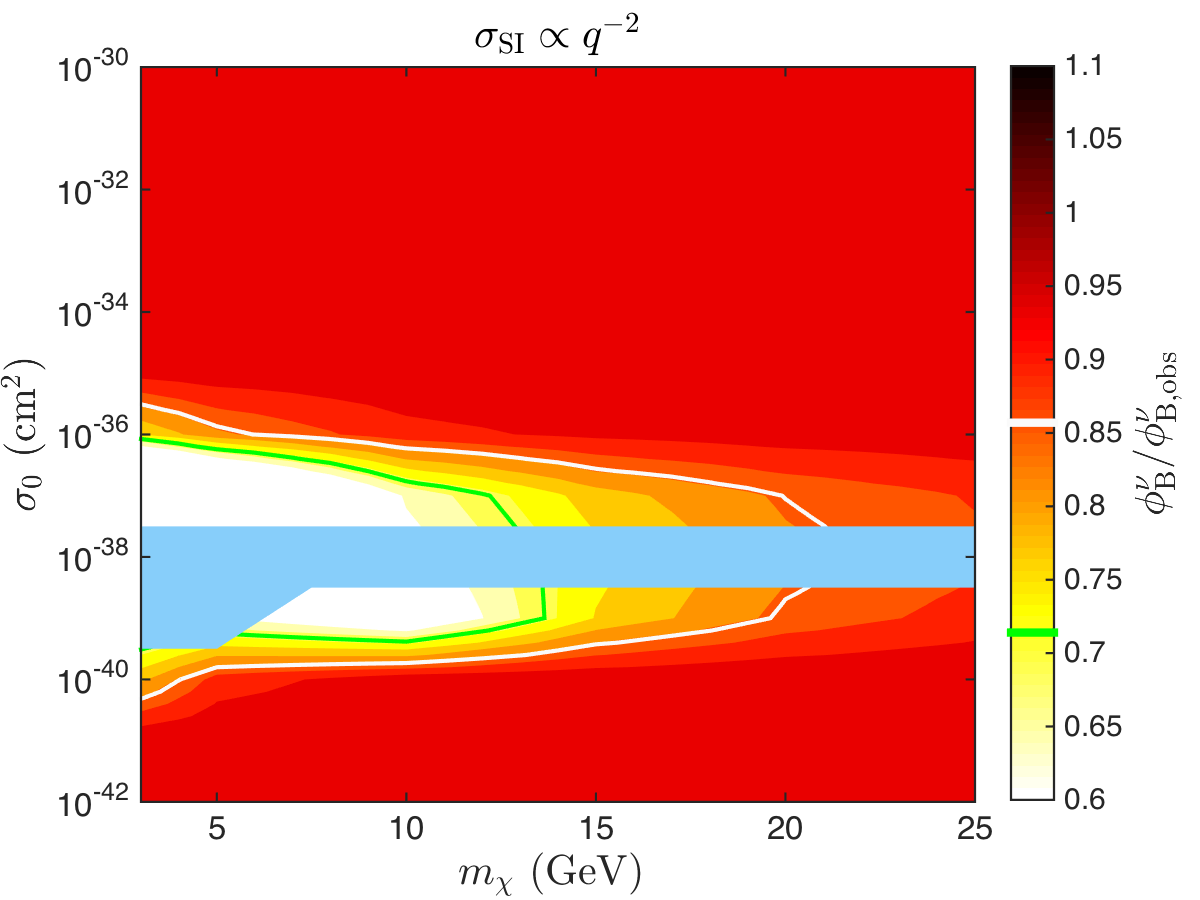} \\
\includegraphics[height = 0.32\textwidth]{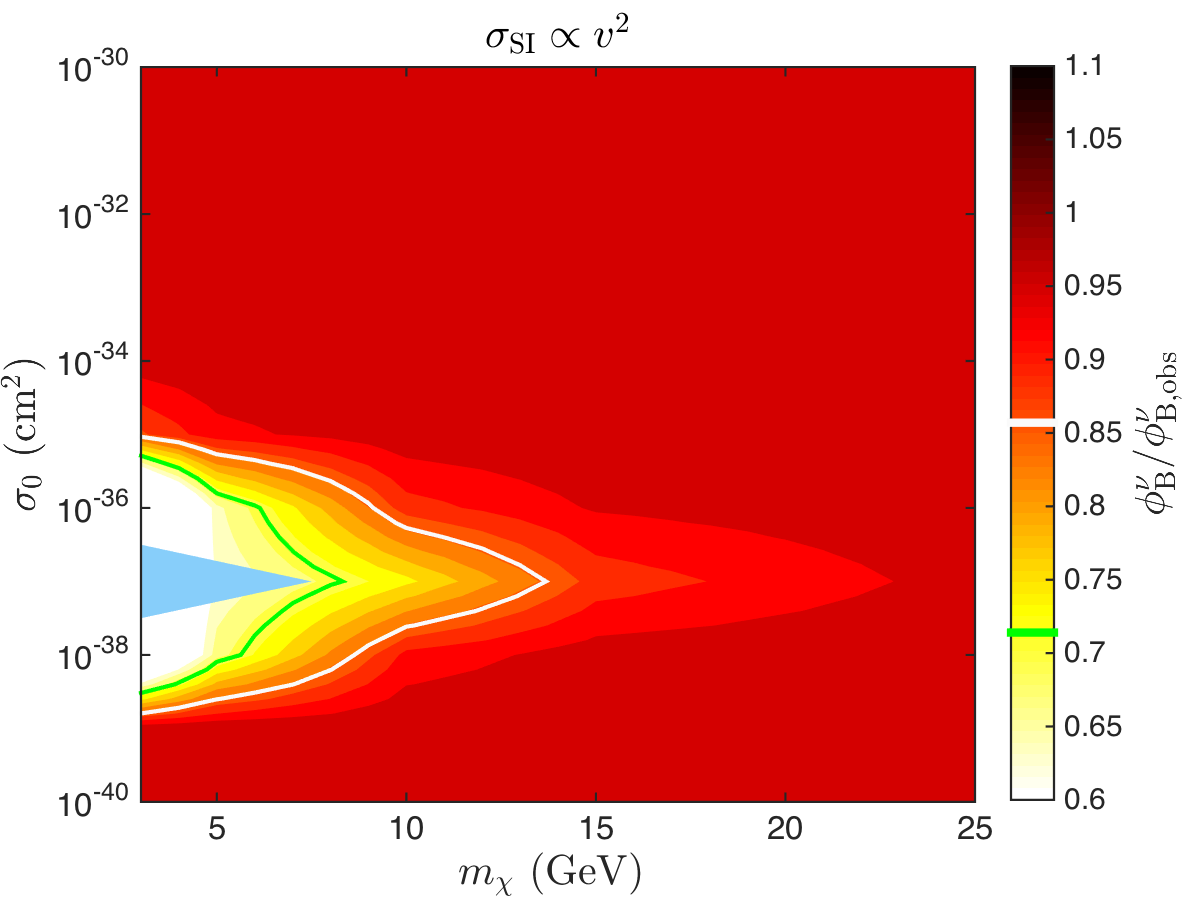} & \includegraphics[height = 0.32\textwidth]{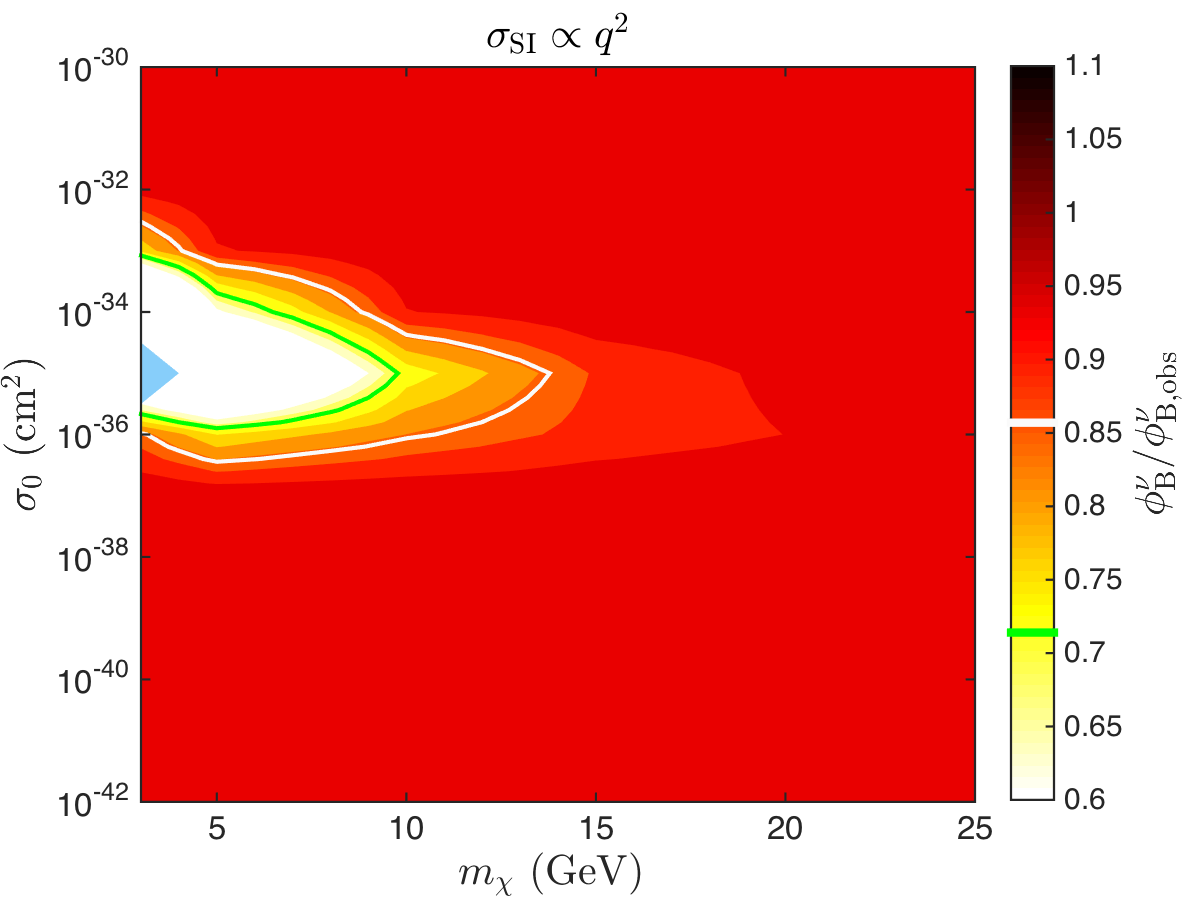} \\
\includegraphics[height = 0.32\textwidth]{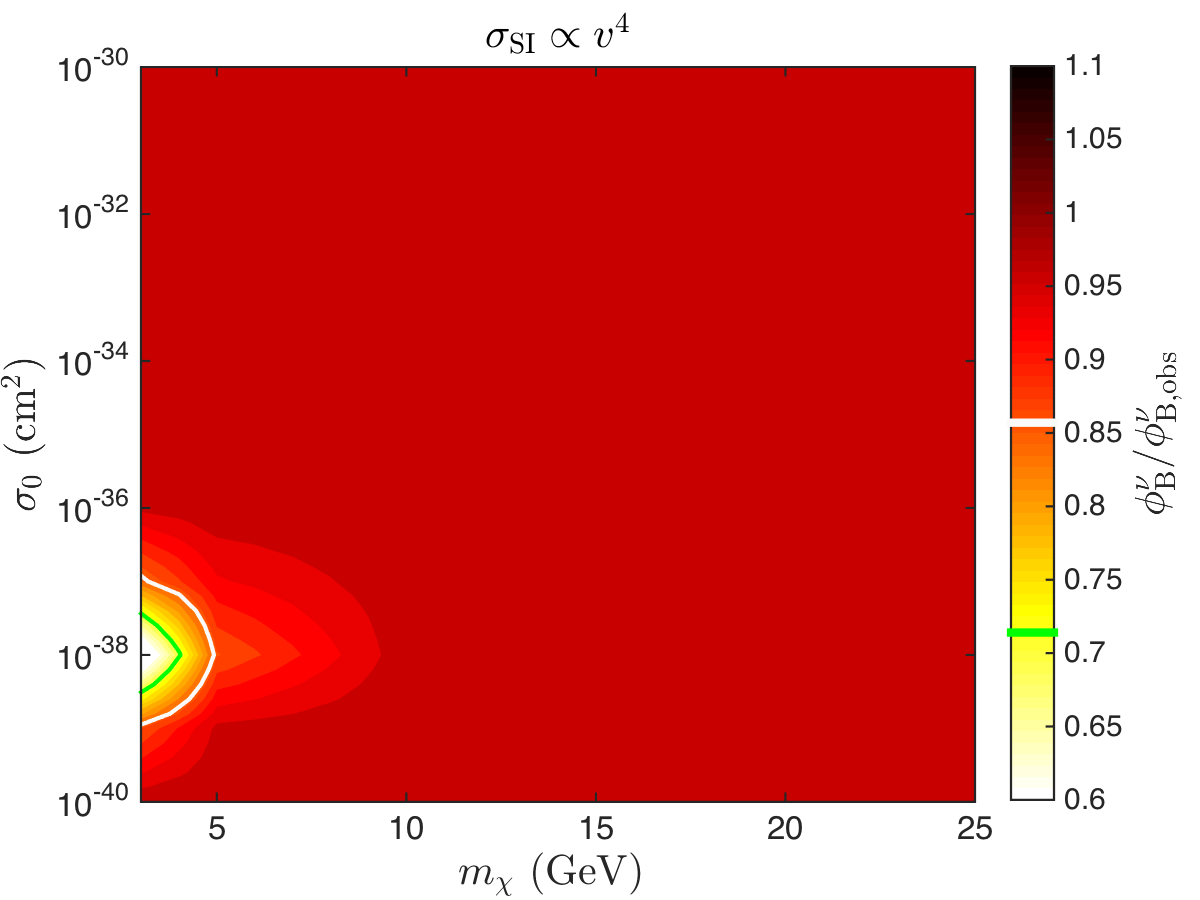} & \includegraphics[height = 0.32\textwidth]{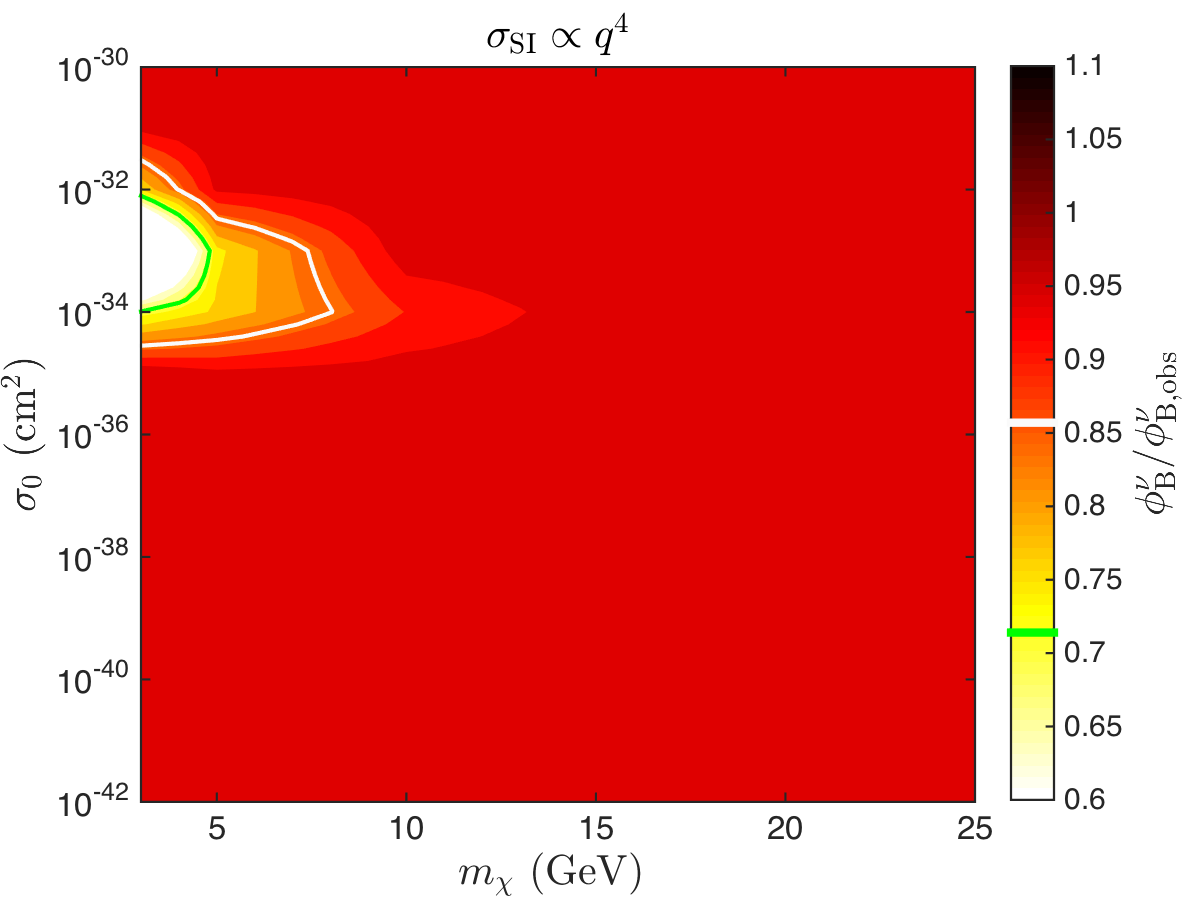} \\
\end{tabular}
\caption{The ratio of the predicted $^8$B neutrino flux to the  measured value $\phi^\nu_{\rm B,obs} = 5.00 \times 10^6$\,cm$^{-2}$s$^{-1}$~\cite{Abe:2010hy}, for each type of spin-independent dark matter coupling defined in Eq.\ \ref{qdepvdep}.  In every case the white and green lines show the isocontours where the flux is respectively 1 and 2$\sigma$ lower than the observed values, based on observational (3\%) and modelling (14\%) errors, added in quadrature. The cross-sections are normalized such that $\sigma = \sigma_0 (v/v_0)^{2n}$ or $\sigma = \sigma_0 (q/q_0)^{2n}$, with $v_0 = 220$\,km\,s$^{-1}$ and $q_0 = 40$\,MeV. Simulations carried out in the masked regions did not converge, due to the significant heat conduction by the DM particles, leading in extreme cases to density inversions in the core. }
\label{SIboronfluxes}
\end{figure}

\begin{figure}[p]
\begin{tabular}{c@{\hspace{0.04\textwidth}}c}
\multicolumn{2}{c}{\includegraphics[height = 0.32\textwidth]{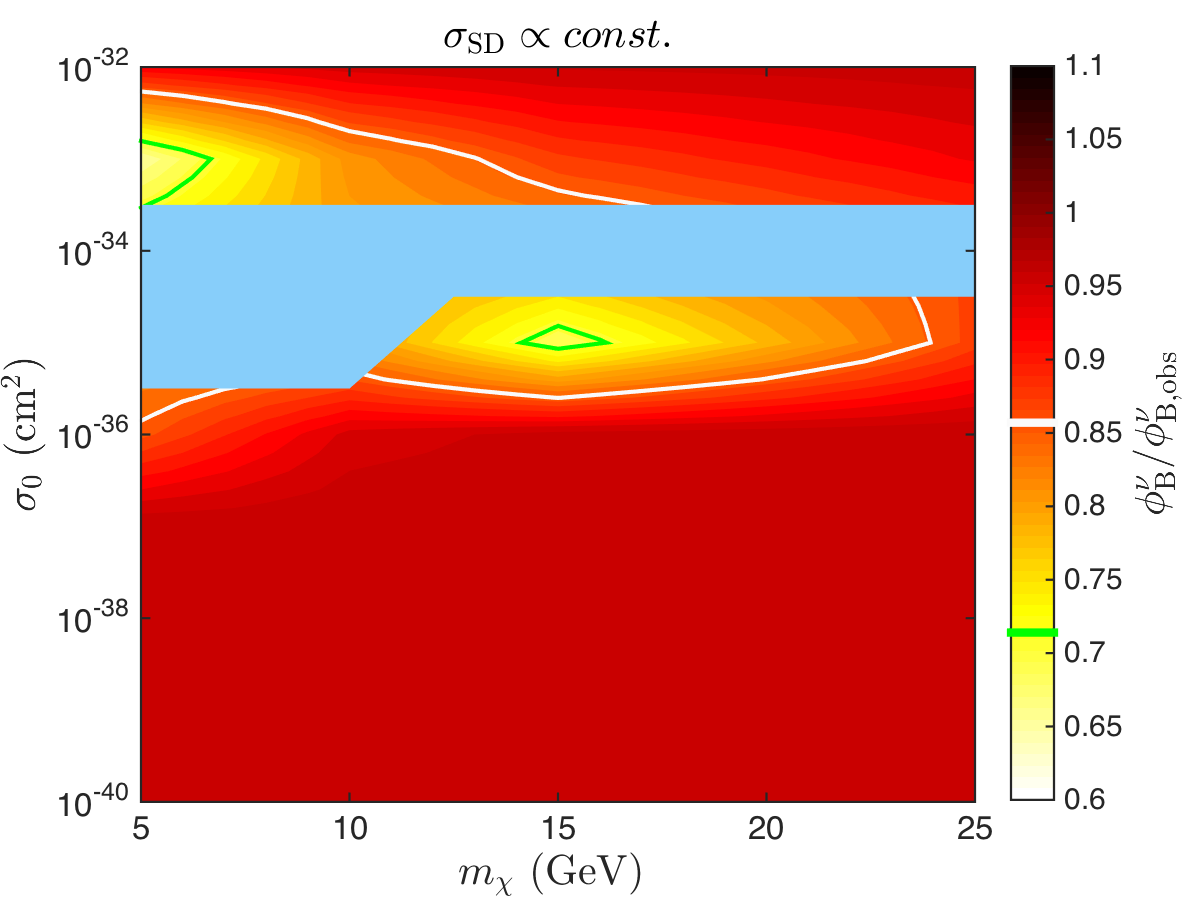}} \\
\includegraphics[height = 0.32\textwidth]{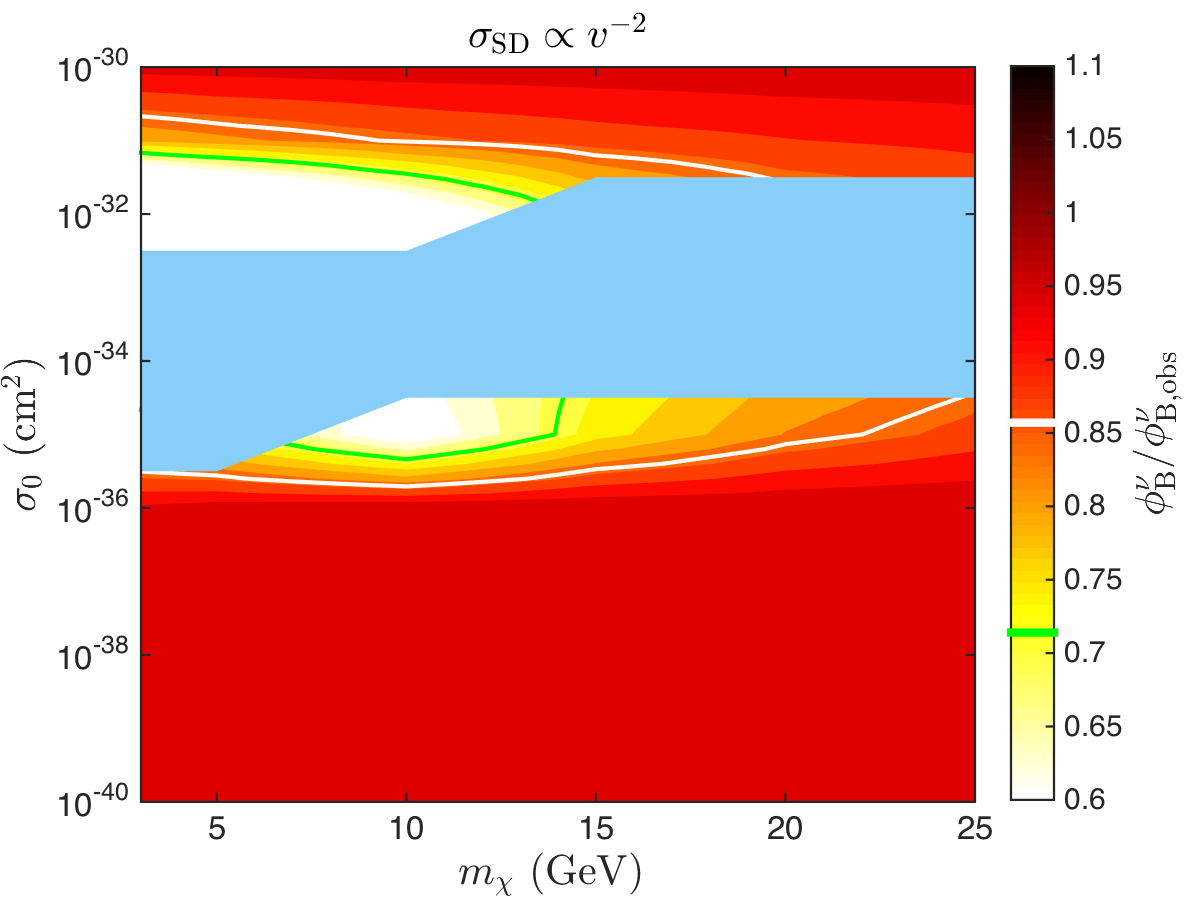} & \includegraphics[height = 0.32\textwidth]{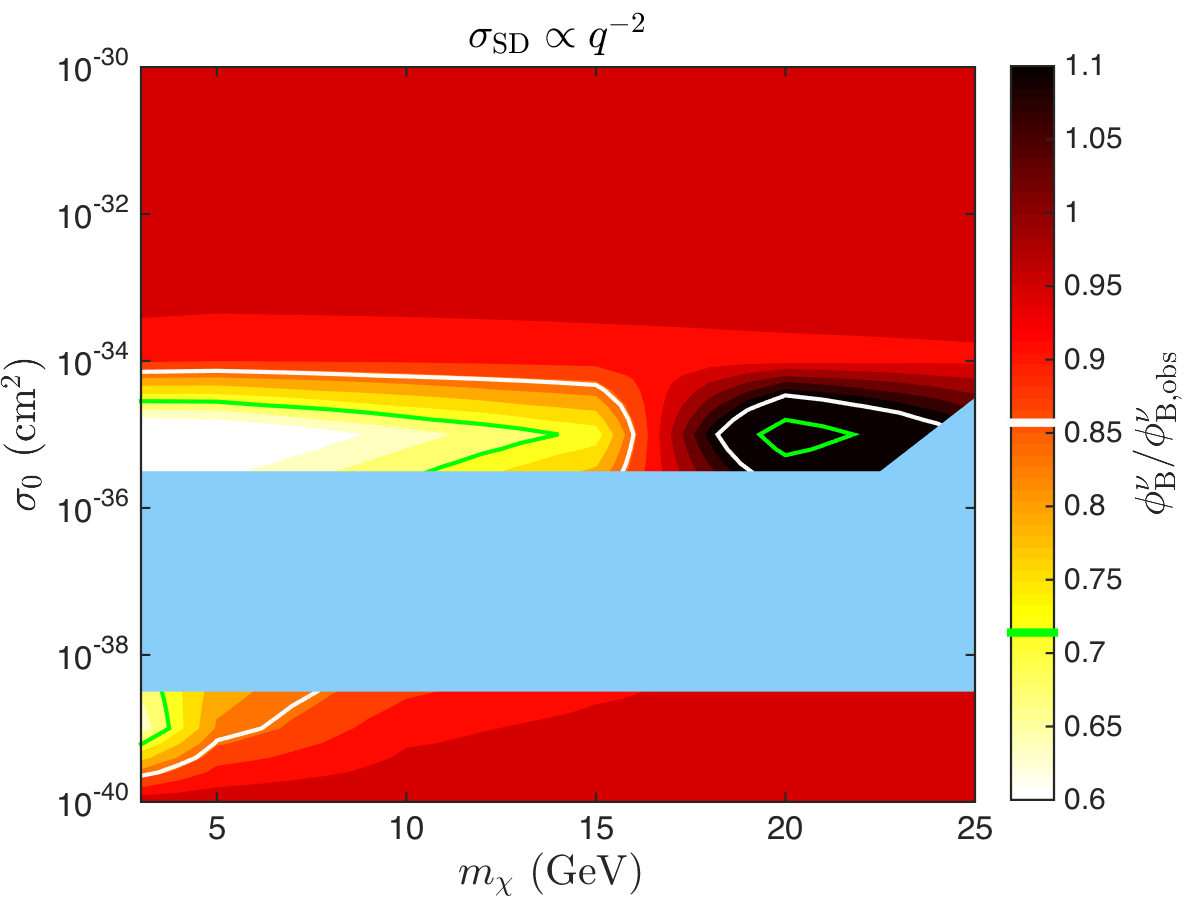} \\
\includegraphics[height = 0.32\textwidth]{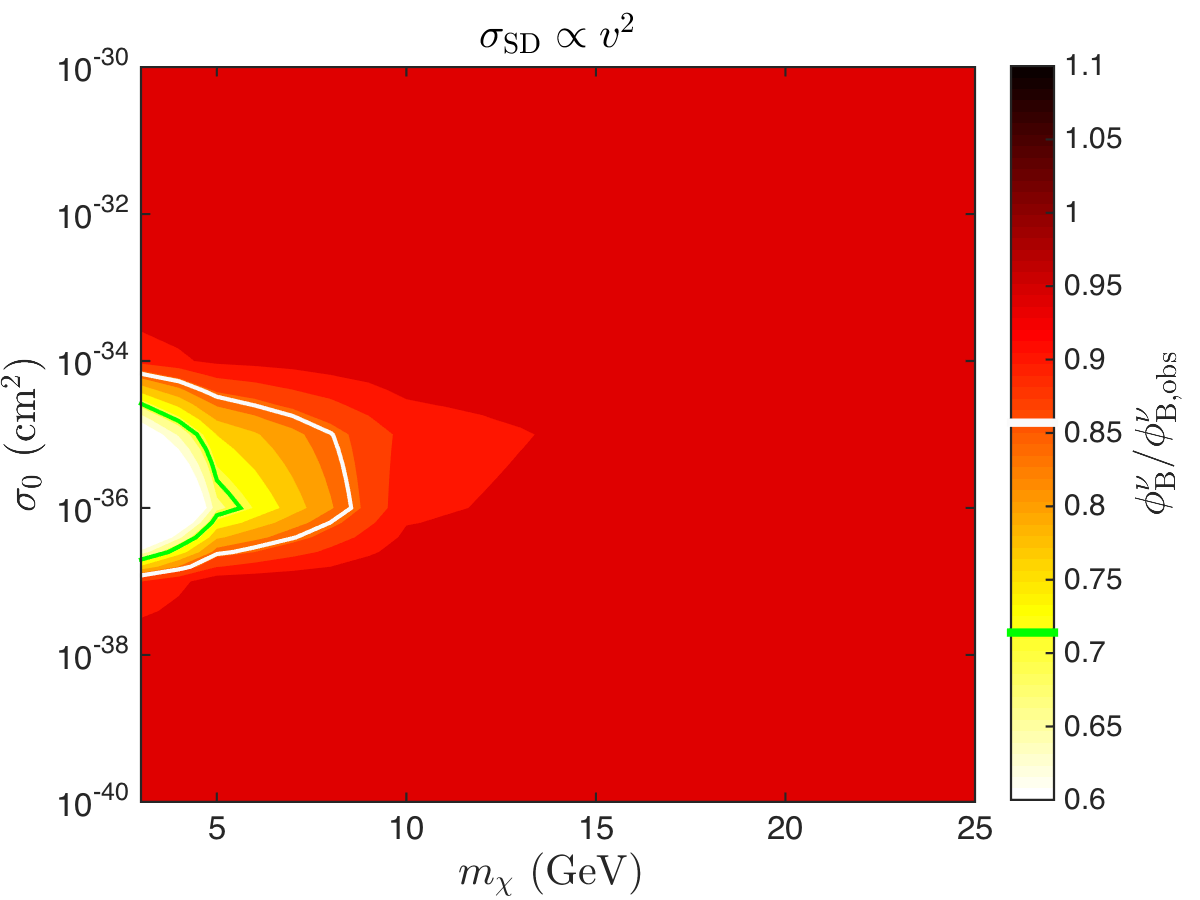} & \includegraphics[height = 0.32\textwidth]{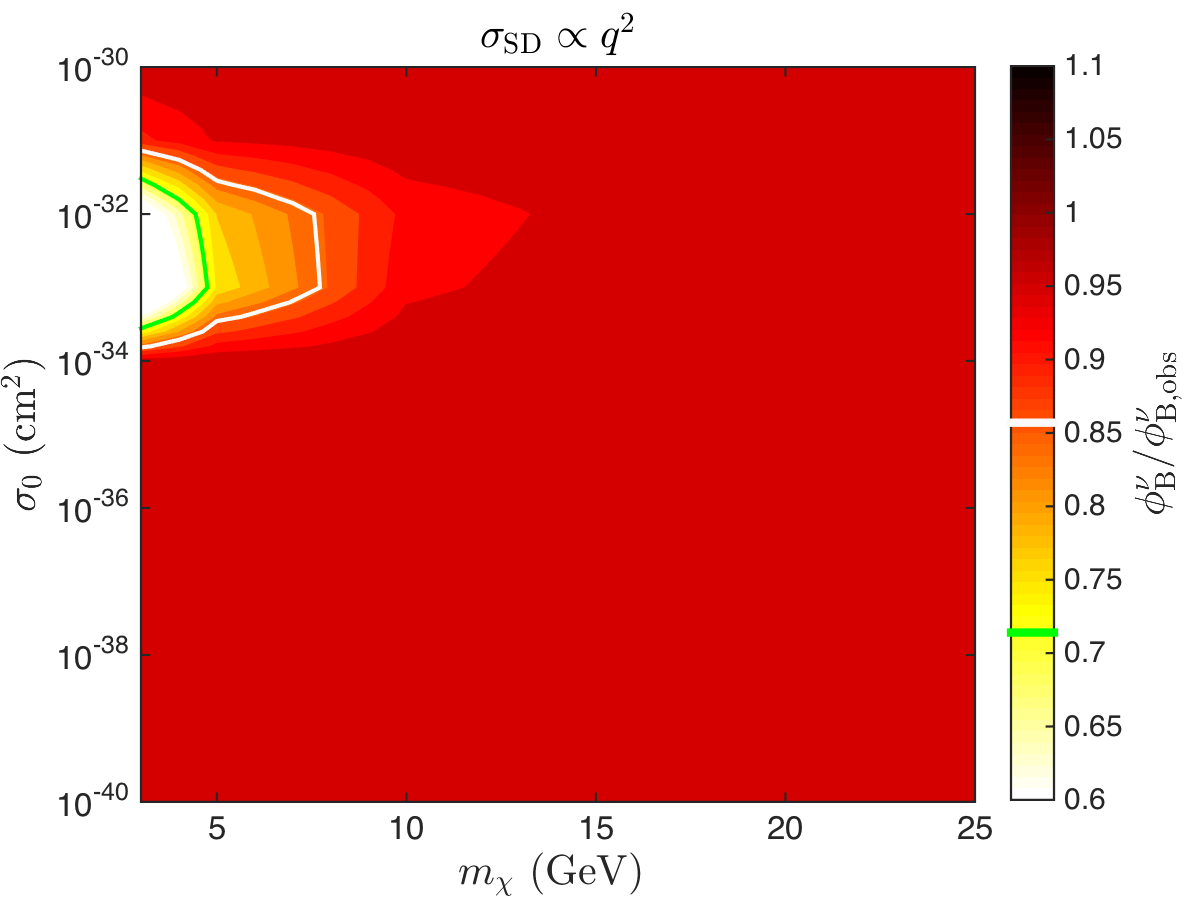} \\
\includegraphics[height = 0.32\textwidth]{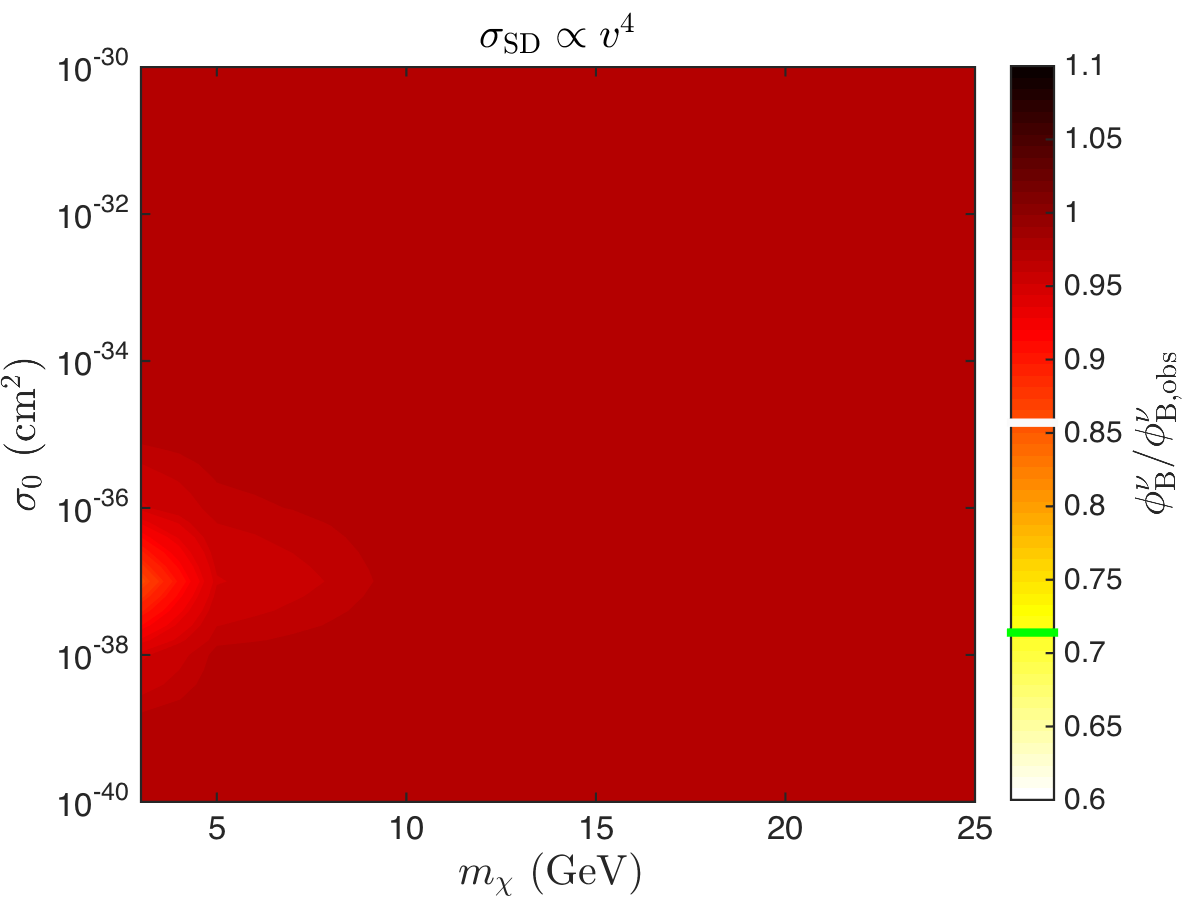} & \includegraphics[height = 0.32\textwidth]{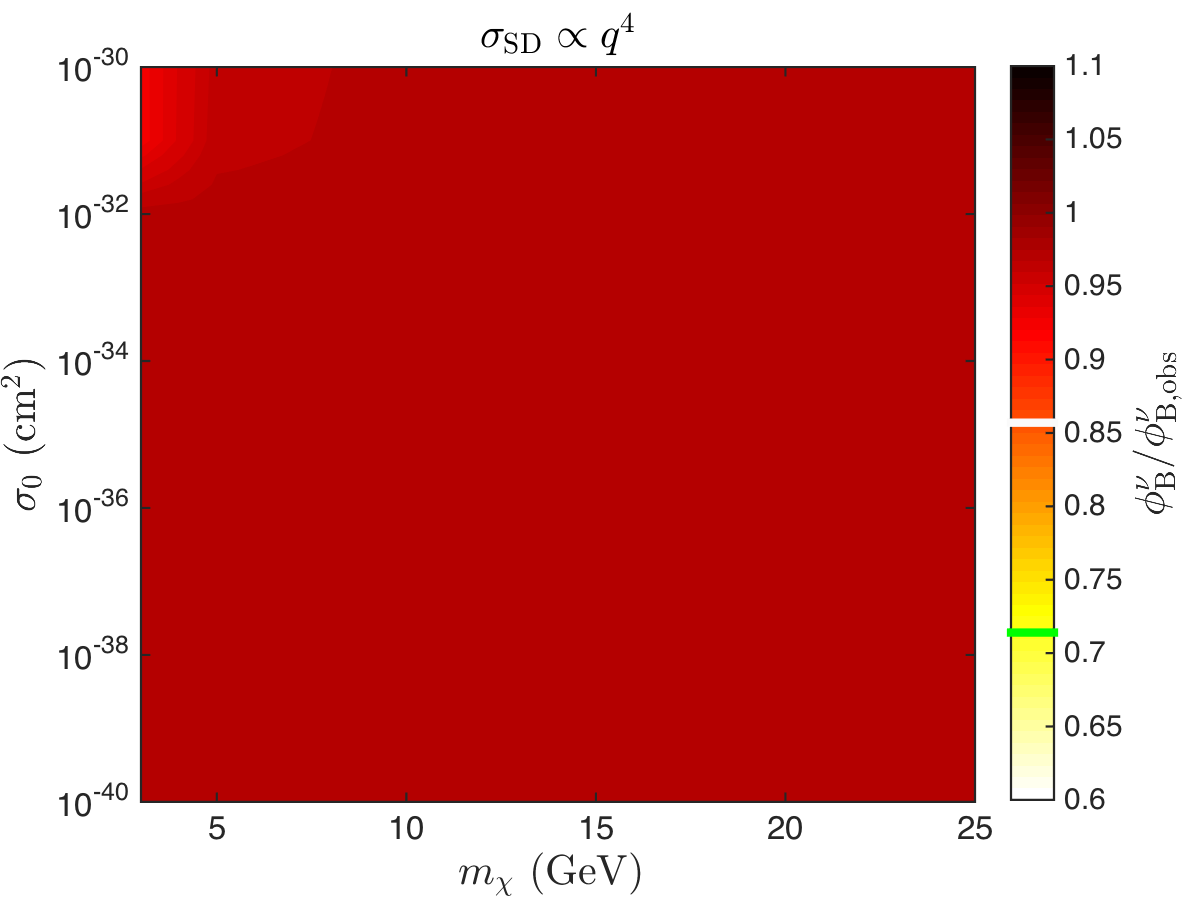} \\
\end{tabular}
\caption{As per Fig.\ \ref{SIboronfluxes}, but for spin-dependent couplings.}
\label{SDboronfluxes}
\end{figure}

\begin{figure}[p]
\begin{tabular}{c@{\hspace{0.04\textwidth}}c}
\multicolumn{2}{c}{\includegraphics[height = 0.32\textwidth]{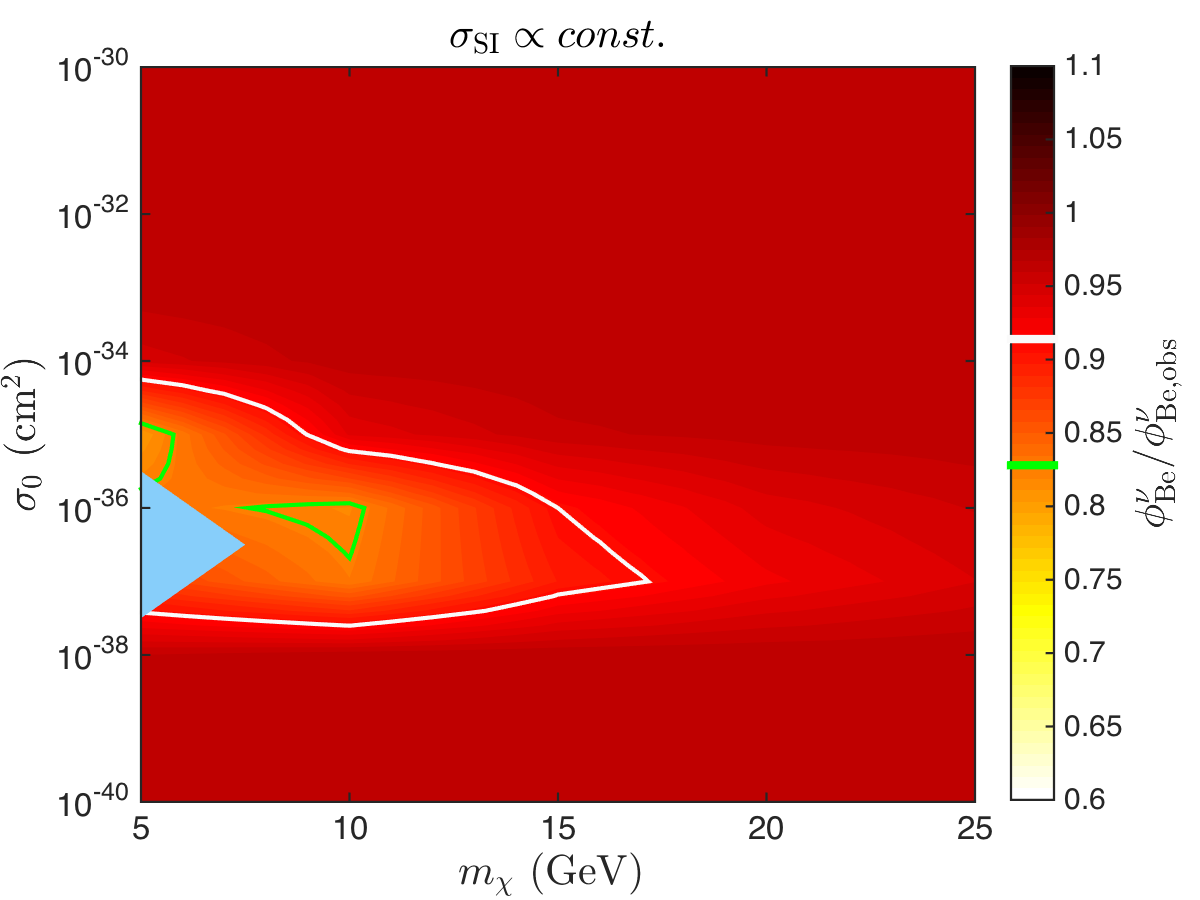}} \\
\includegraphics[height = 0.32\textwidth]{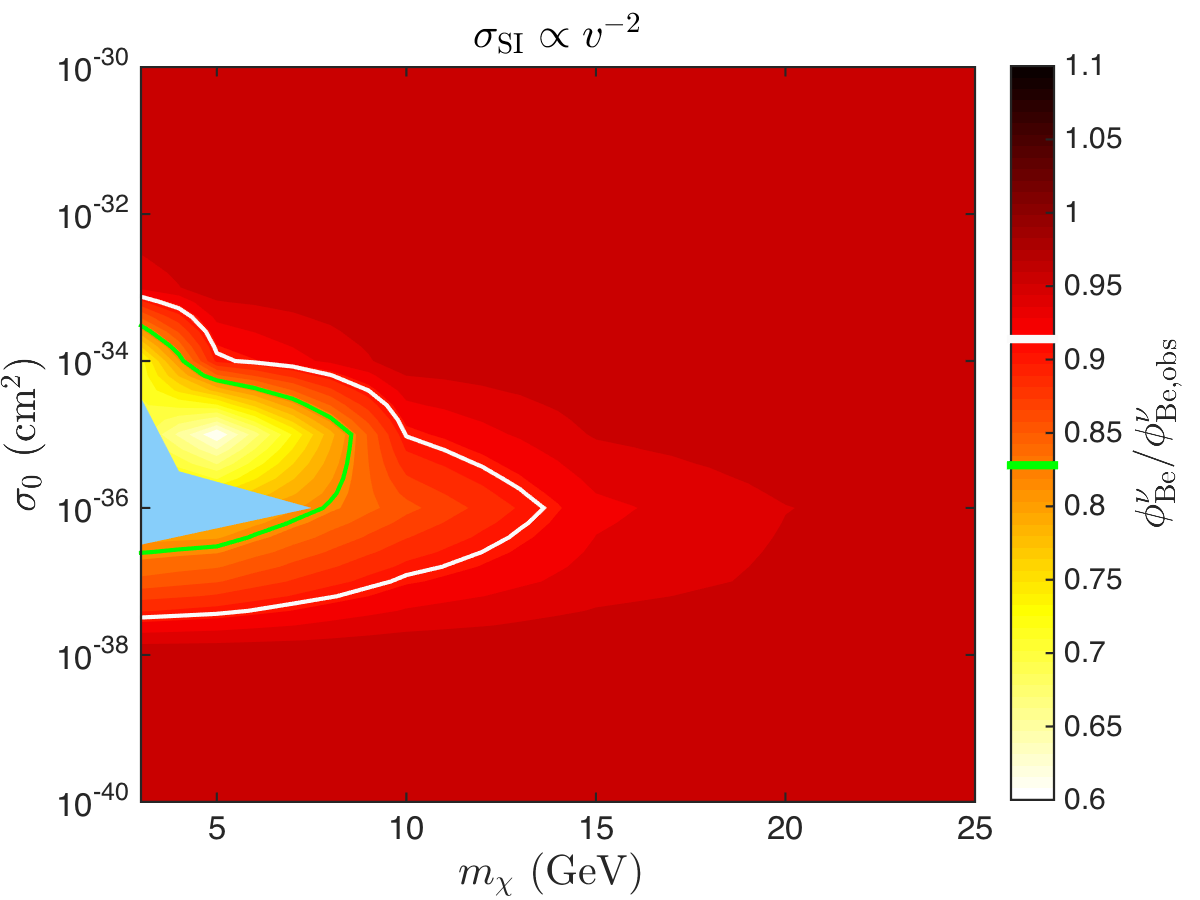} & \includegraphics[height = 0.32\textwidth]{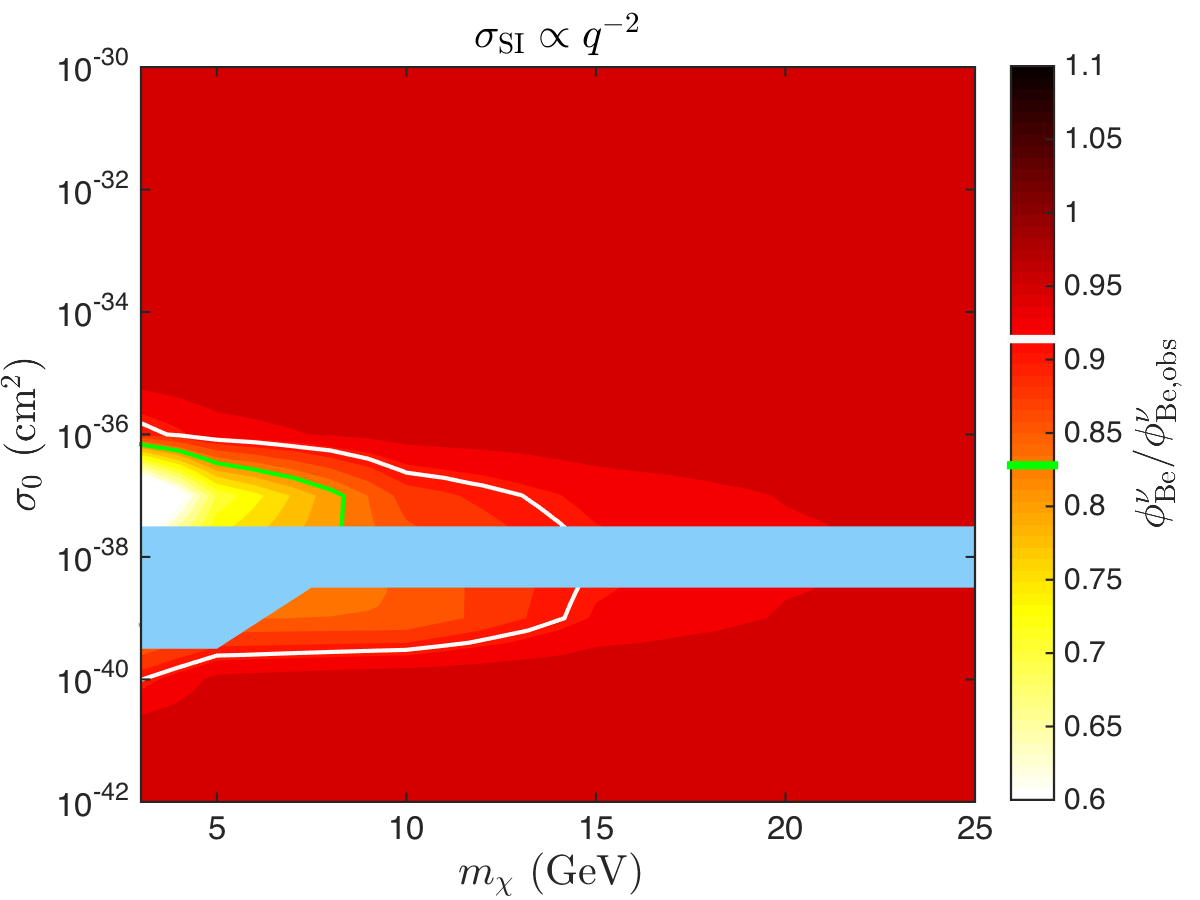} \\
\includegraphics[height = 0.32\textwidth]{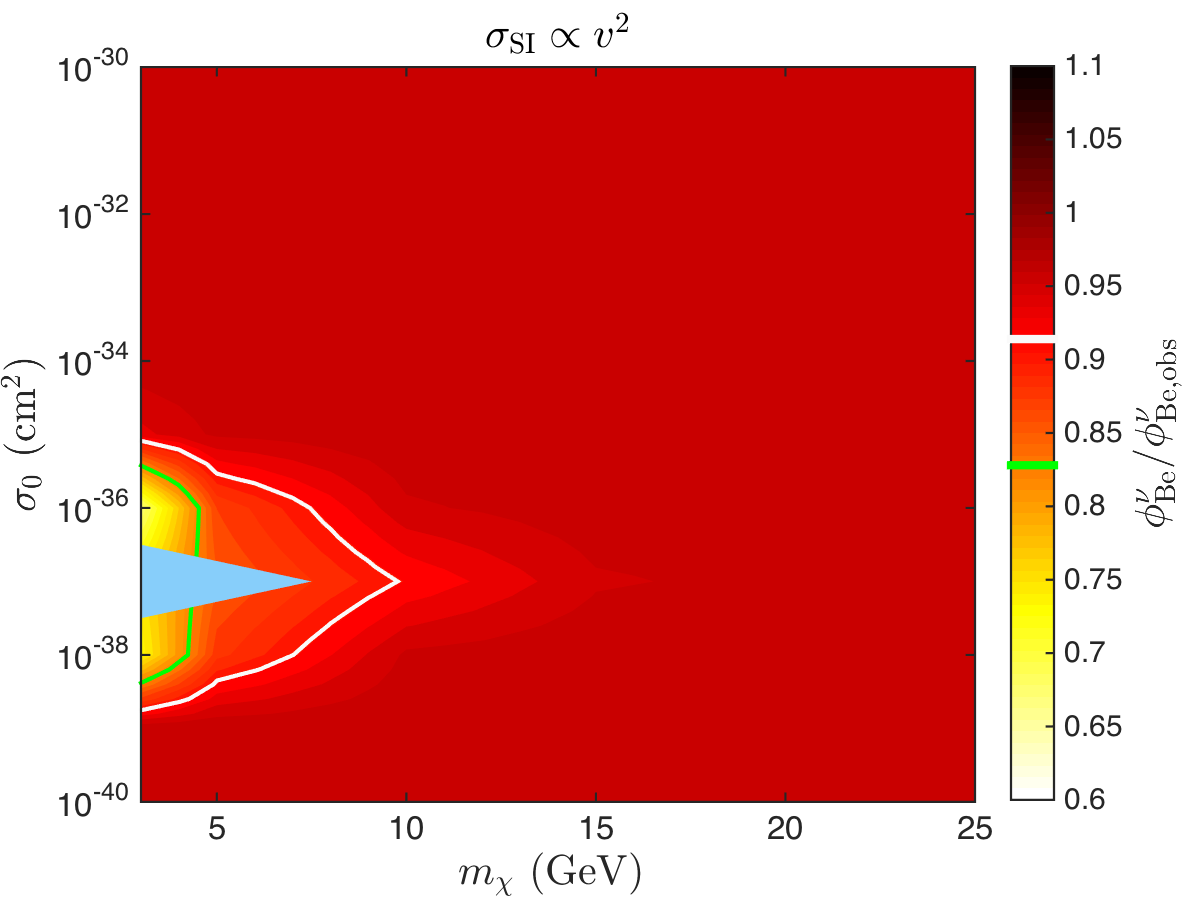} & \includegraphics[height = 0.32\textwidth]{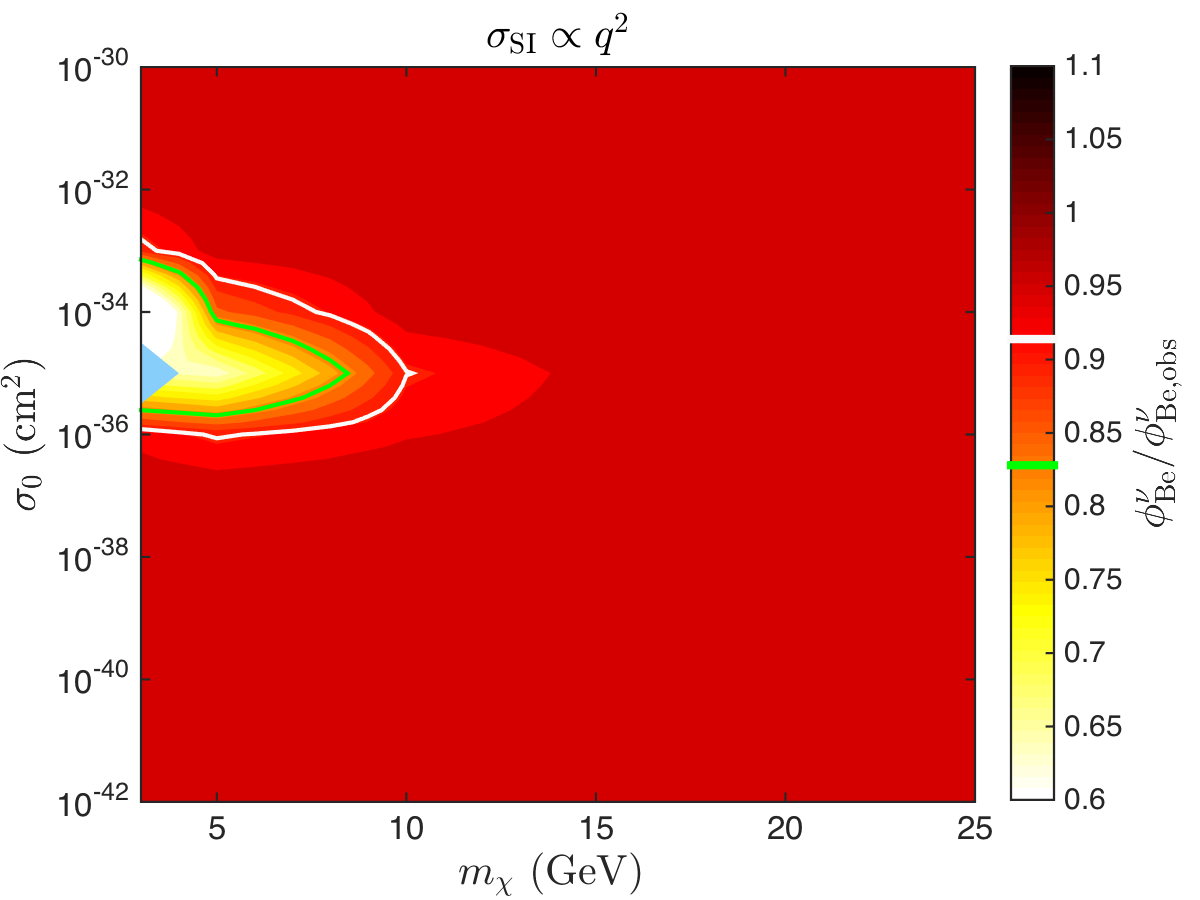} \\
\includegraphics[height = 0.32\textwidth]{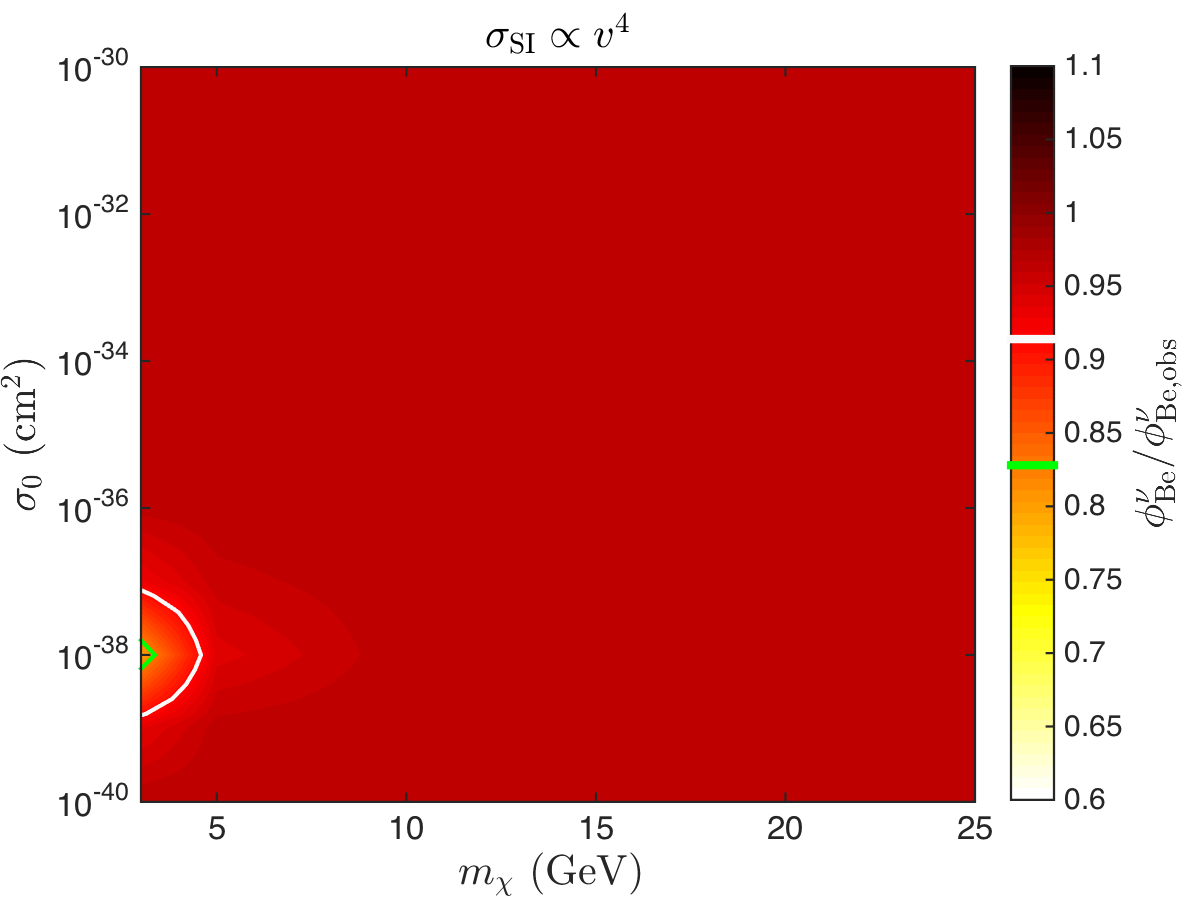} & \includegraphics[height = 0.32\textwidth]{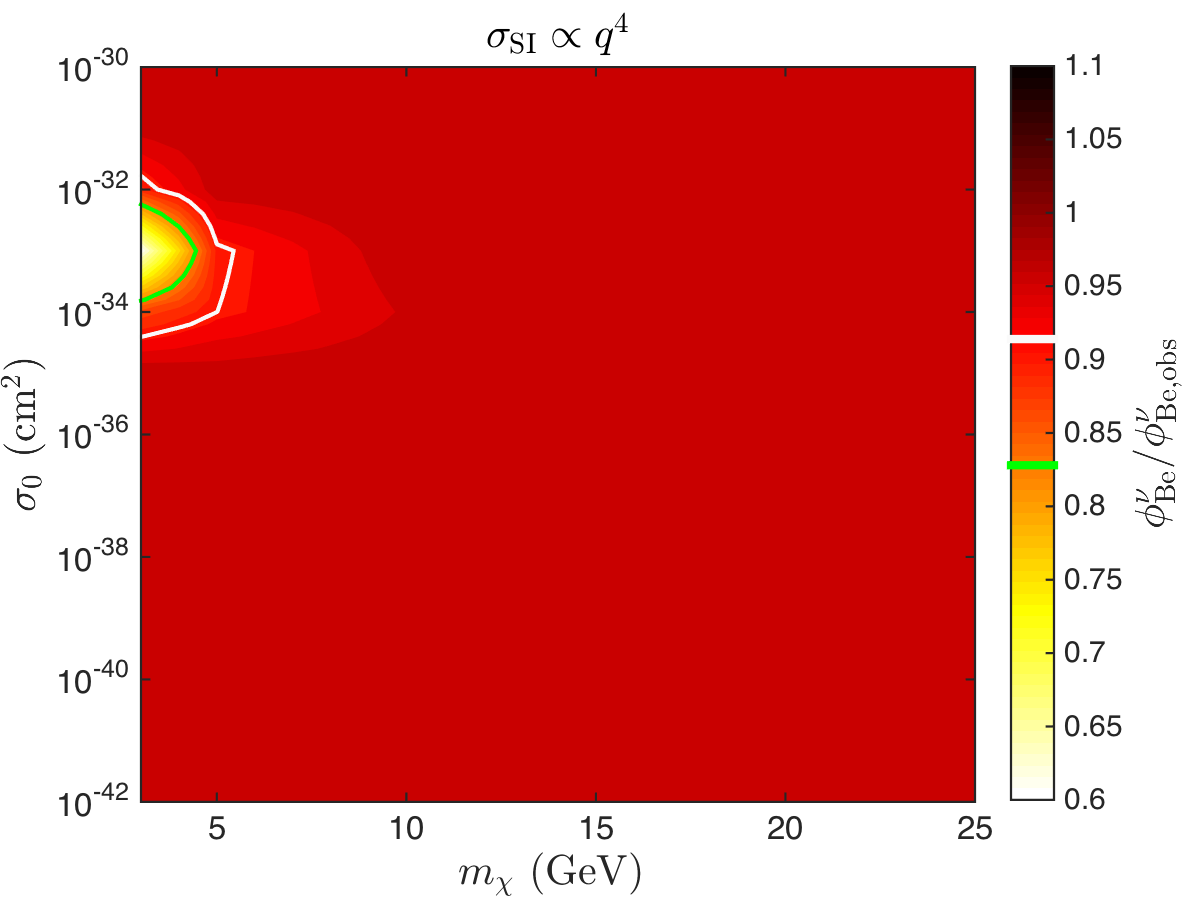} \\
\end{tabular}
\caption{The ratio of the predicted $^7$Be neutrino flux to the  measured value $\phi^\nu_{\rm Be,obs} = 4.82 \times 10^9$\,cm$^{-2}$s$^{-1}$~\cite{Abe:2010hy}, for each type of spin-independent dark matter coupling defined in Eq.\ \ref{qdepvdep}.  In every case the white and green lines show the isocontours where the flux is respectively 1 and 2$\sigma$ lower than the observed values, based on observational (5\%) and modelling (7\%) errors, added in quadrature. The cross-sections are normalized such that $\sigma = \sigma_0 (v/v_0)^{2n}$ or $\sigma = \sigma_0 (q/q_0)^{2n}$, with $v_0 = 220$\,km\,s$^{-1}$ and $q_0 = 40$\,MeV.  Regions masked in light blue correspond to parameter combinations where models did not converge.}
\label{SIBerylliumfluxes}
\end{figure}

\begin{figure}[p]
\begin{tabular}{c@{\hspace{0.04\textwidth}}c}
\multicolumn{2}{c}{\includegraphics[height = 0.32\textwidth]{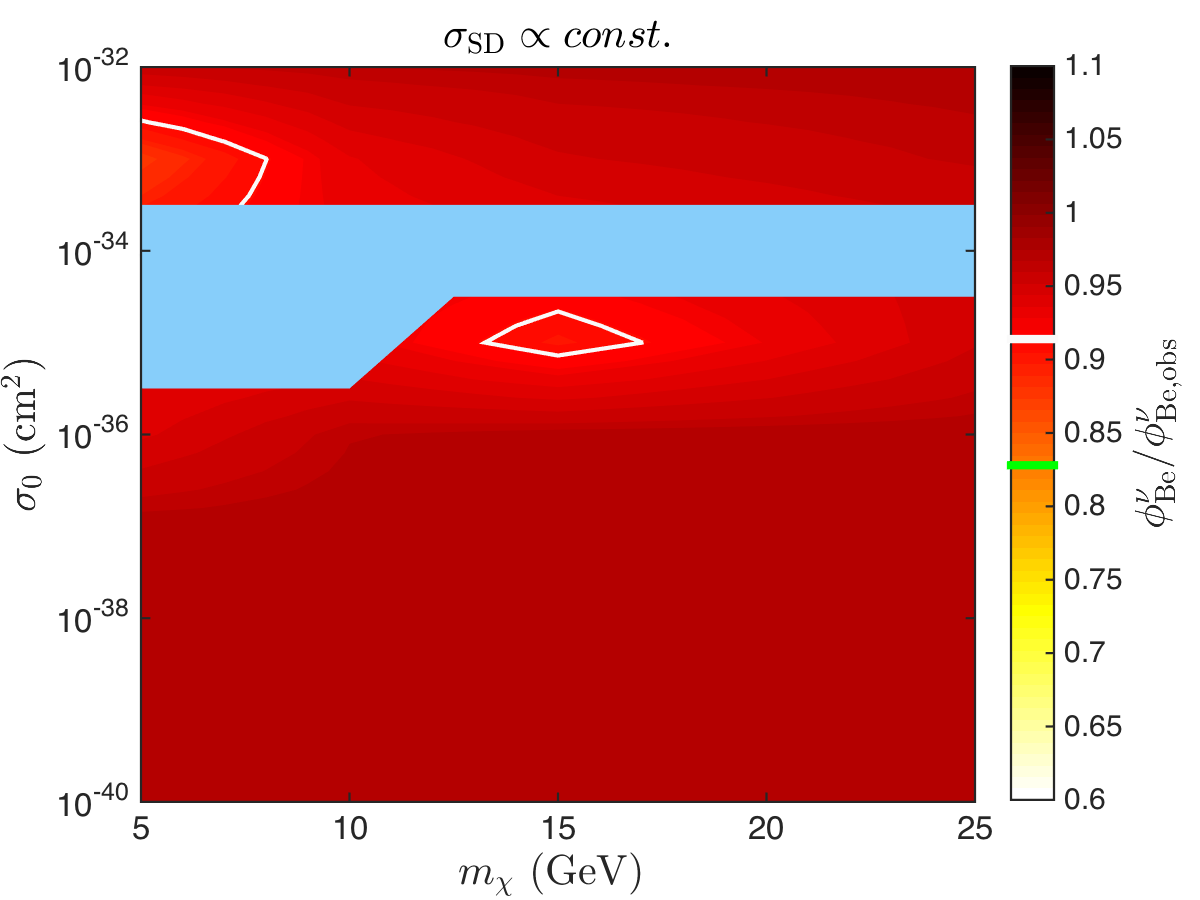}} \\
\includegraphics[height = 0.32\textwidth]{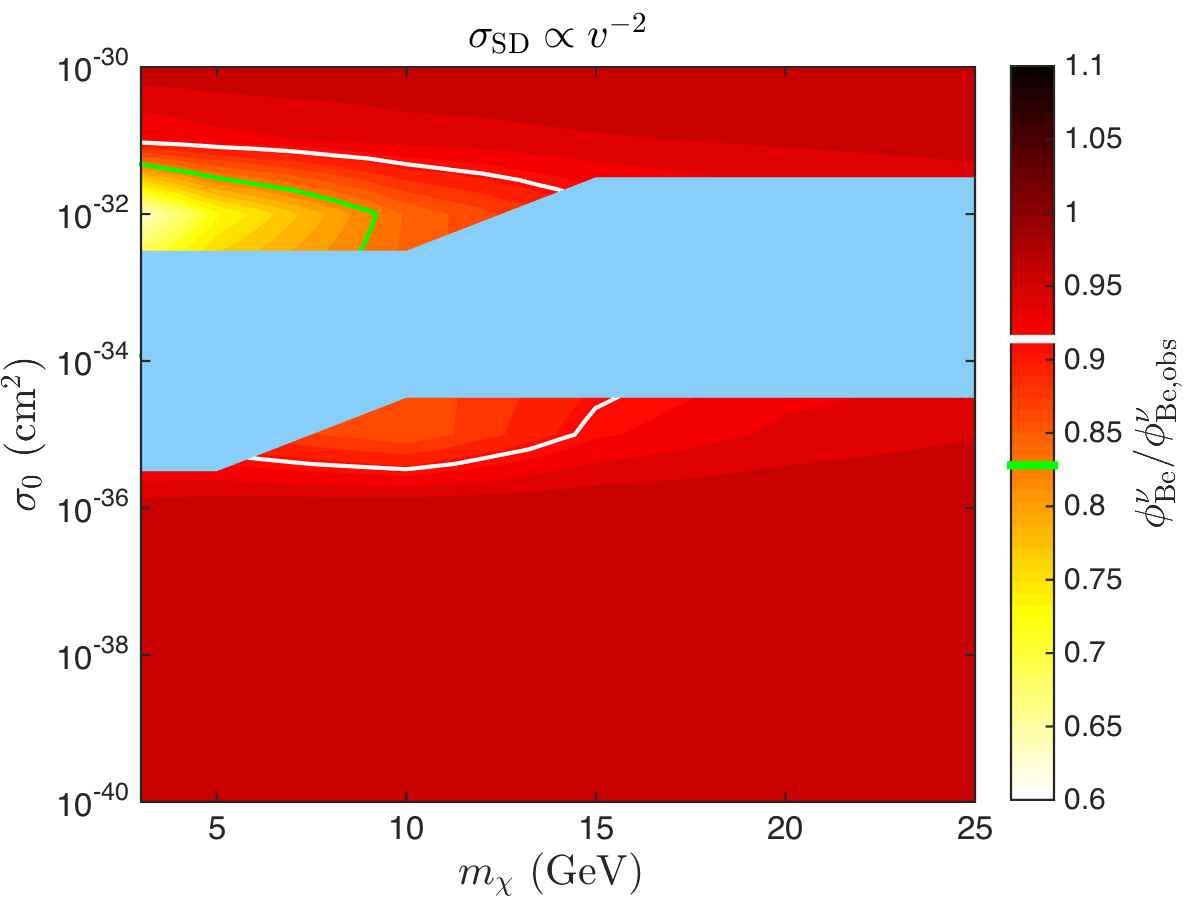} & \includegraphics[height = 0.32\textwidth]{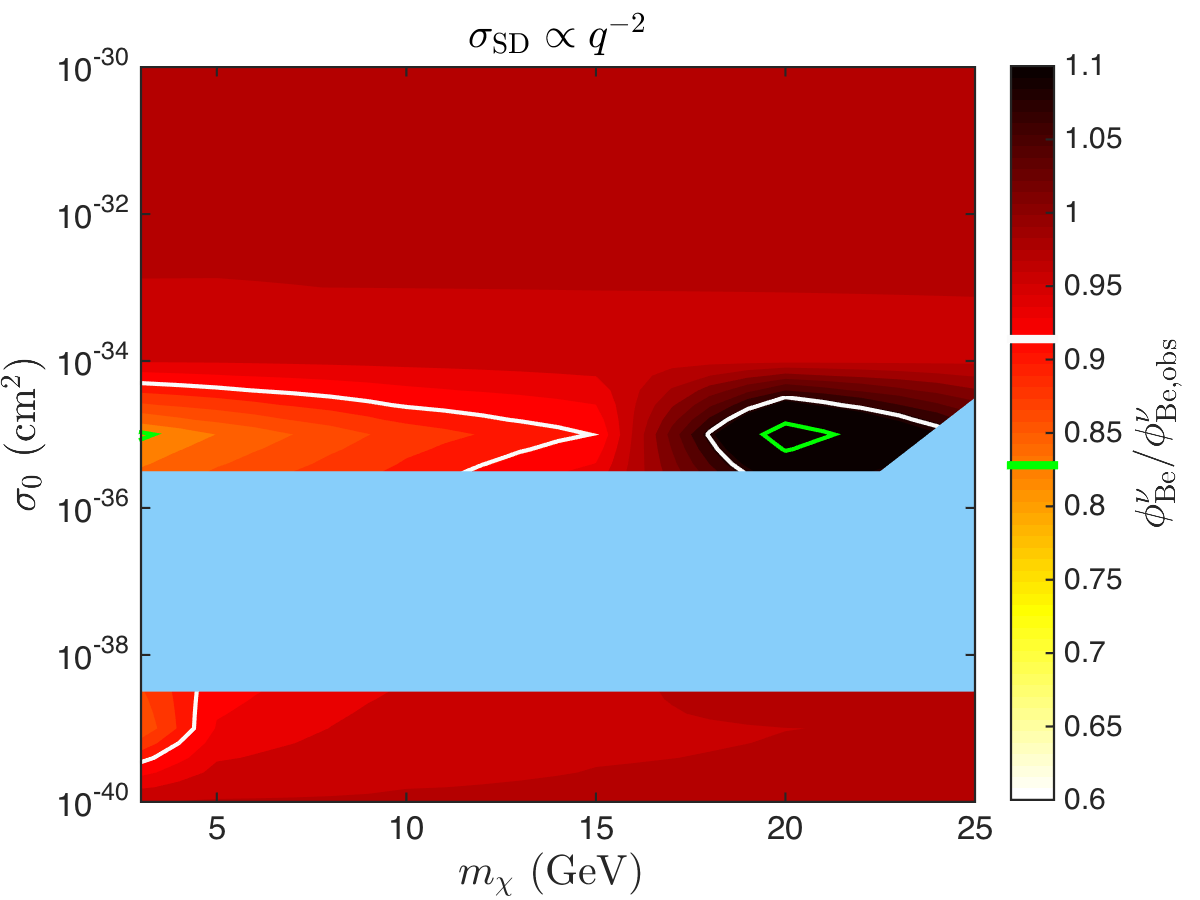} \\
\includegraphics[height = 0.32\textwidth]{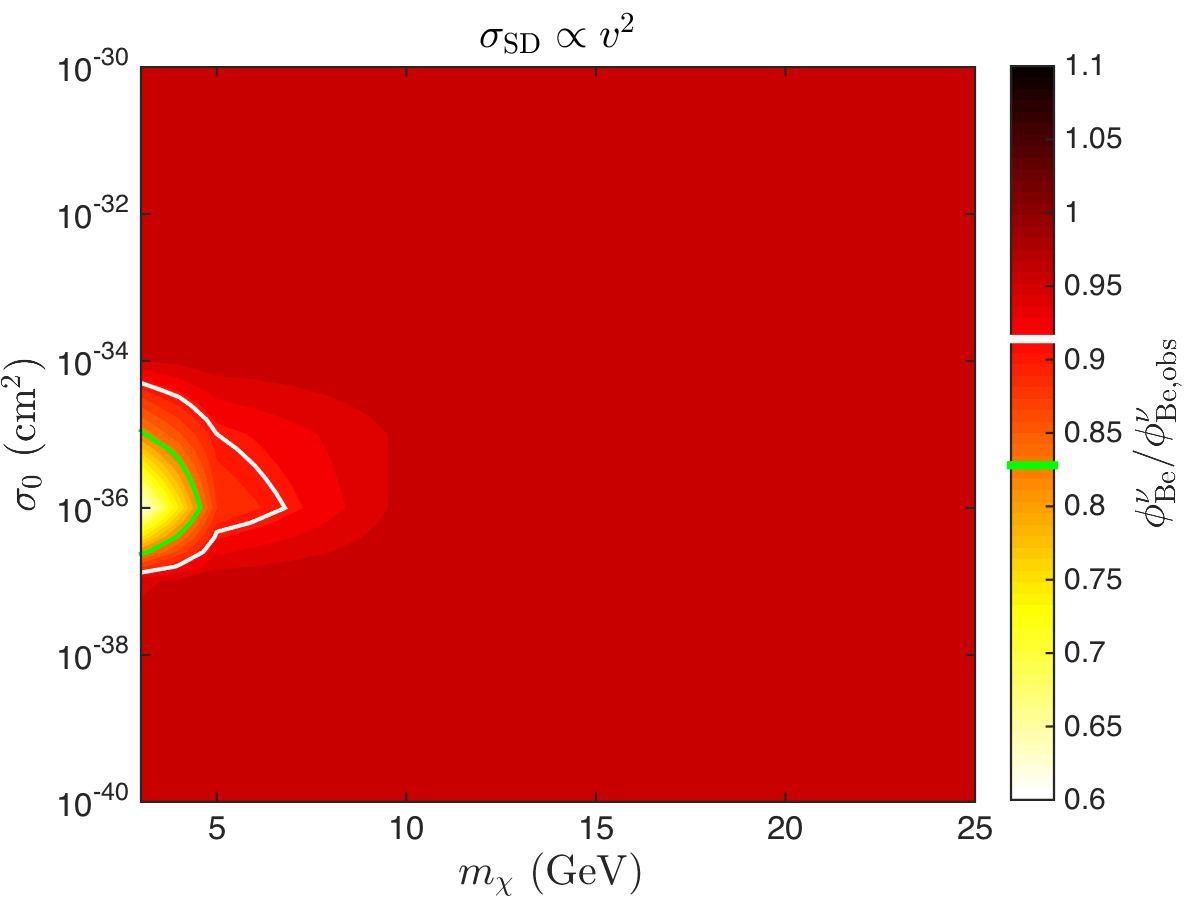} & \includegraphics[height = 0.32\textwidth]{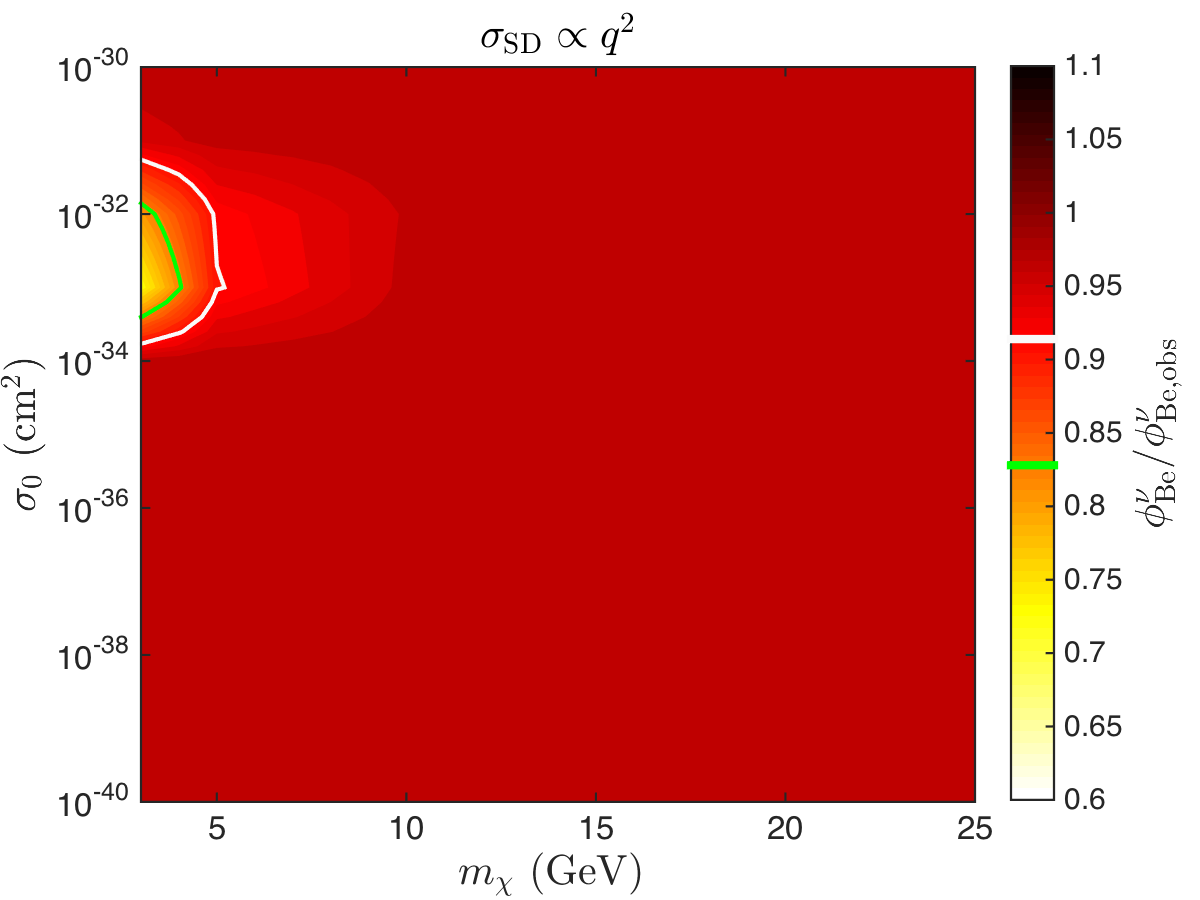} \\
\includegraphics[height = 0.32\textwidth]{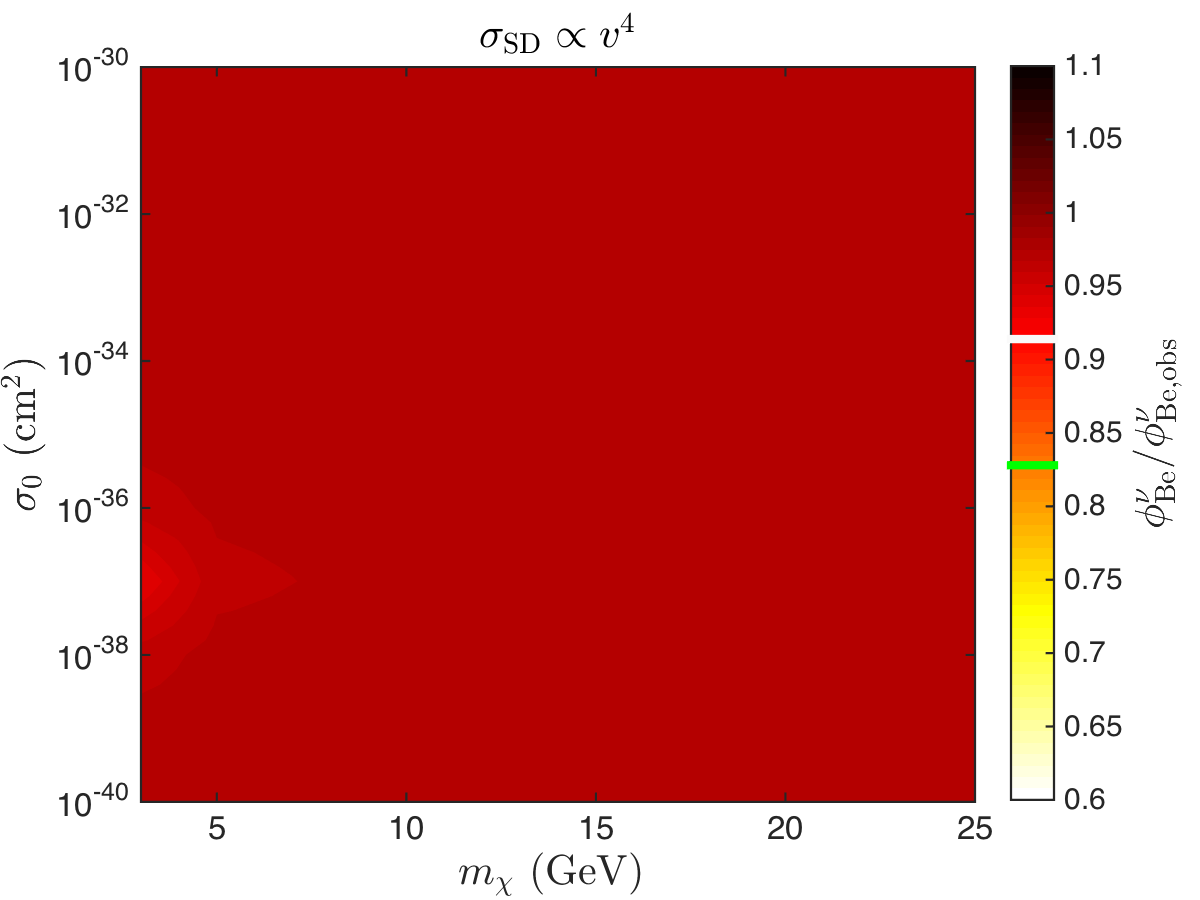} & \includegraphics[height = 0.32\textwidth]{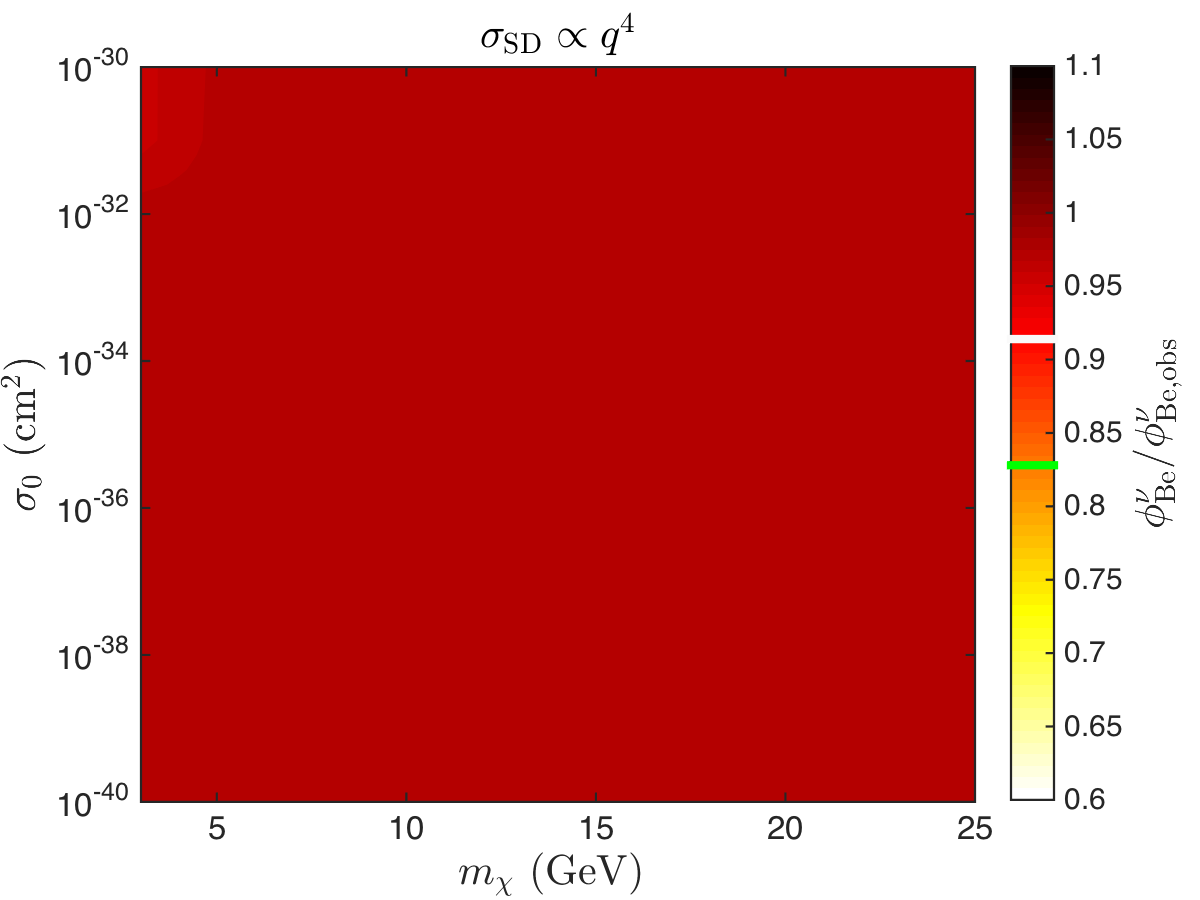} \\
\end{tabular}
\caption{As per Fig.\ \ref{SIBerylliumfluxes}, but for spin-dependent couplings.}
\label{SDBerylliumfluxes}
\end{figure}

\begin{figure}[!p]
\begin{tabular}{c@{\hspace{0.04\textwidth}}c}
\multicolumn{2}{c}{\includegraphics[height = 0.32\textwidth]{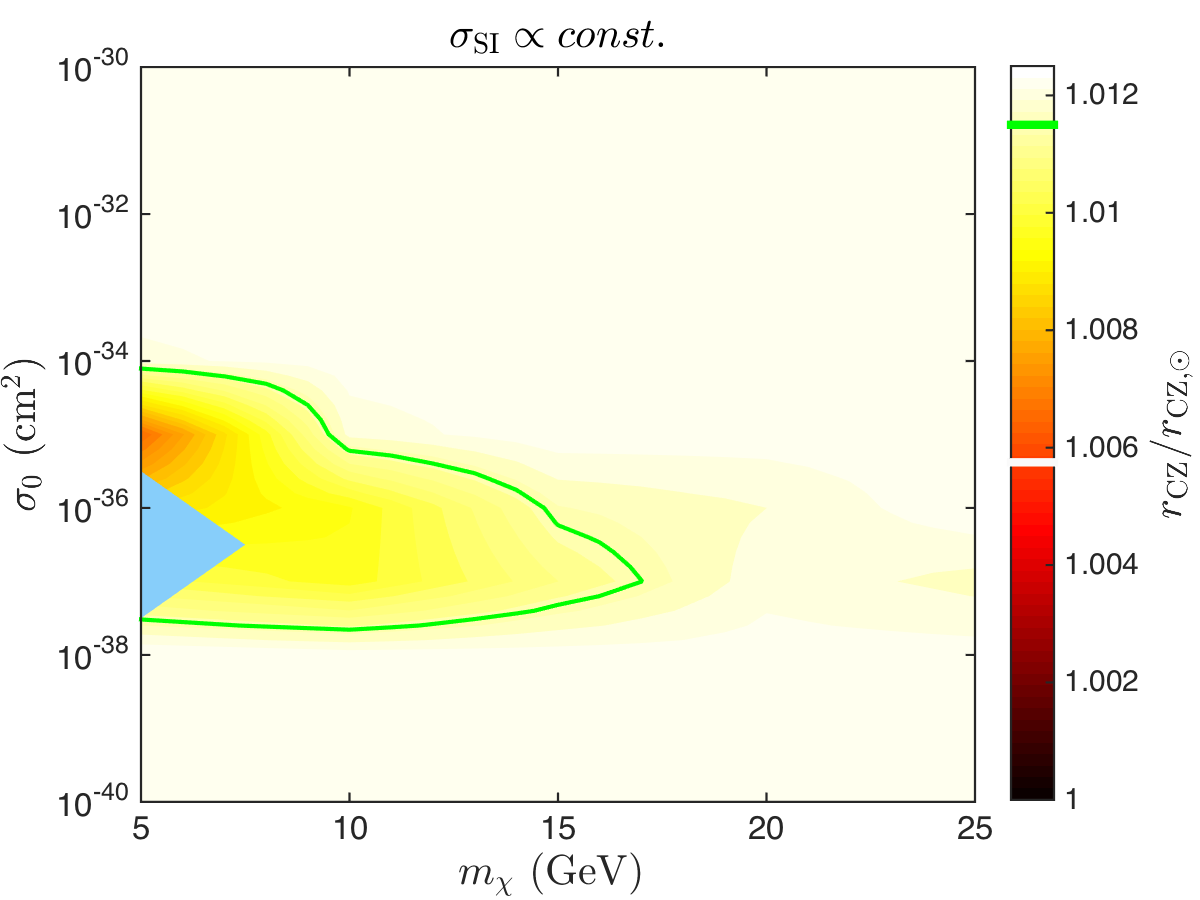}} \\
\includegraphics[height = 0.32\textwidth]{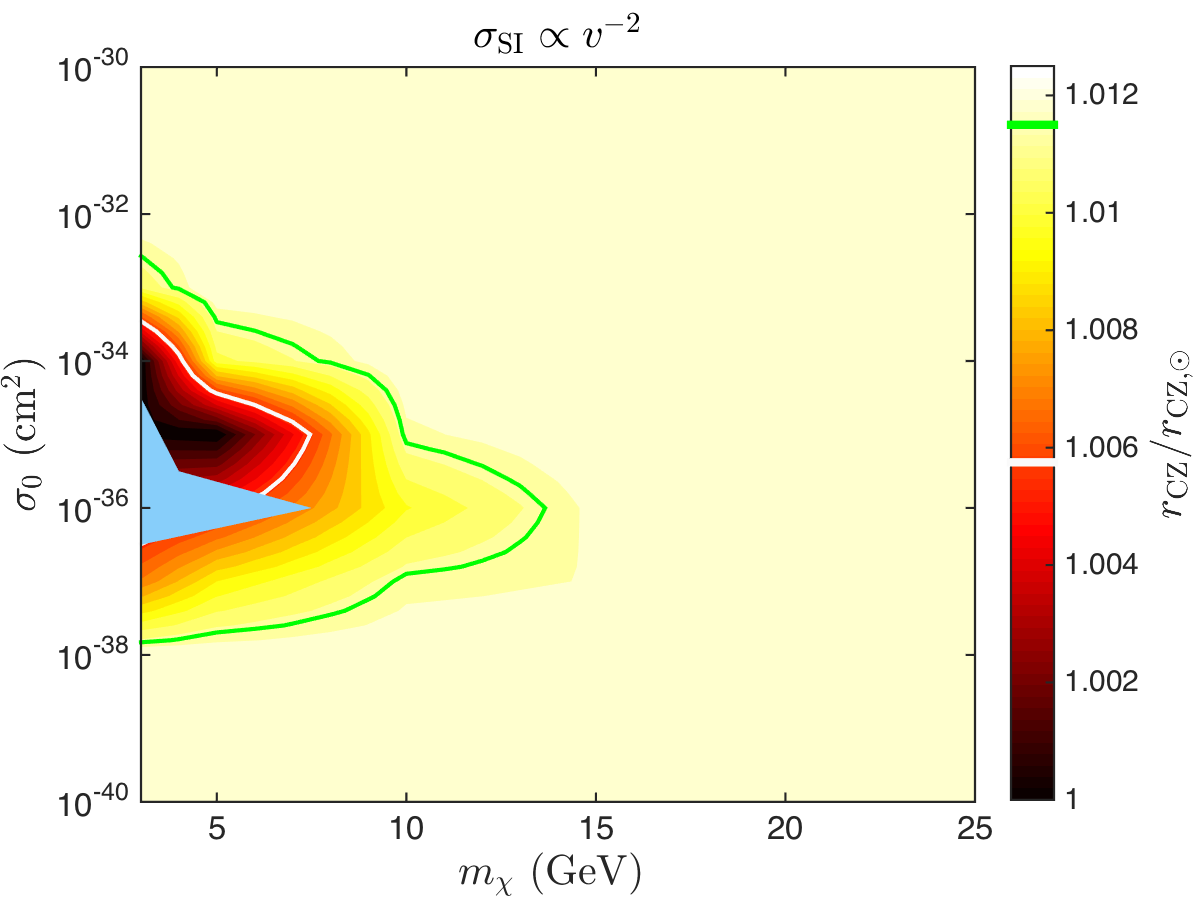} & \includegraphics[height = 0.32\textwidth]{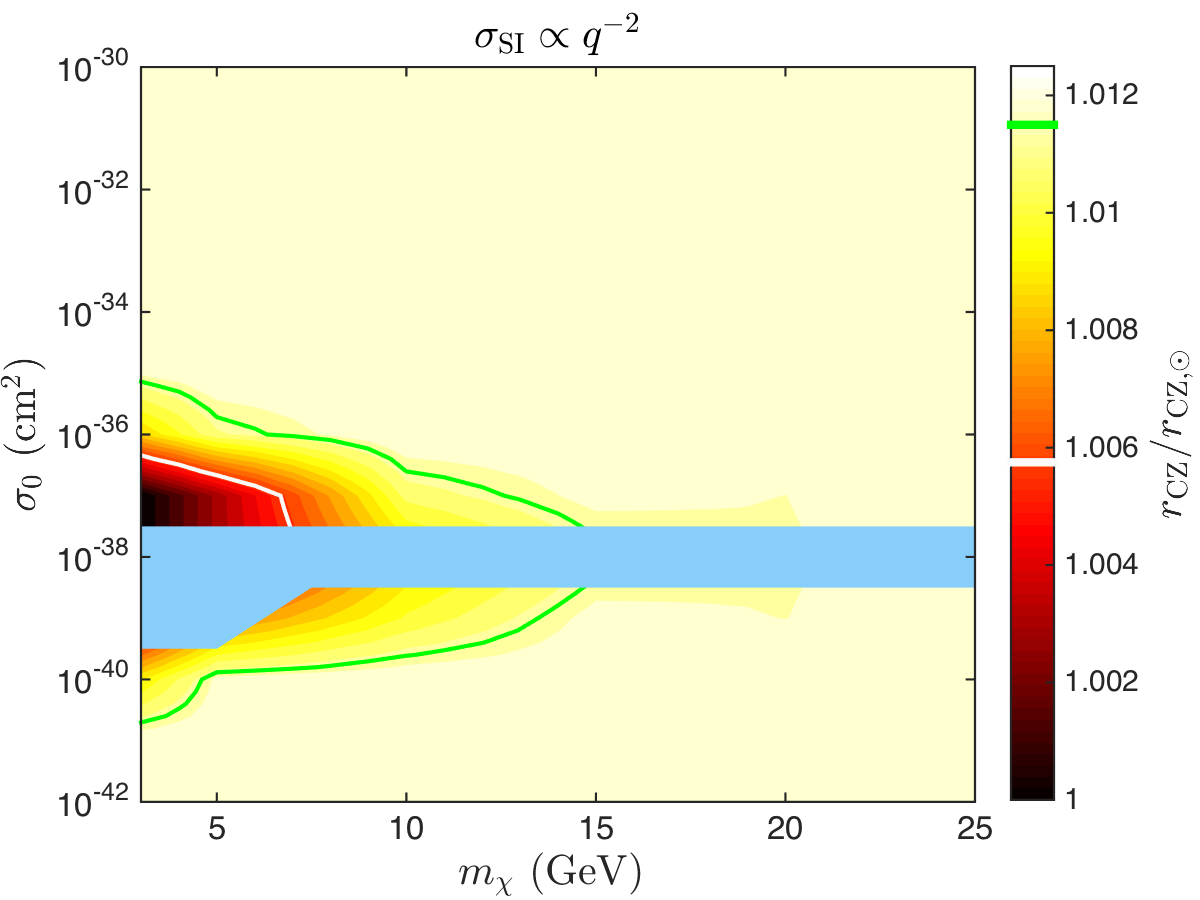} \\
\includegraphics[height = 0.32\textwidth]{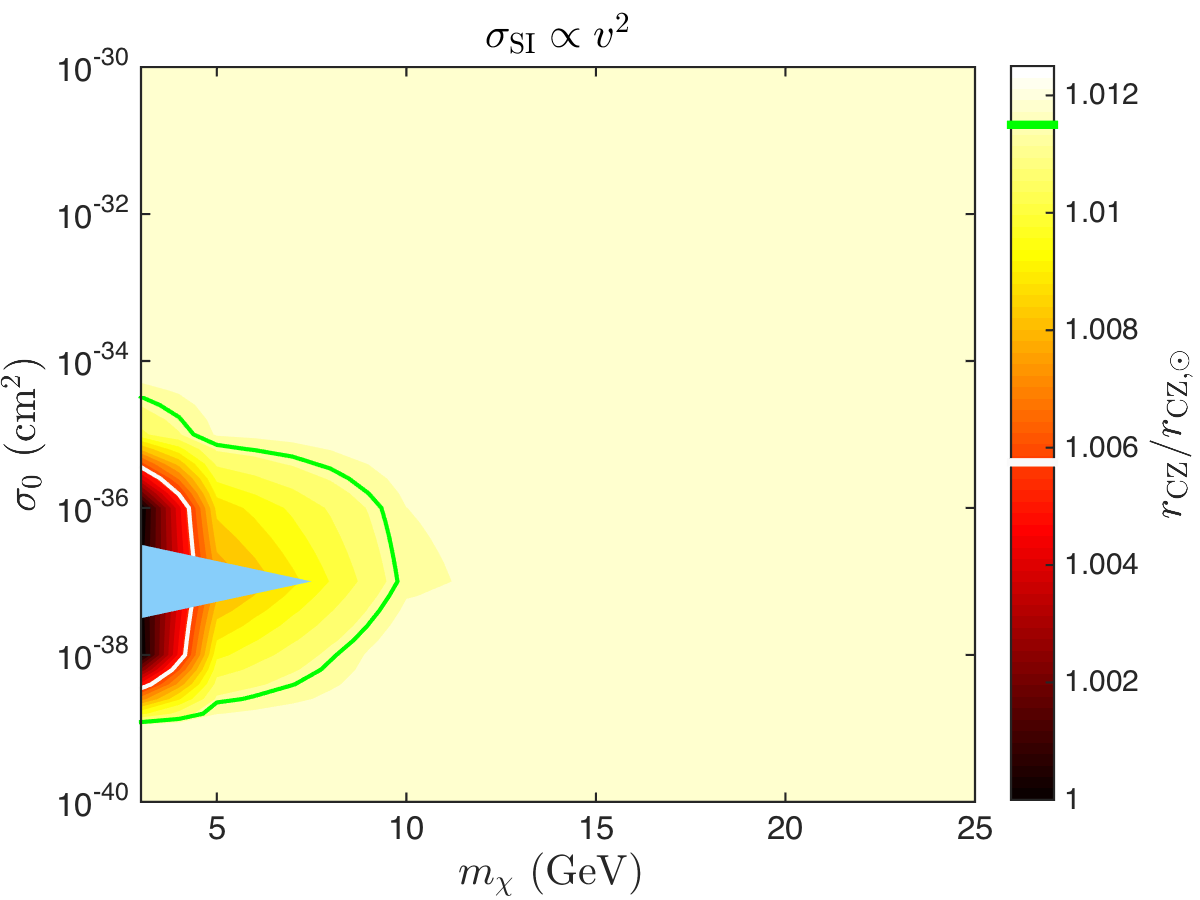} & \includegraphics[height = 0.32\textwidth]{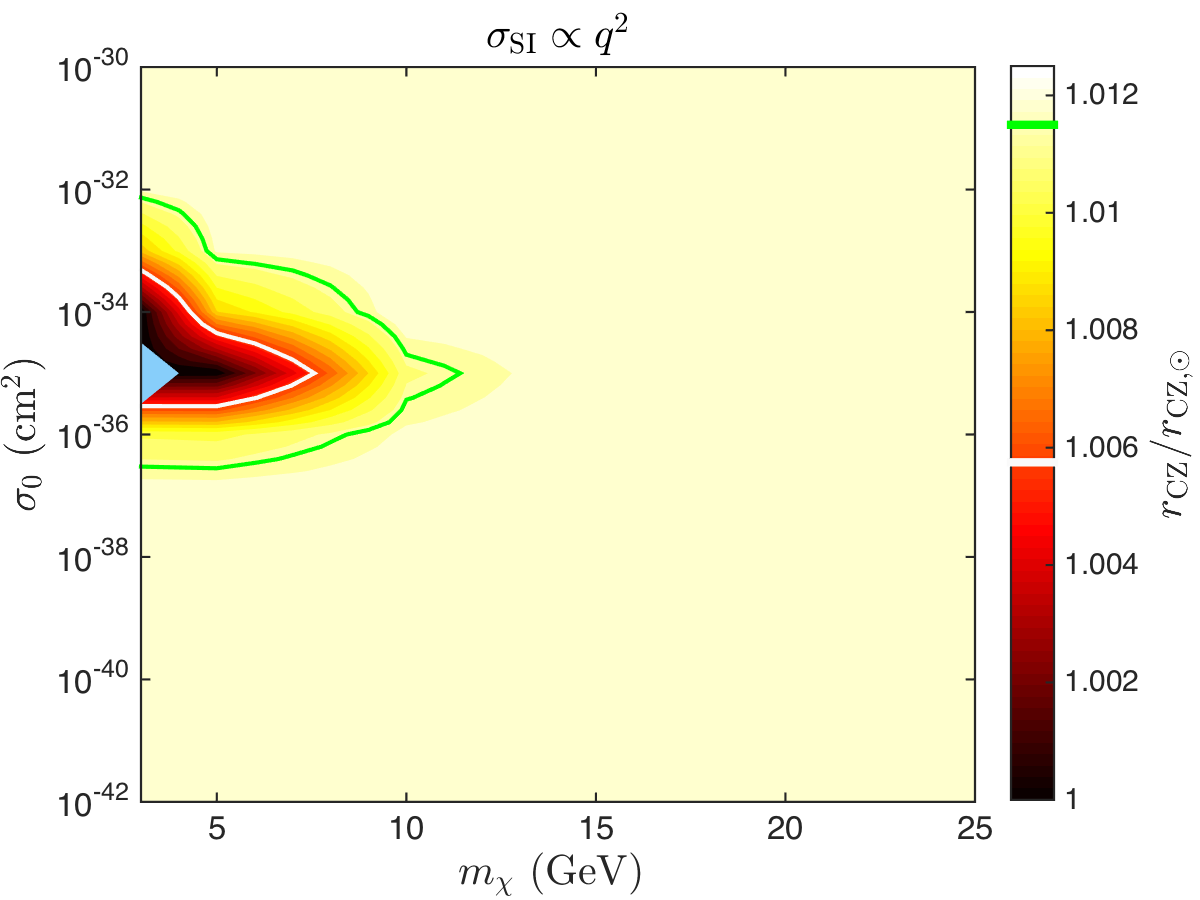} \\
\includegraphics[height = 0.32\textwidth]{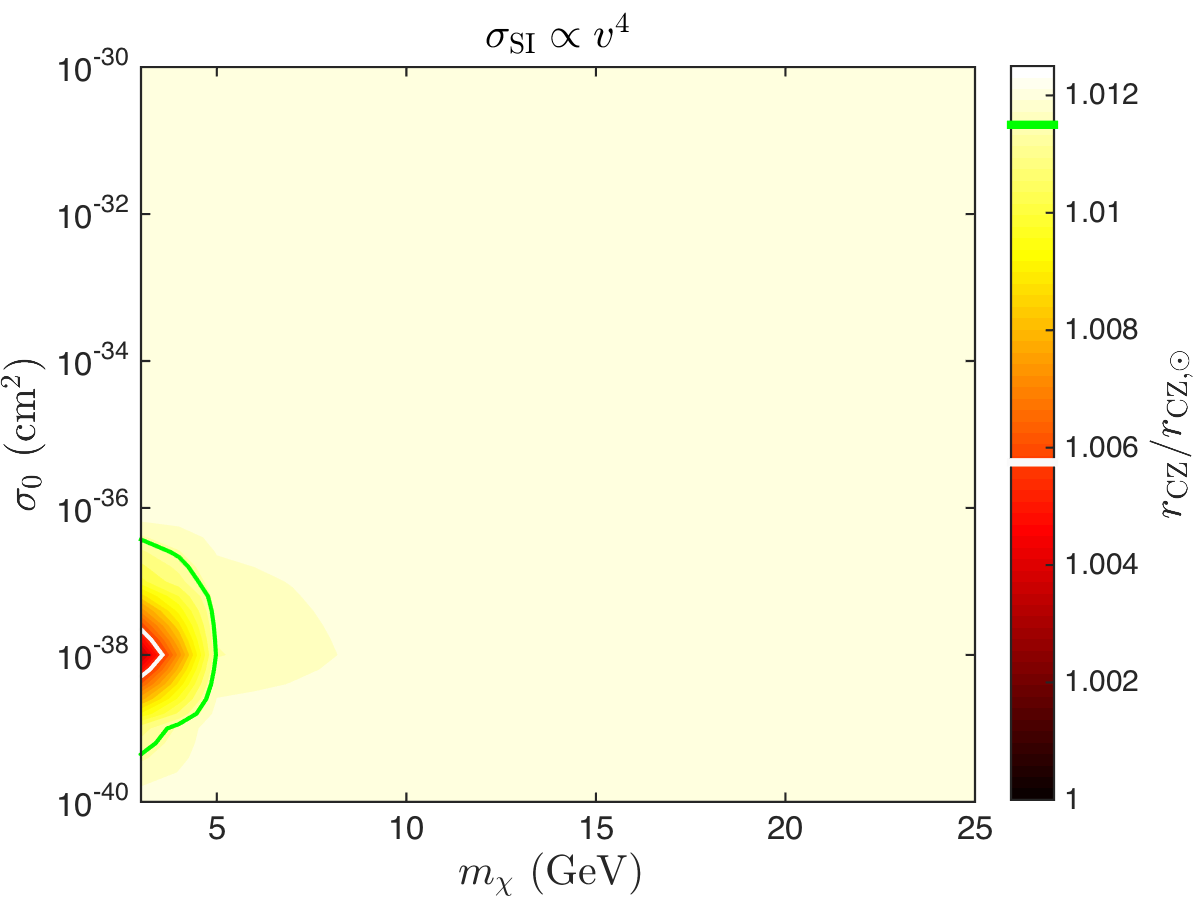} & \includegraphics[height = 0.32\textwidth]{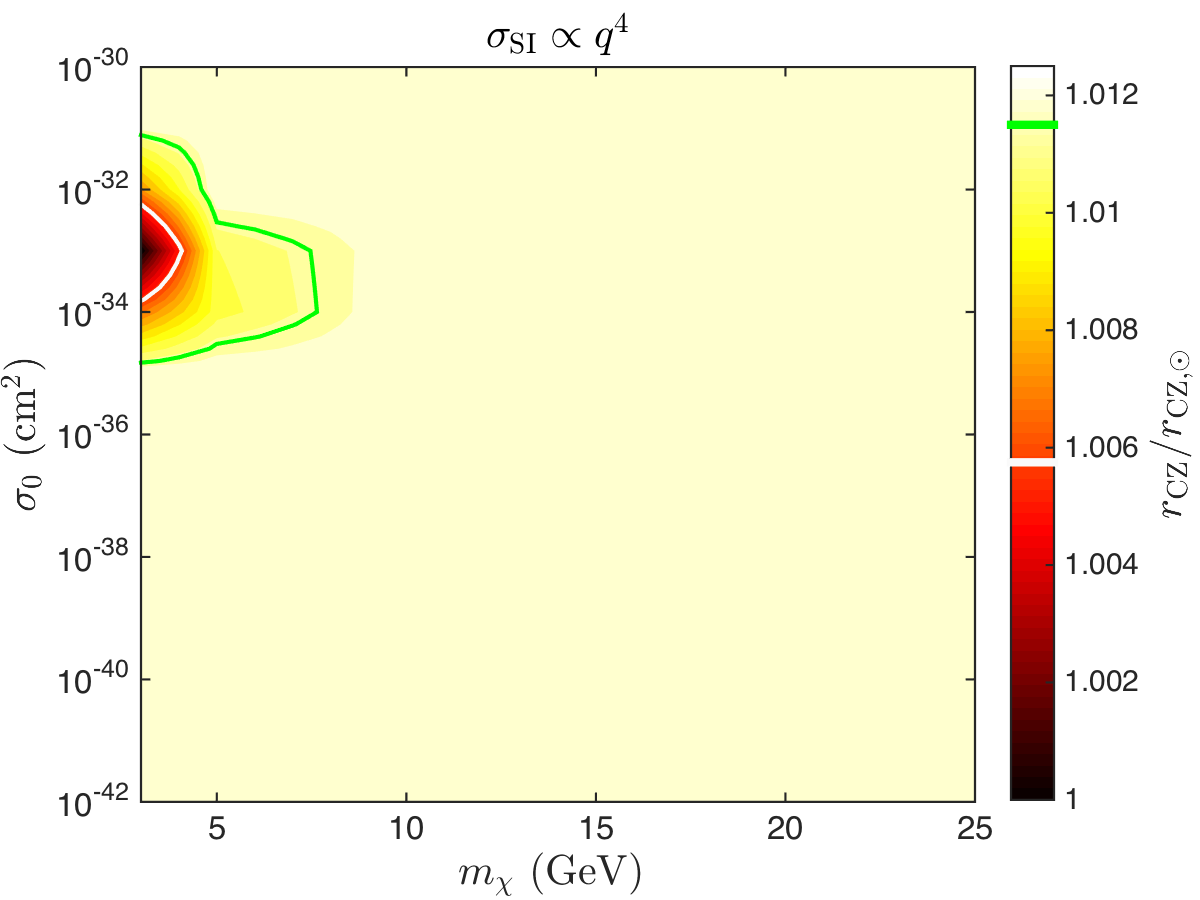} \\
\end{tabular}
\caption{Ratio between the modelled and measured location of the bottom of the convection zone $r_{\rm CZ}$, for spin-independent couplings. Darker regions represent a better fit to the observed value than the SSM. The white and green lines represent the contours at which the predicted value falls within $1\sigma$ and $2\sigma$ of the measured value, respectively. The theoretical uncertainty on $r_{\rm CZ}$ (0.004\,$R_\odot$) is much larger than the experimental error (0.001\,$R_\odot$), so the former dominates when we add them in quadrature. Regions masked in light blue correspond to parameter combinations where models did not converge.}
\label{SIrc}
\end{figure}

\begin{figure}[!p]
\begin{tabular}{c@{\hspace{0.04\textwidth}}c}
\multicolumn{2}{c}{\includegraphics[height = 0.32\textwidth]{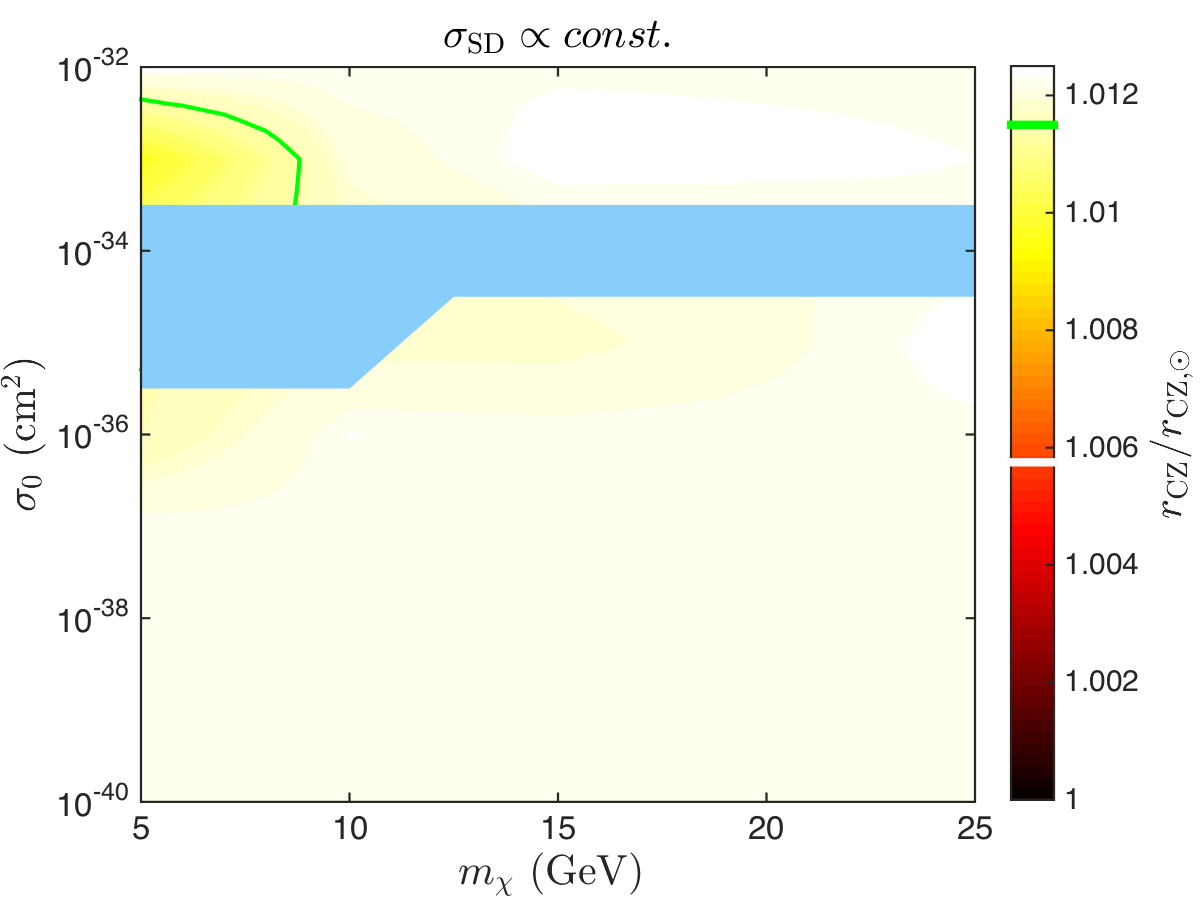}} \\
\includegraphics[height = 0.32\textwidth]{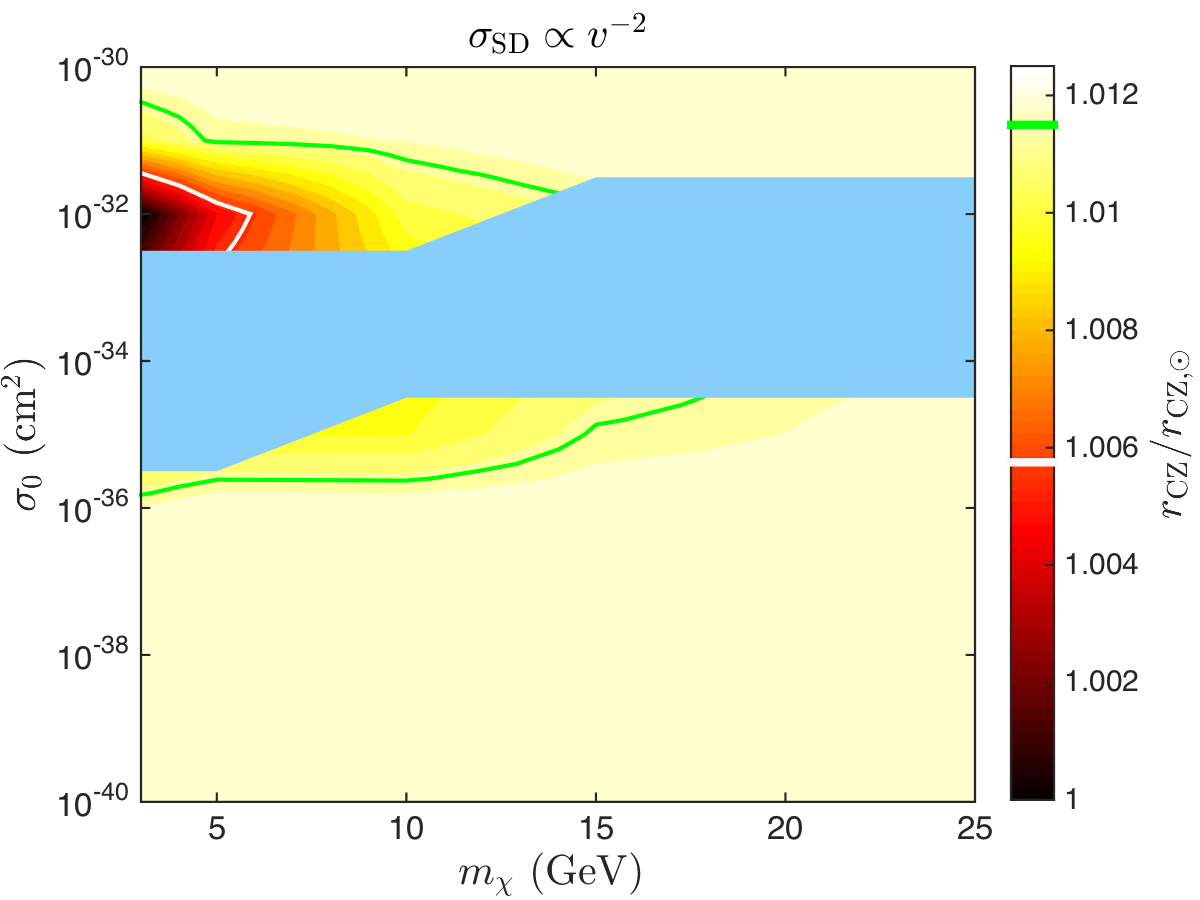} & \includegraphics[height = 0.32\textwidth]{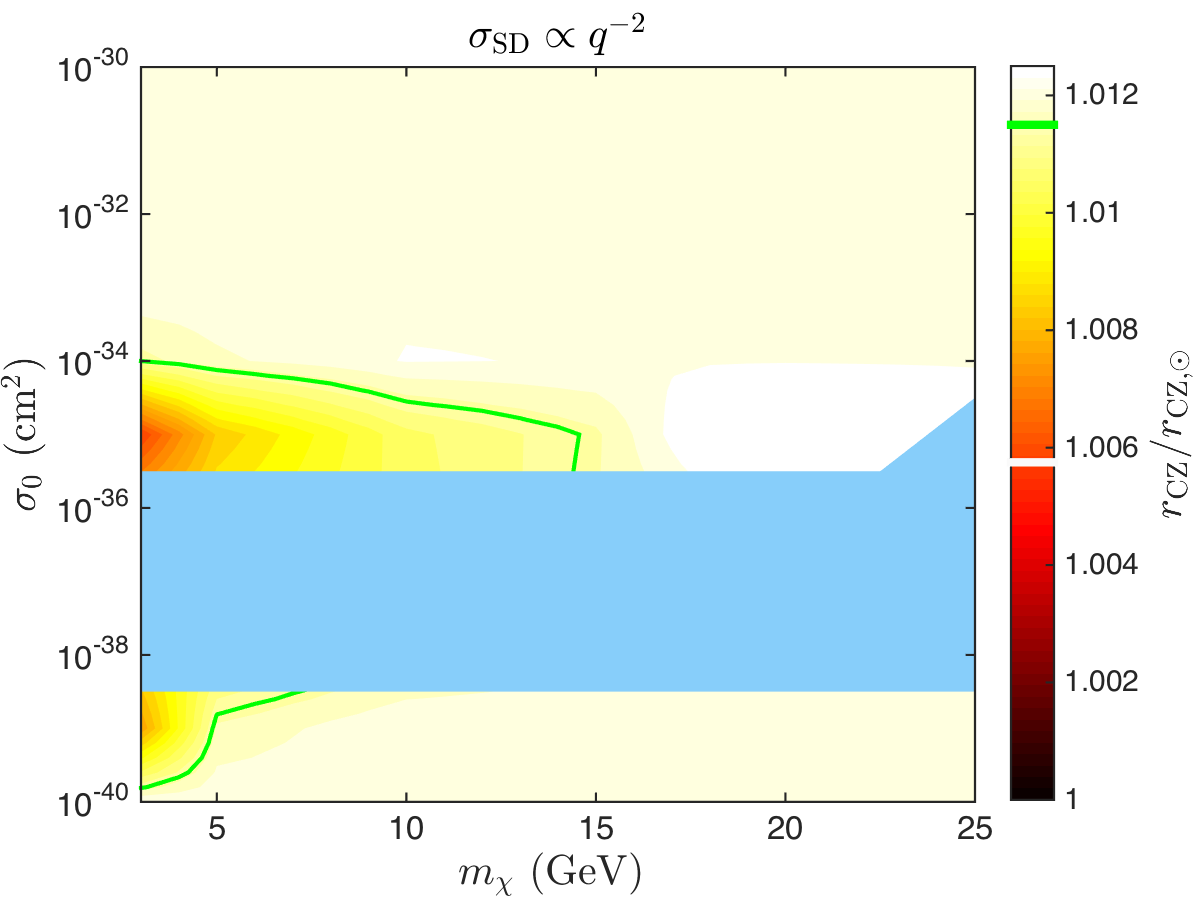} \\
\includegraphics[height = 0.32\textwidth]{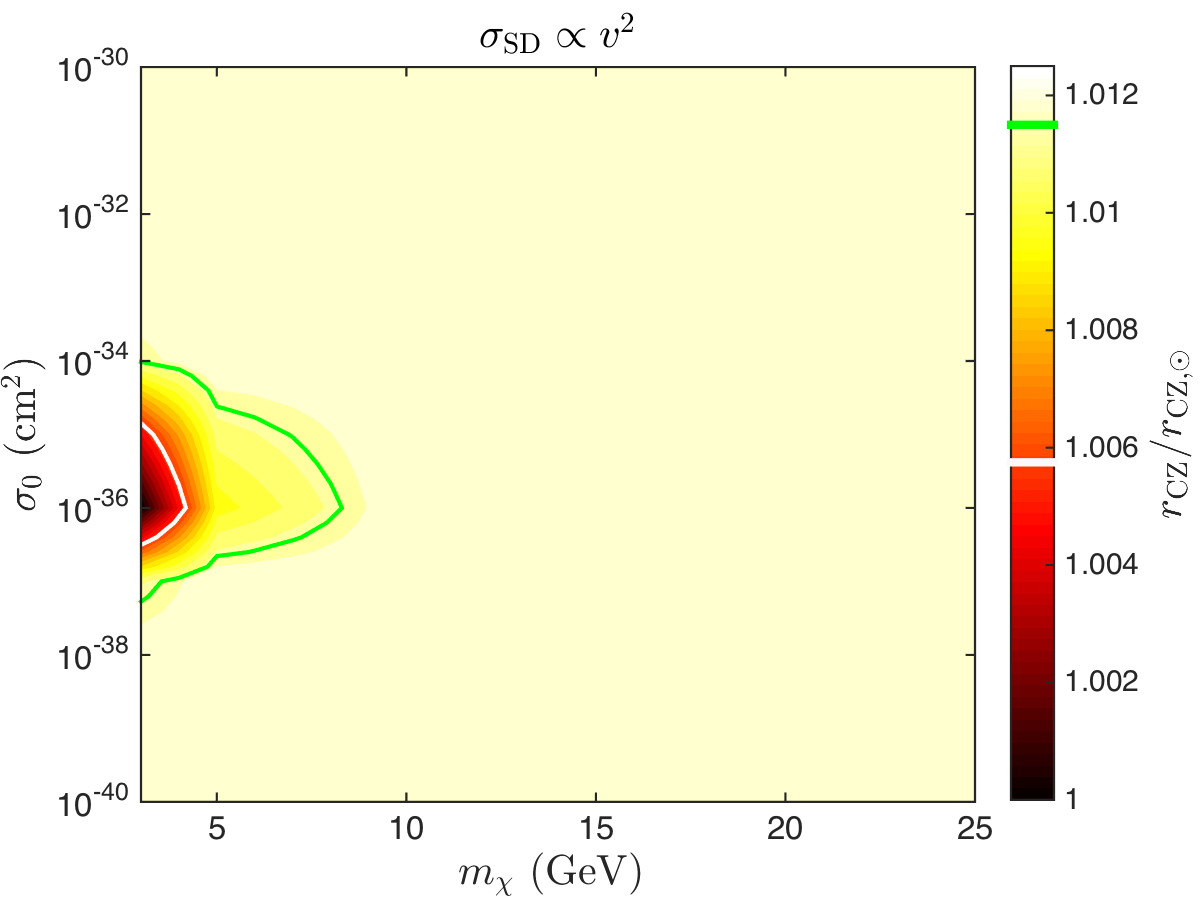} & \includegraphics[height = 0.32\textwidth]{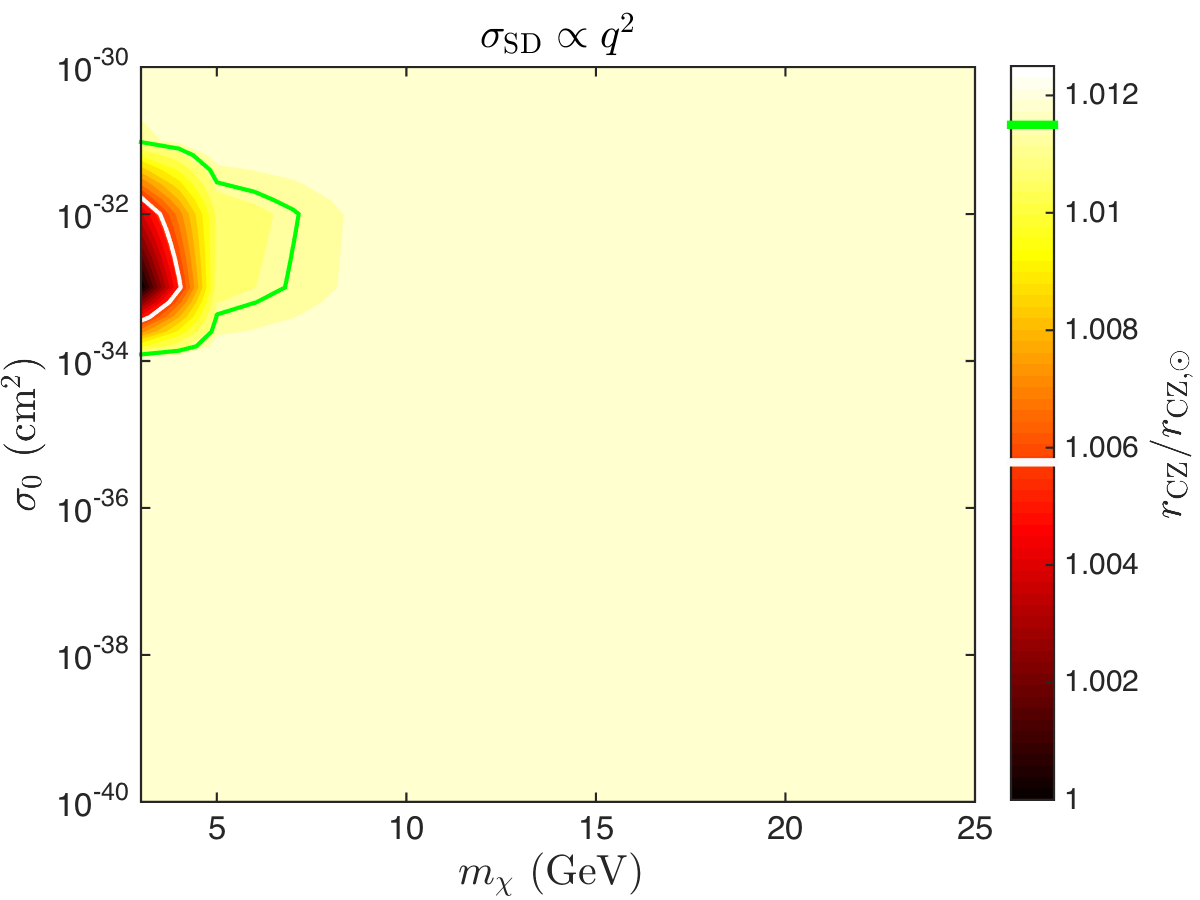} \\
\includegraphics[height = 0.32\textwidth]{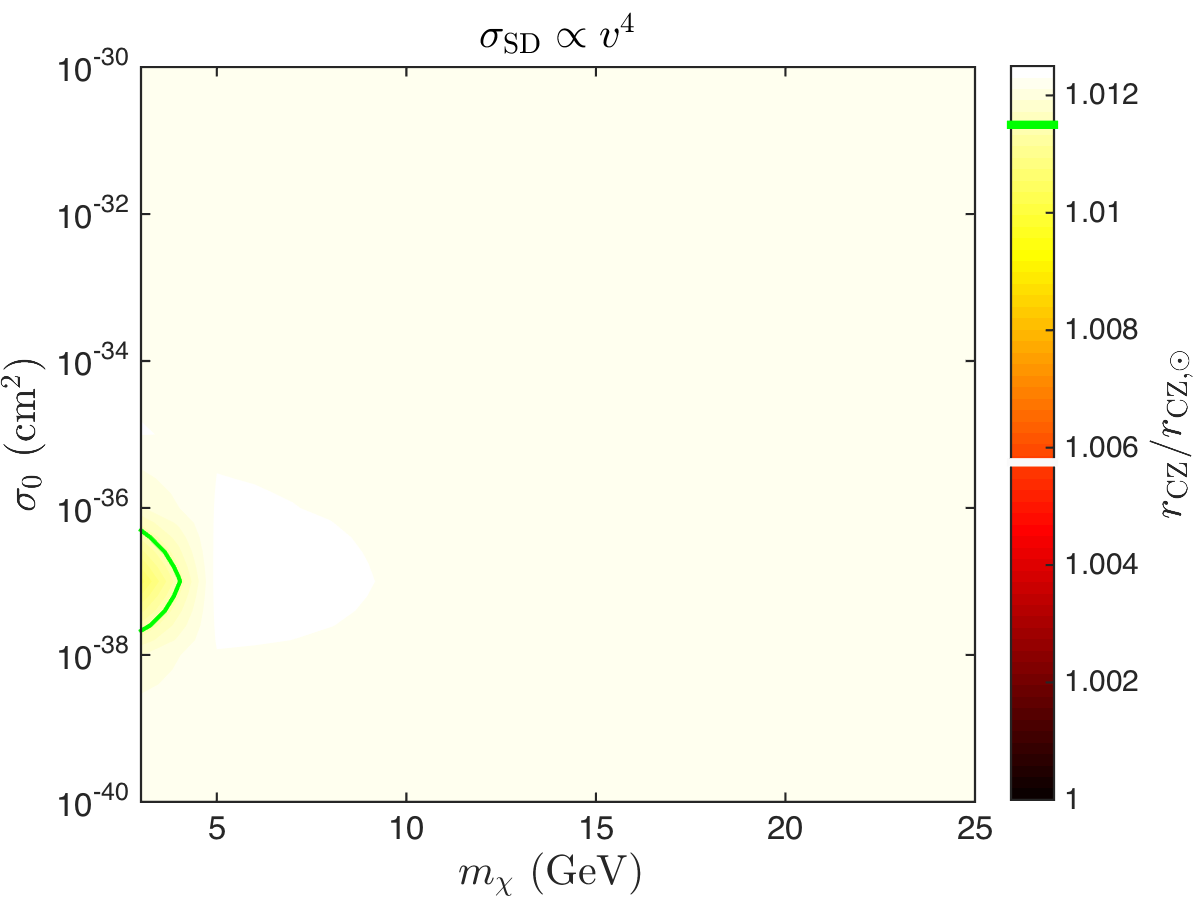} & \includegraphics[height = 0.32\textwidth]{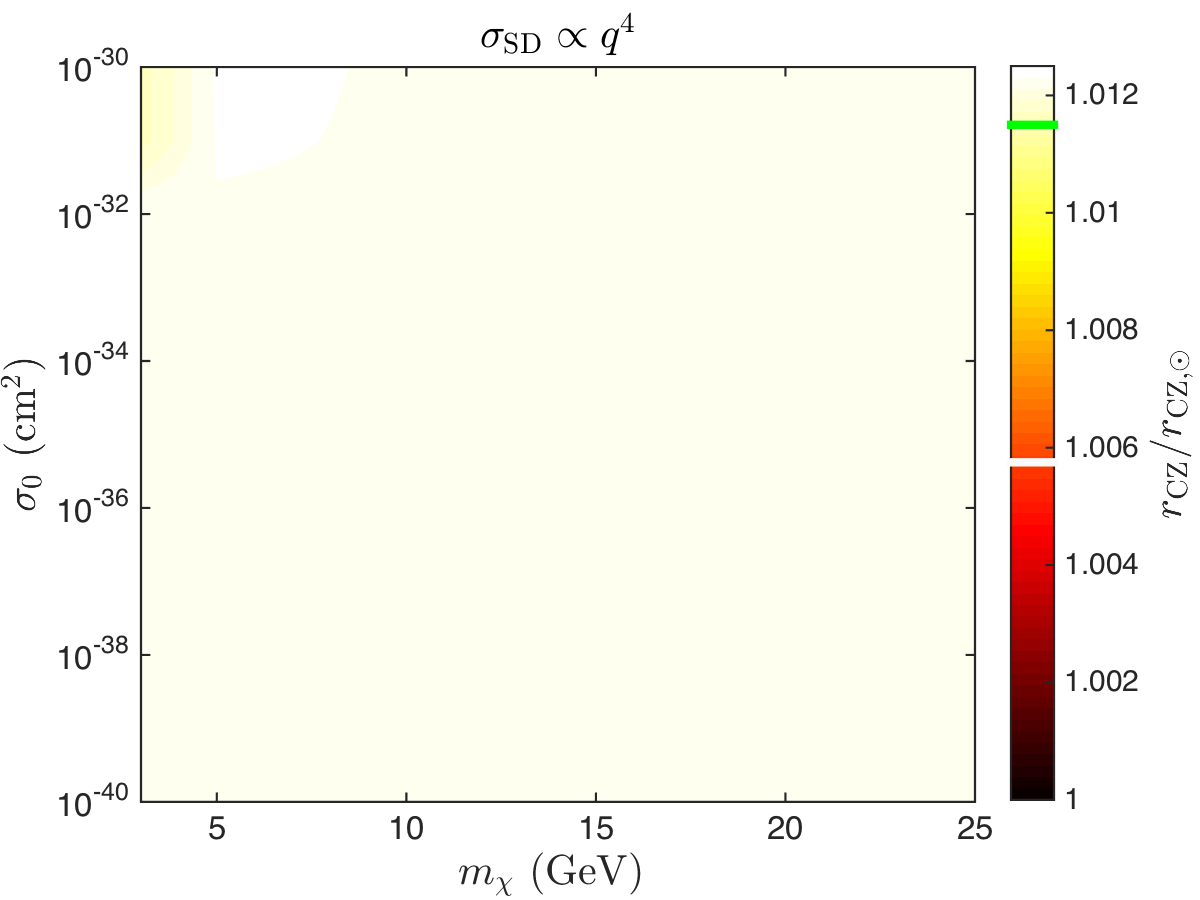} \\
\end{tabular}
\caption{As per Fig.\ \ref{SIrc}, but for spin-dependent couplings.}
\label{SDrc}
\end{figure}

\begin{figure}[!p]
\begin{tabular}{c@{\hspace{0.04\textwidth}}c}
\multicolumn{2}{c}{\includegraphics[height = 0.32\textwidth]{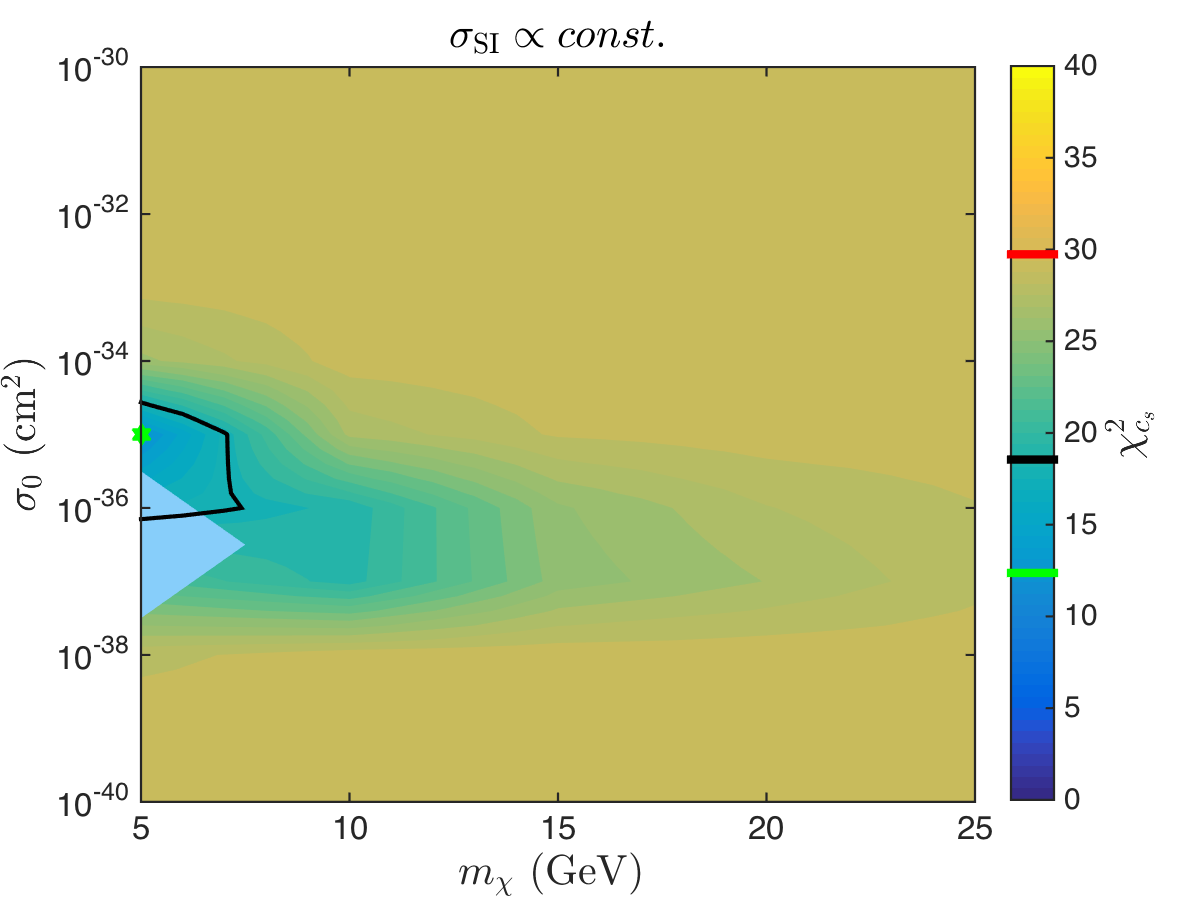}} \\
\includegraphics[height = 0.32\textwidth]{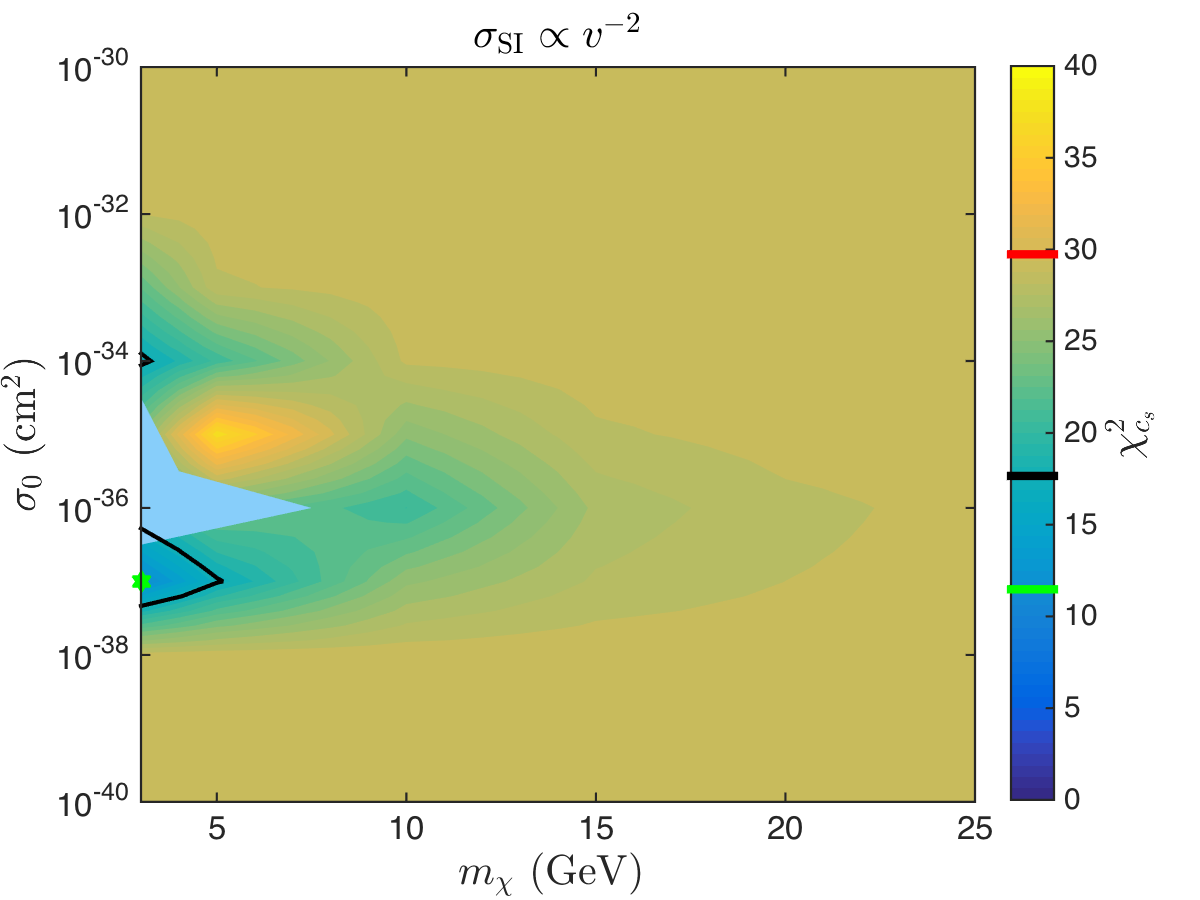} & \includegraphics[height = 0.32\textwidth]{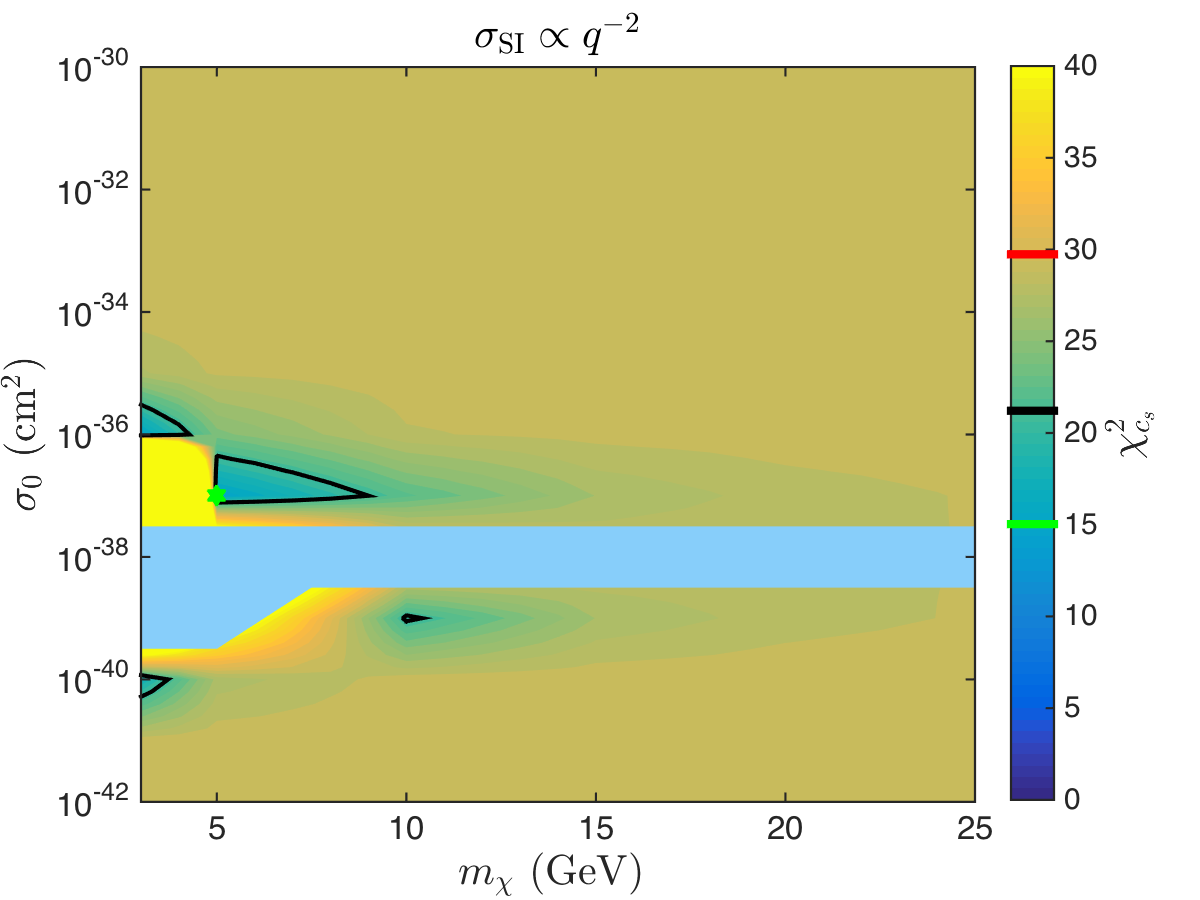} \\
\includegraphics[height = 0.32\textwidth]{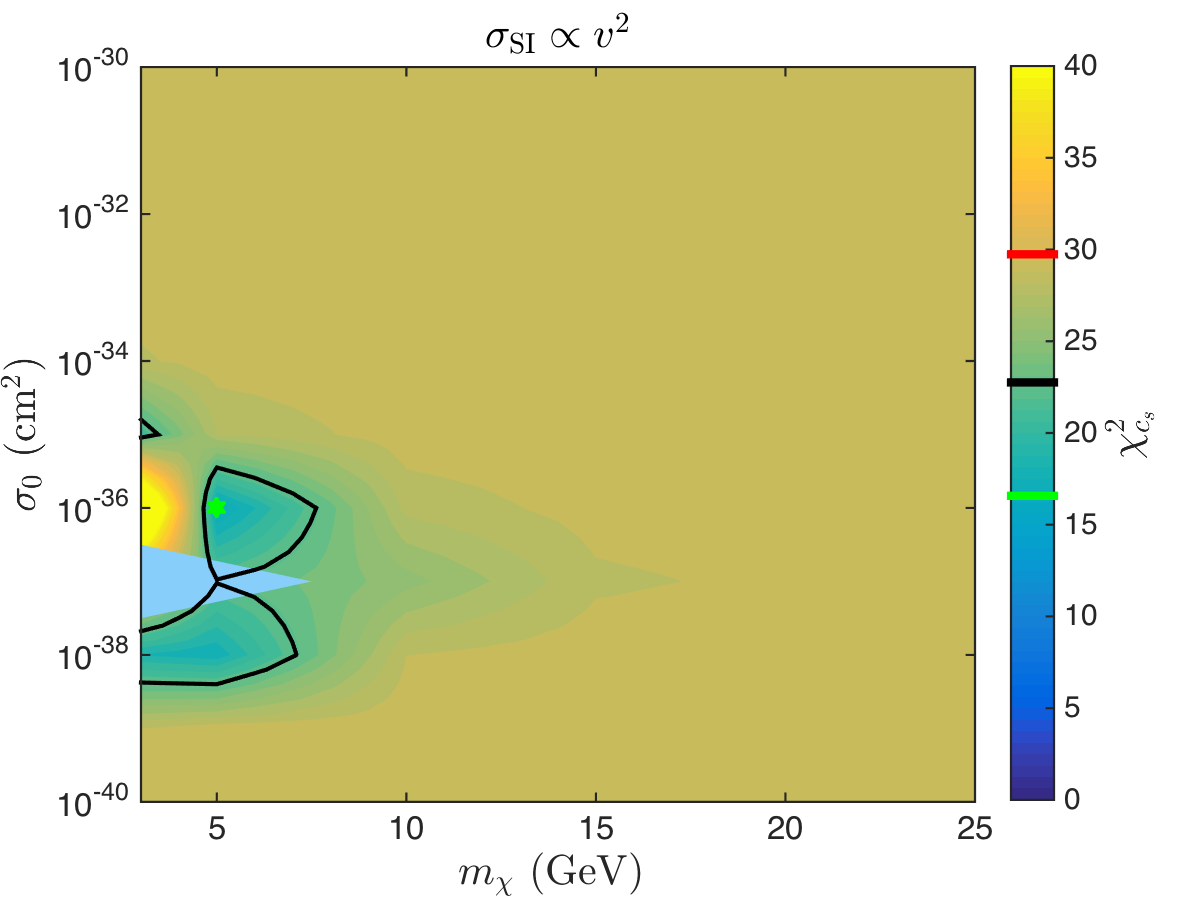} & \includegraphics[height = 0.32\textwidth]{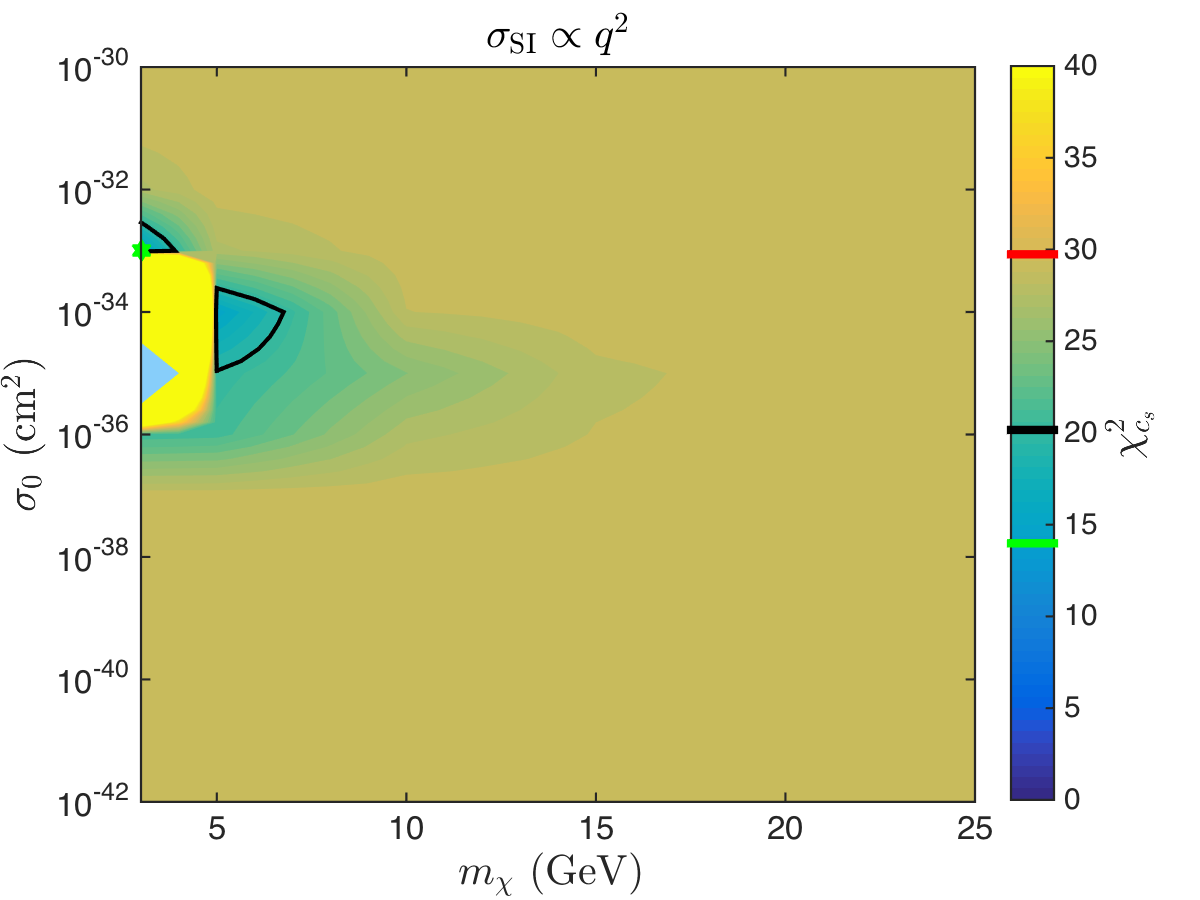} \\
\includegraphics[height = 0.32\textwidth]{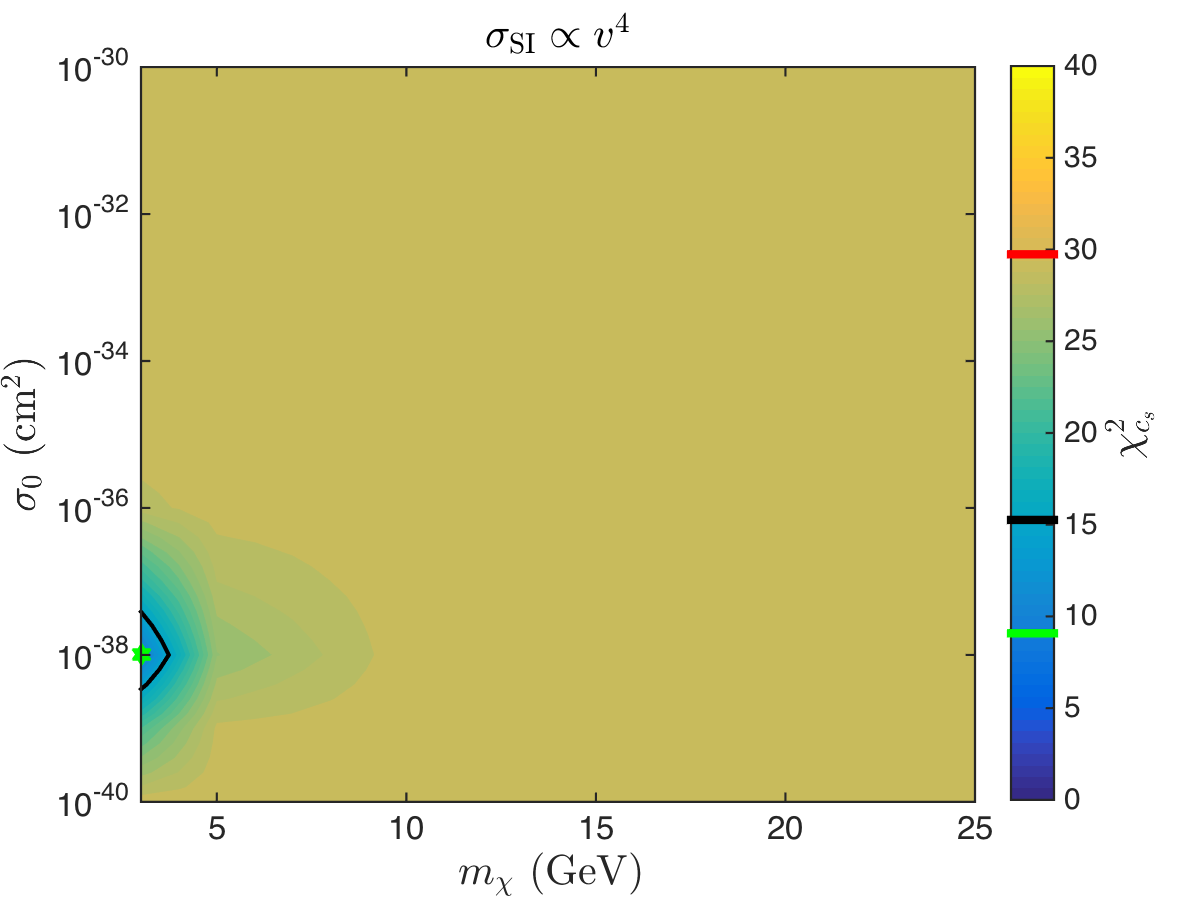} & \includegraphics[height = 0.32\textwidth]{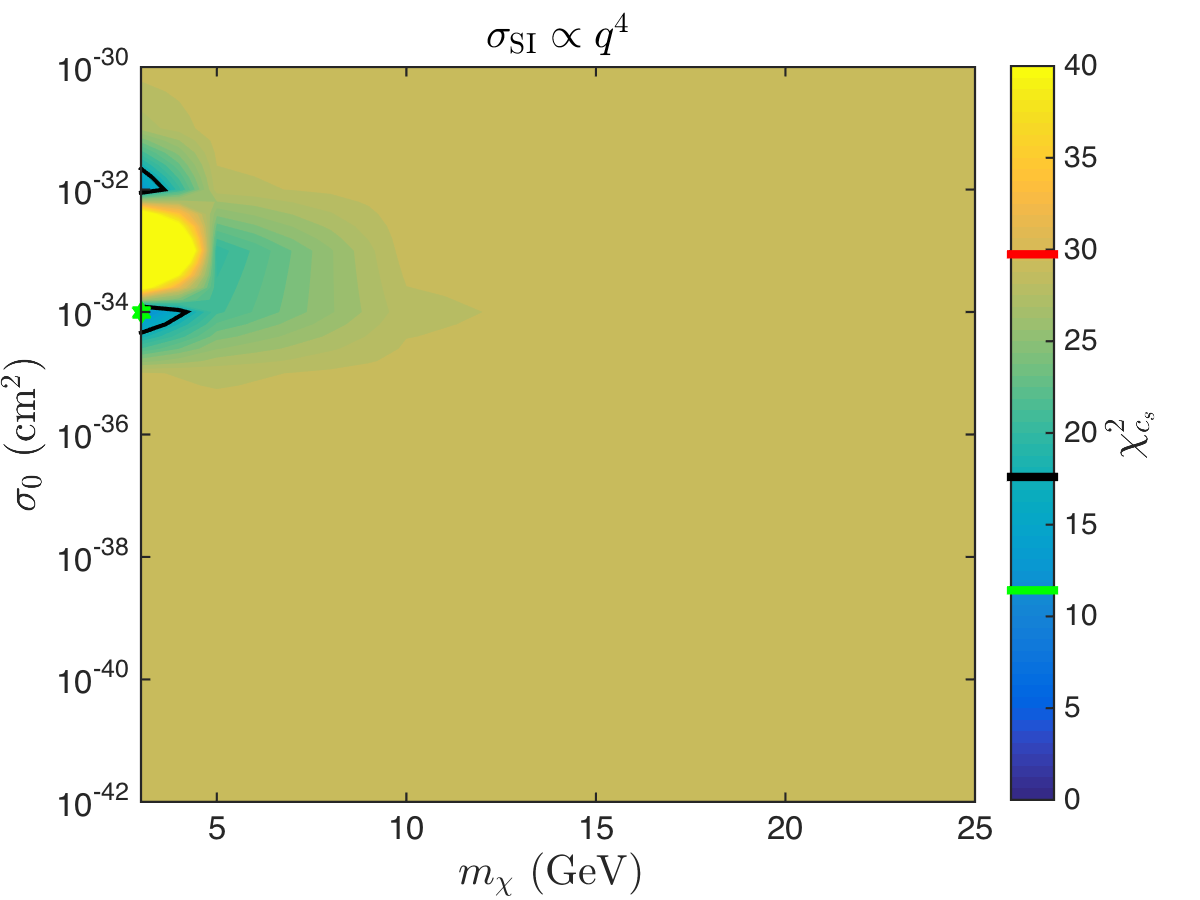} \\
\end{tabular}
\caption{Sound speed $\chi^2$. Darker regions represent a better fit to the observed value than the SSM. Best fits are shown as green stars and green lines on colour bars. 3$\sigma$ deviations from the best fits are marked in black, and red lines show the $\chi^2$ value of the SSM. Regions masked in light blue correspond to parameter combinations where models did not converge.}
\label{SIcschsq}
\end{figure}

\begin{figure}[!p]
\begin{tabular}{c@{\hspace{0.04\textwidth}}c}
\multicolumn{2}{c}{\includegraphics[height = 0.32\textwidth]{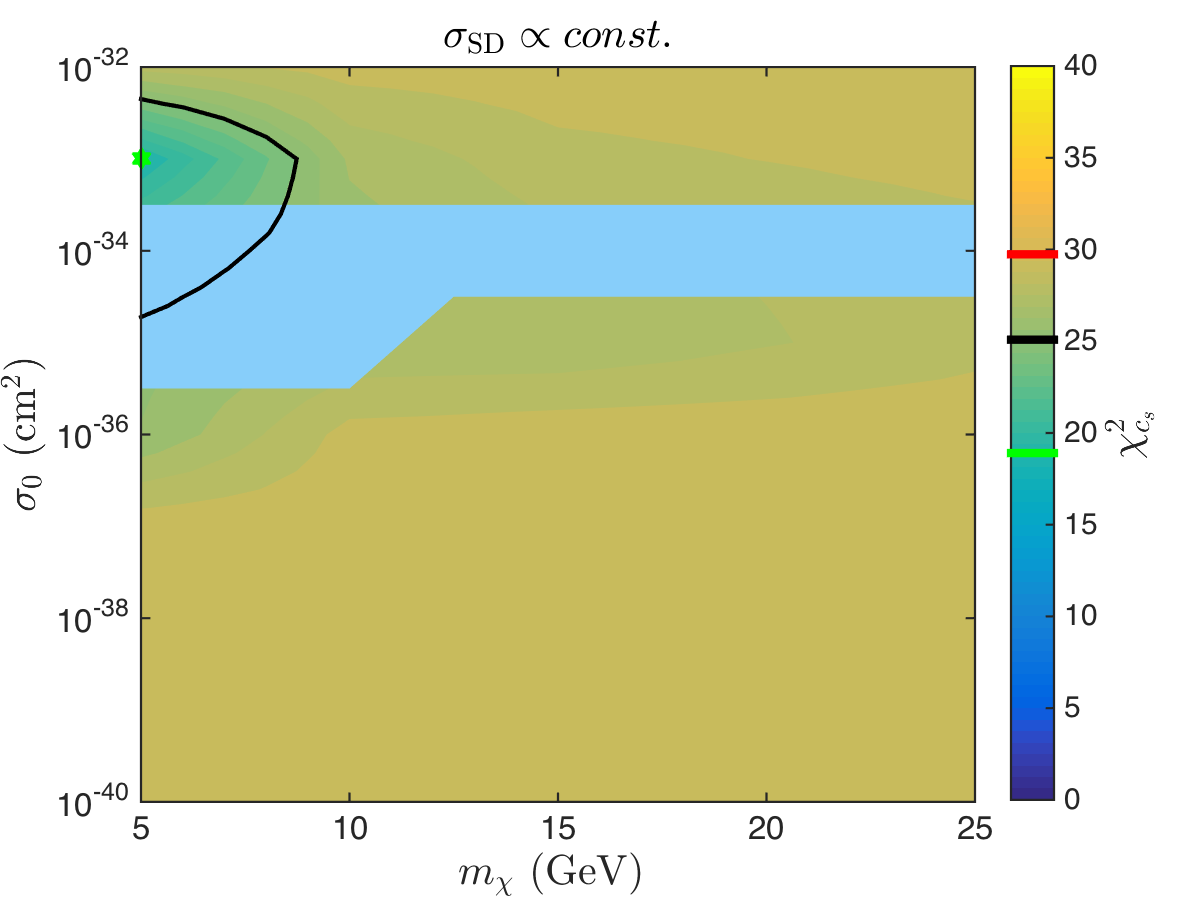}} \\
\includegraphics[height = 0.32\textwidth]{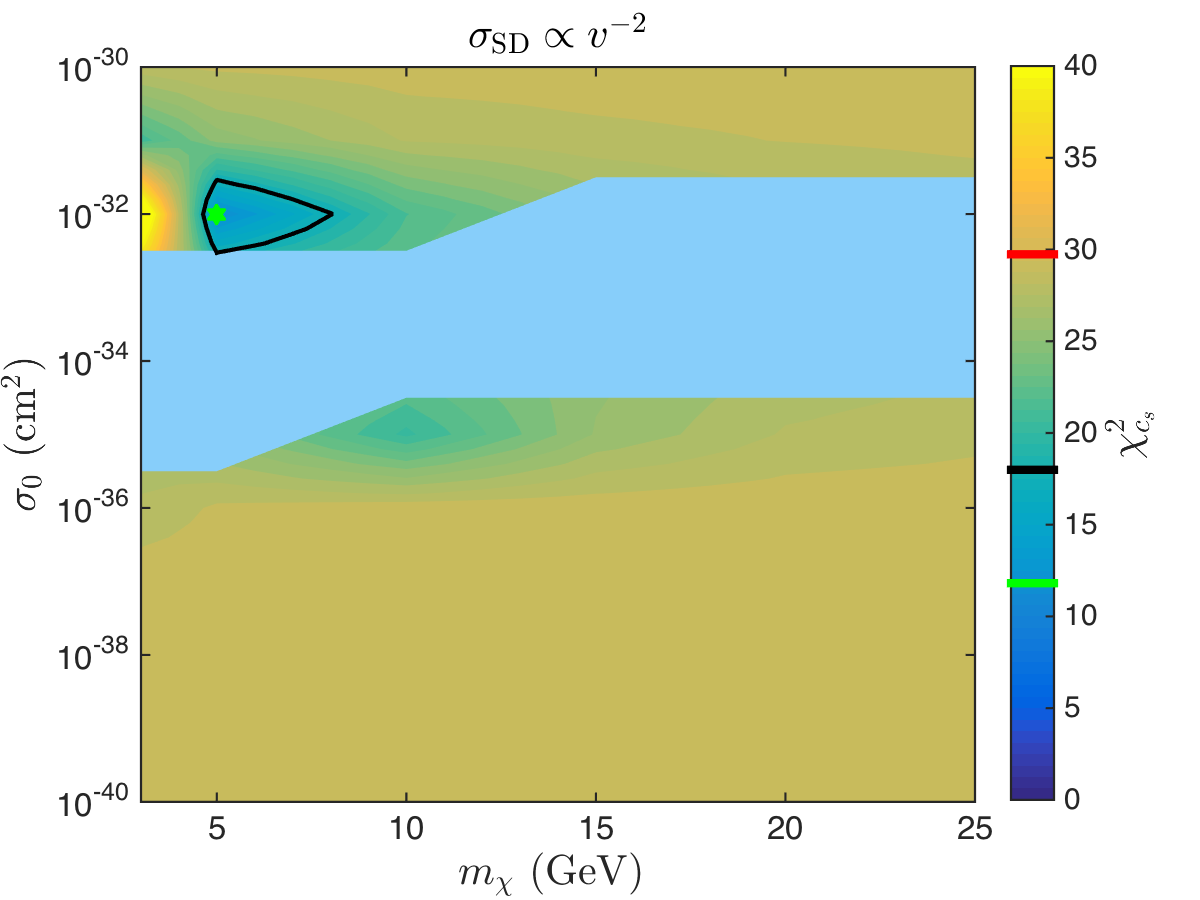} & \includegraphics[height = 0.32\textwidth]{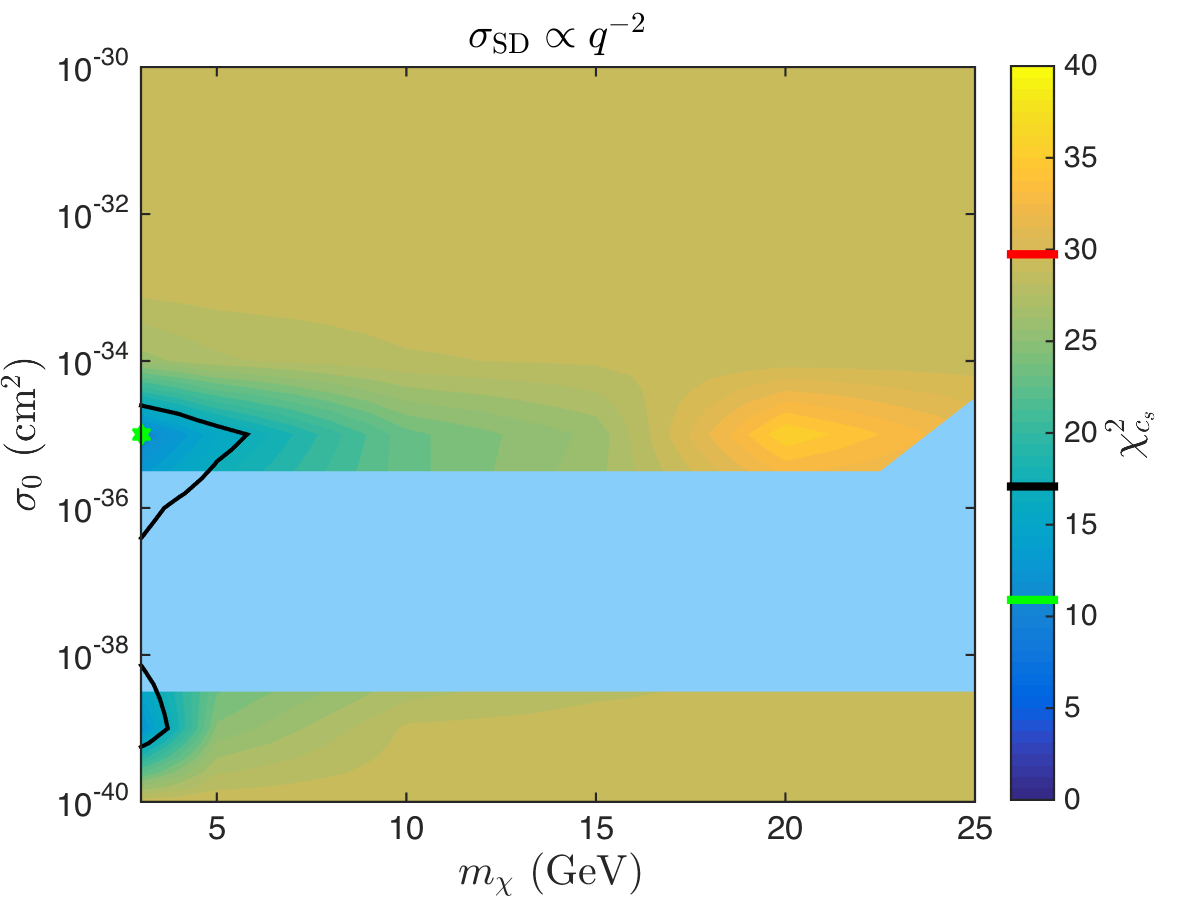} \\
\includegraphics[height = 0.32\textwidth]{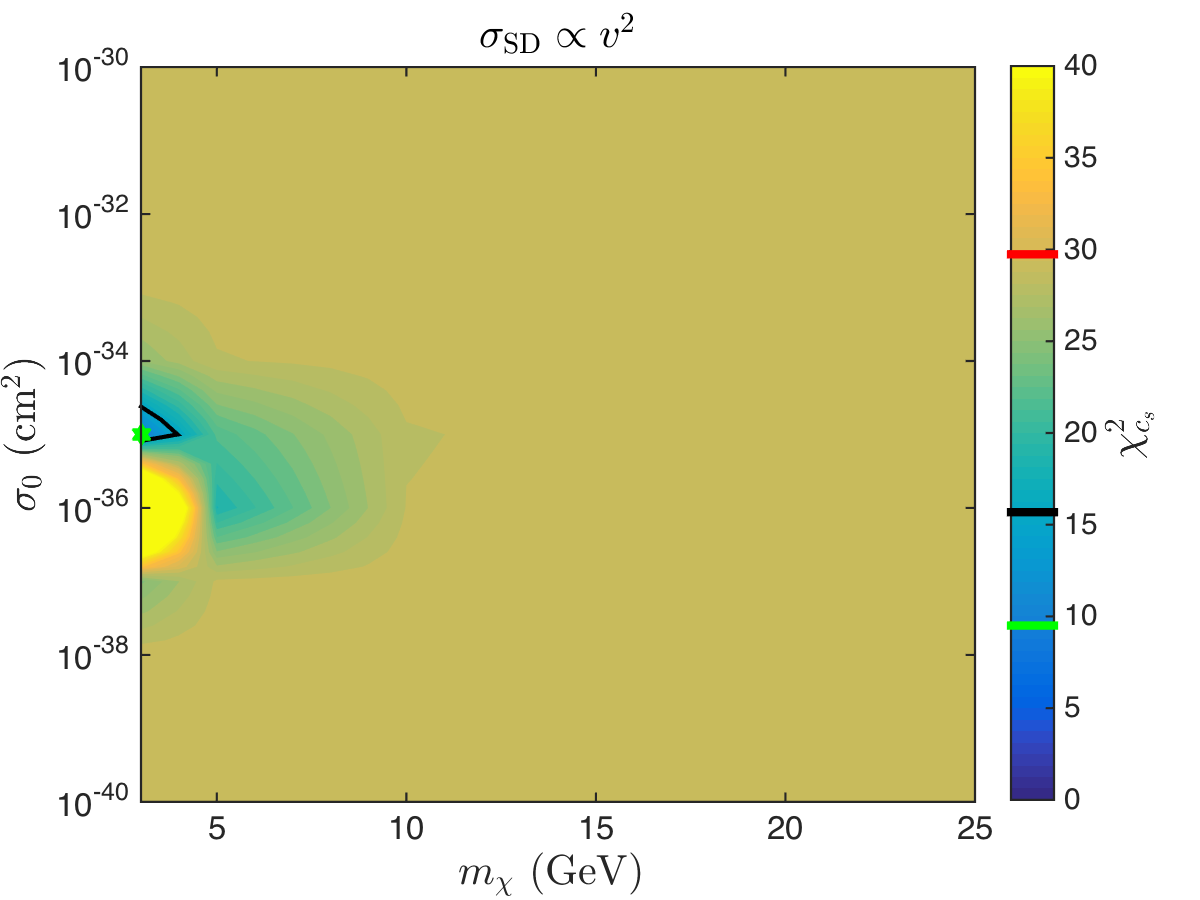} & \includegraphics[height = 0.32\textwidth]{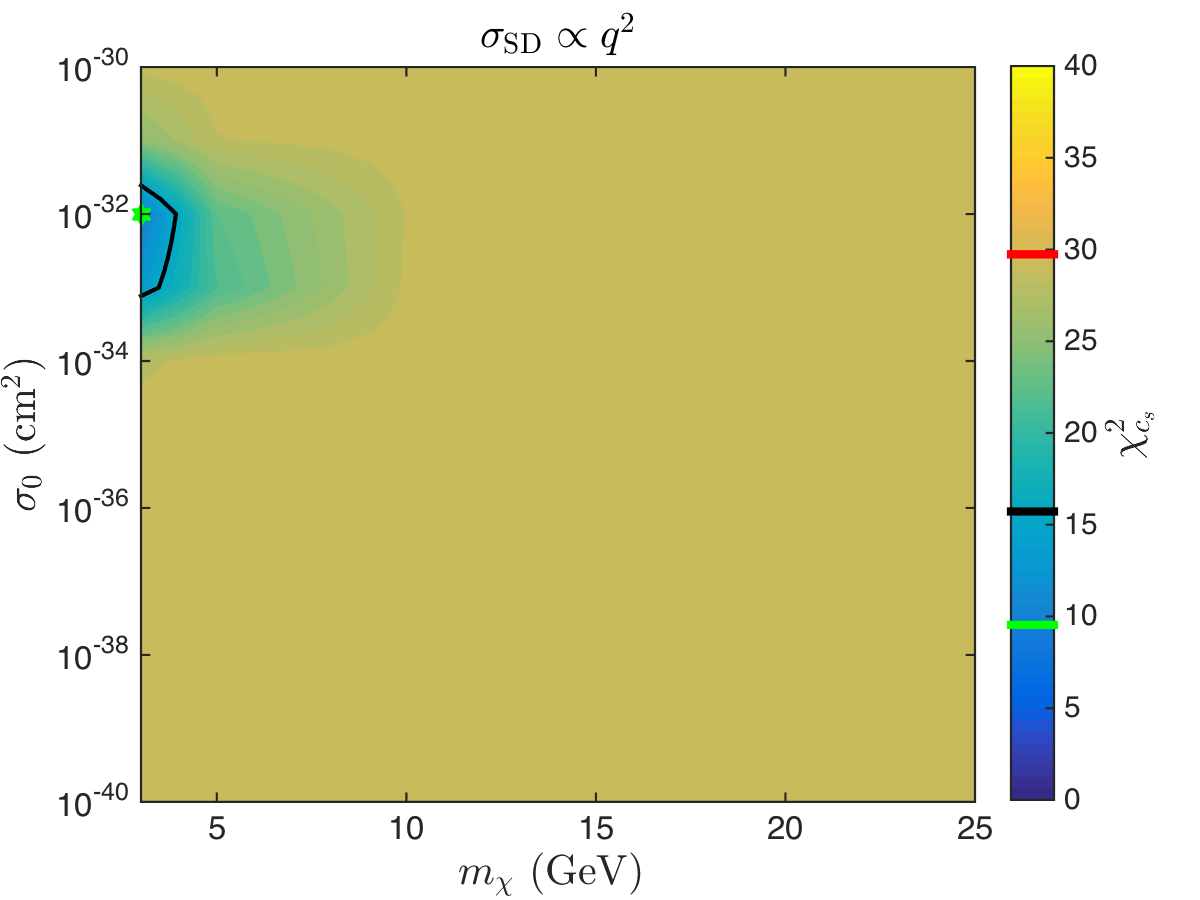} \\
\includegraphics[height = 0.32\textwidth]{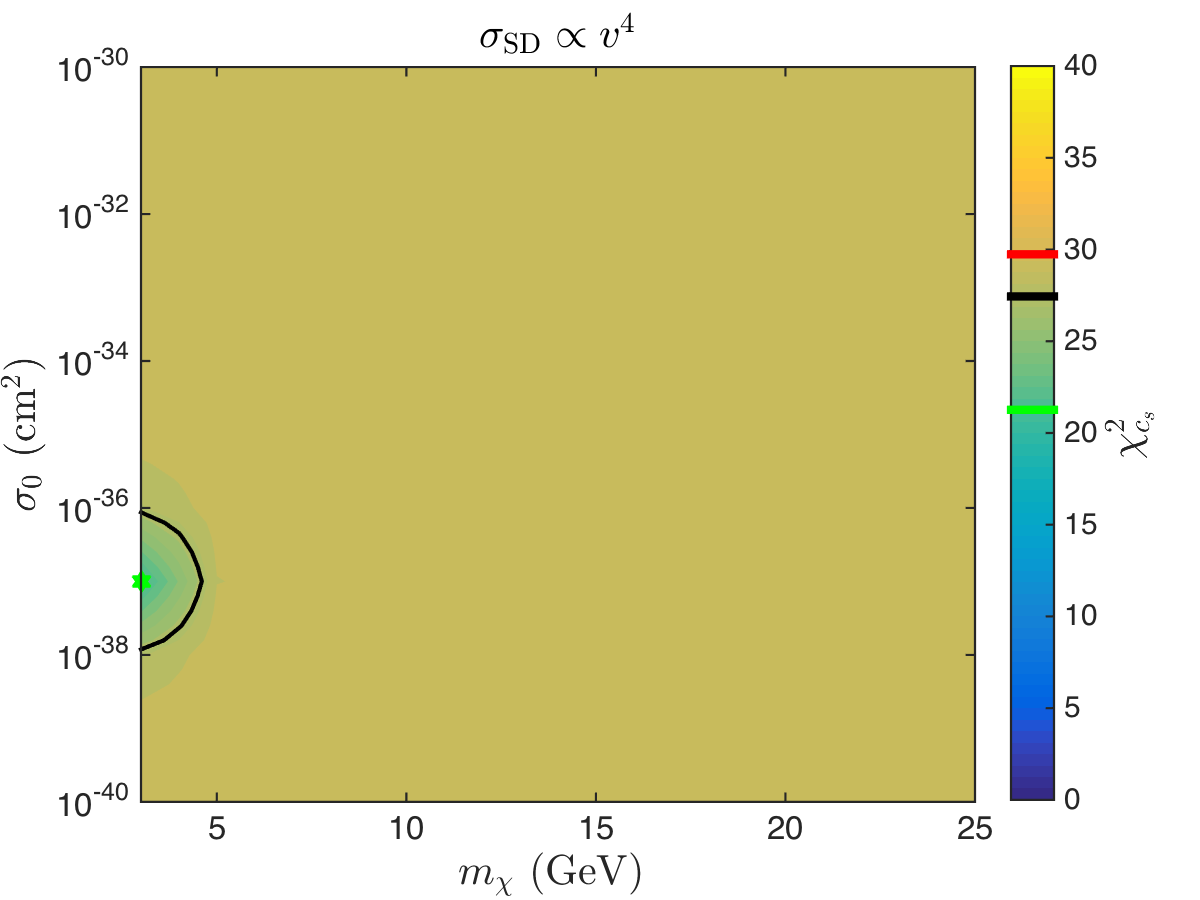} & \includegraphics[height = 0.32\textwidth]{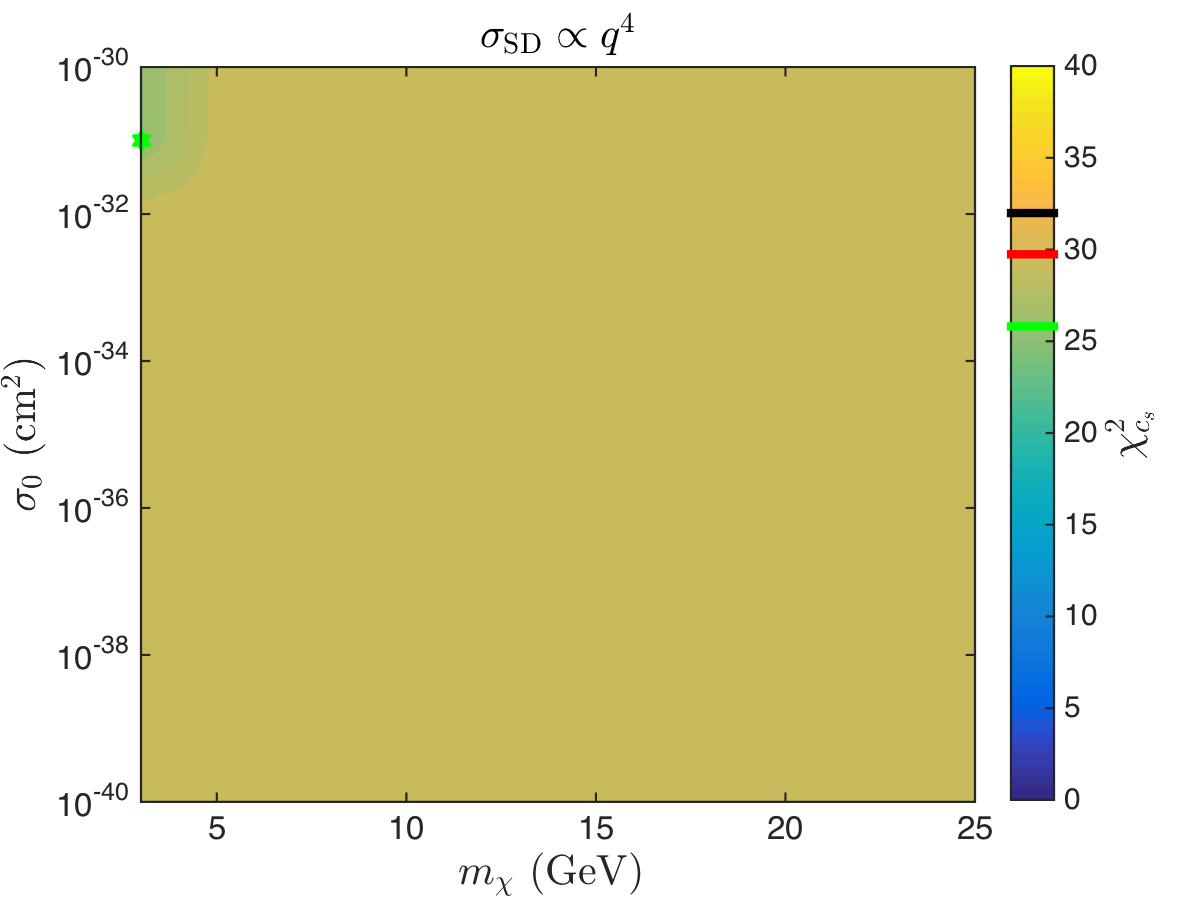} \\
\end{tabular}
\caption{As per Fig.\ \ref{SIcschsq}, but for spin-dependent couplings.}
\label{SDcschsq}
\end{figure}

\begin{figure}[!p]
\begin{tabular}{c@{\hspace{0.04\textwidth}}c}
\multicolumn{2}{c}{\includegraphics[height = 0.32\textwidth]{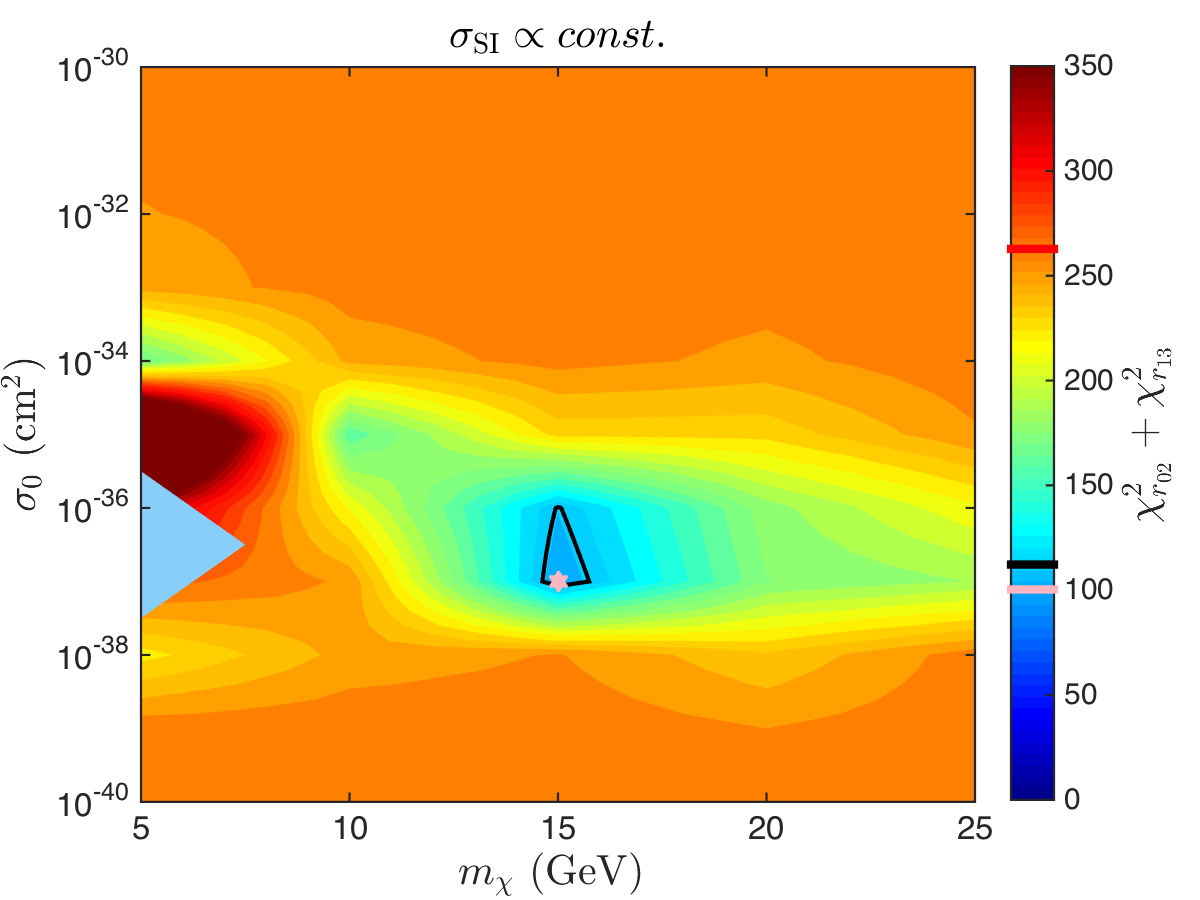}} \\
\includegraphics[height = 0.32\textwidth]{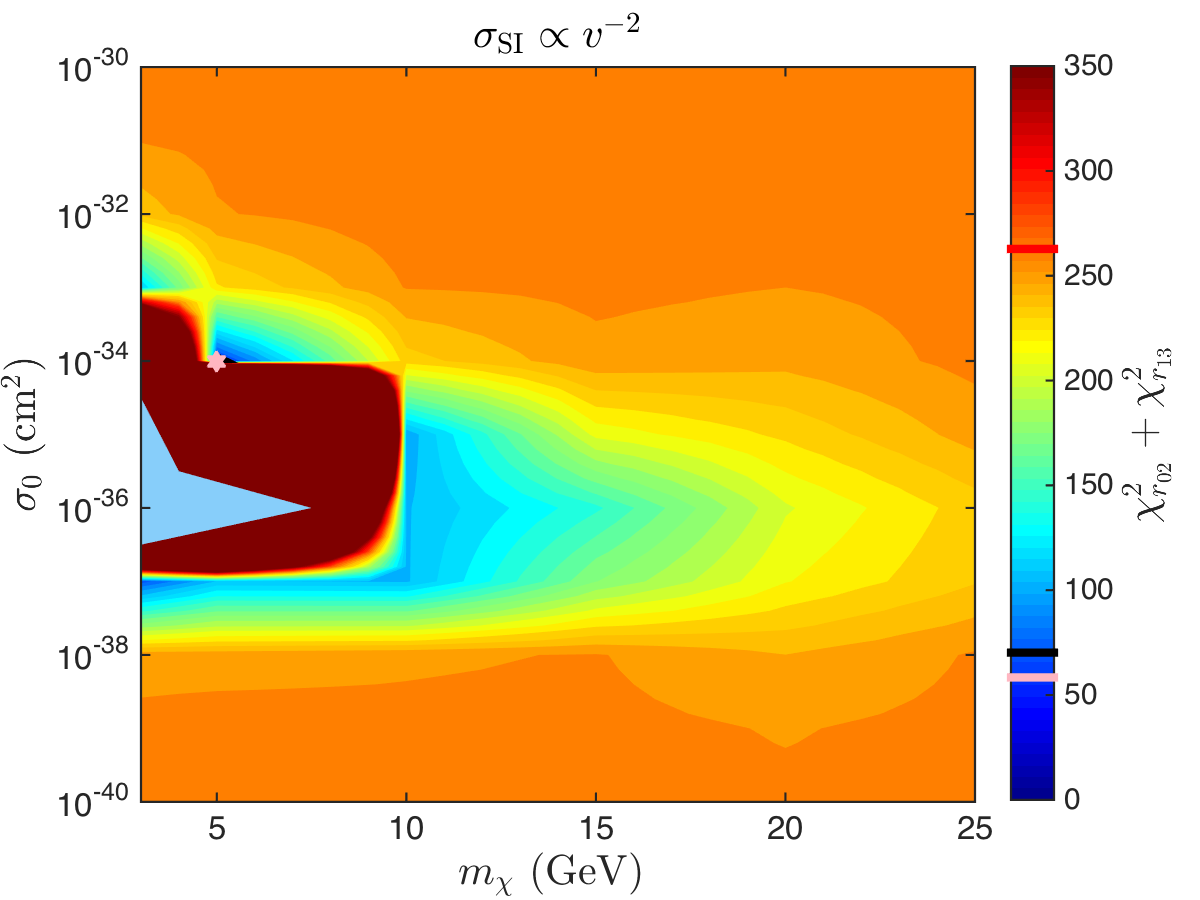} & \includegraphics[height = 0.32\textwidth]{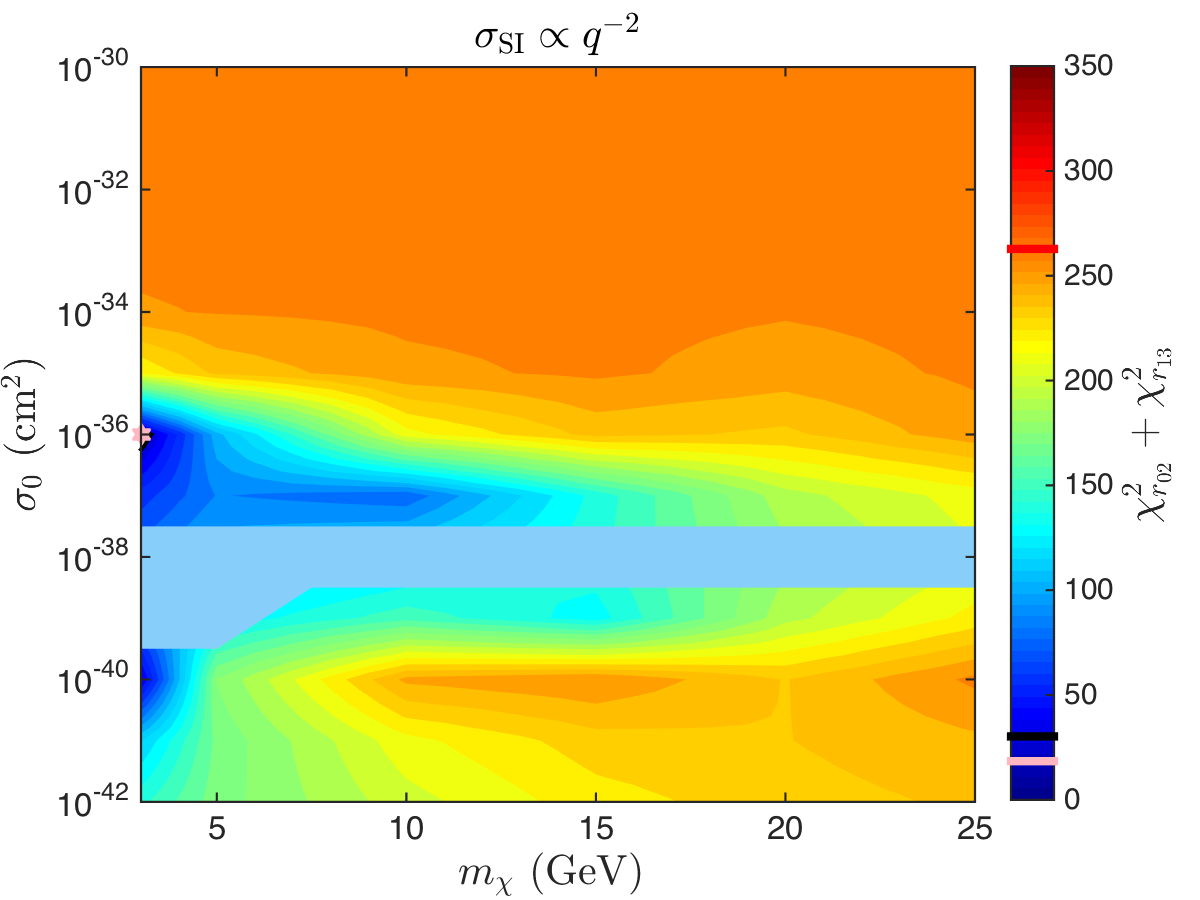} \\
\includegraphics[height = 0.32\textwidth]{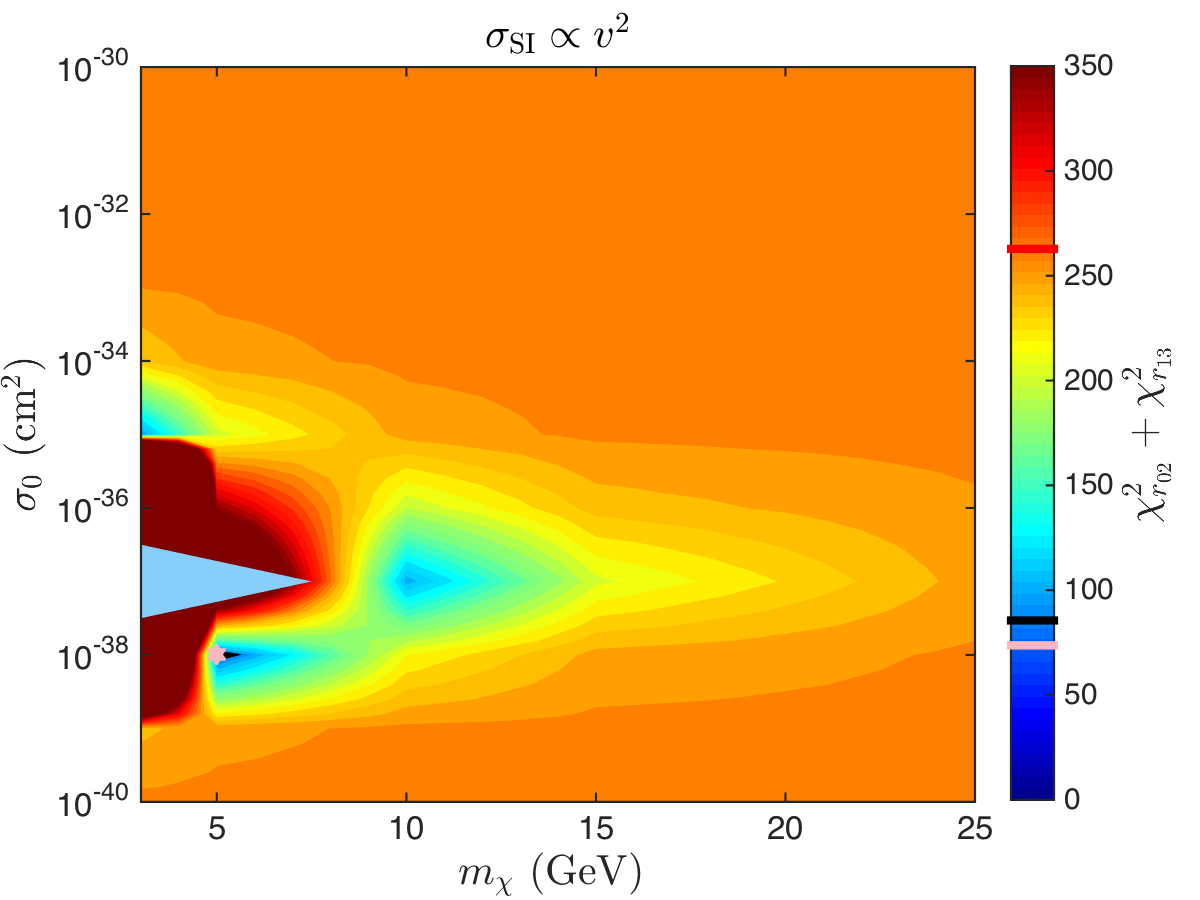} & \includegraphics[height = 0.32\textwidth]{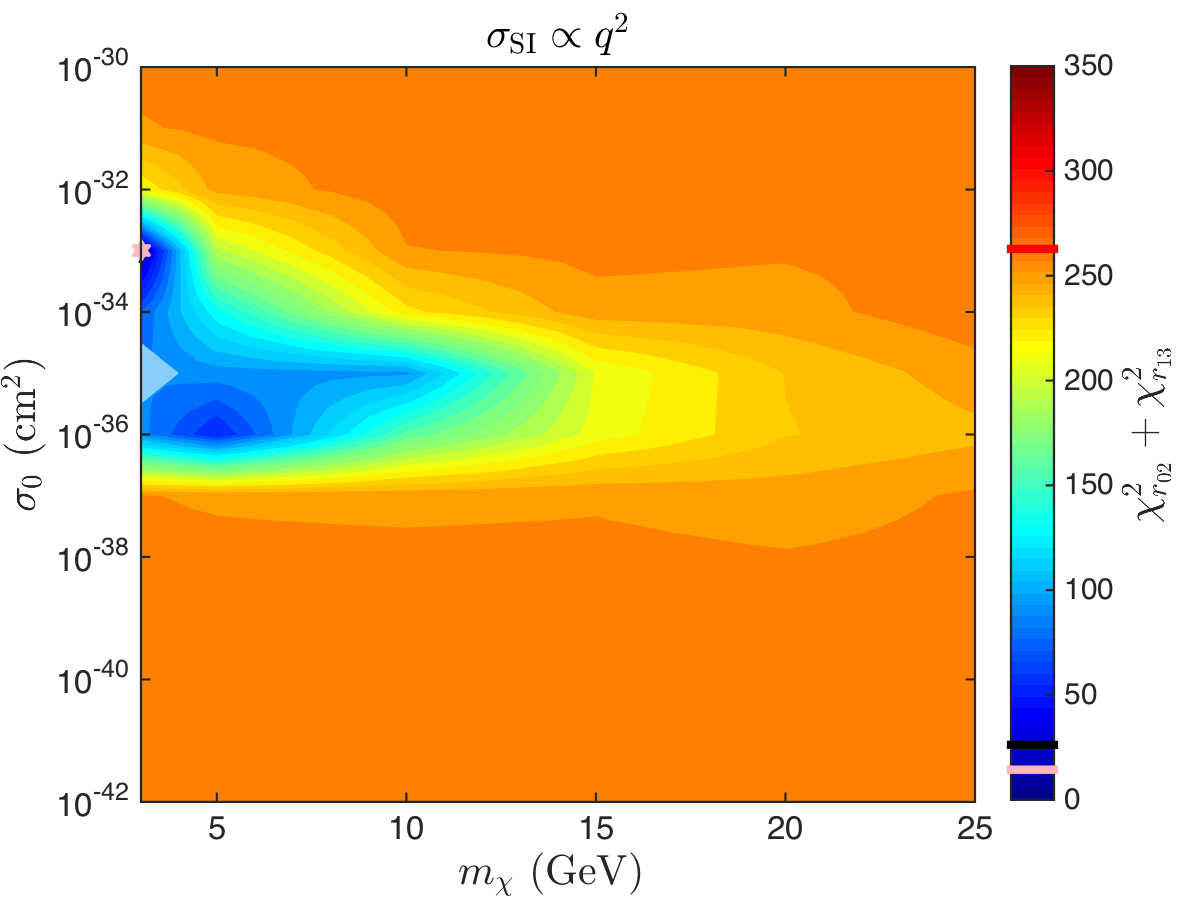} \\
\includegraphics[height = 0.32\textwidth]{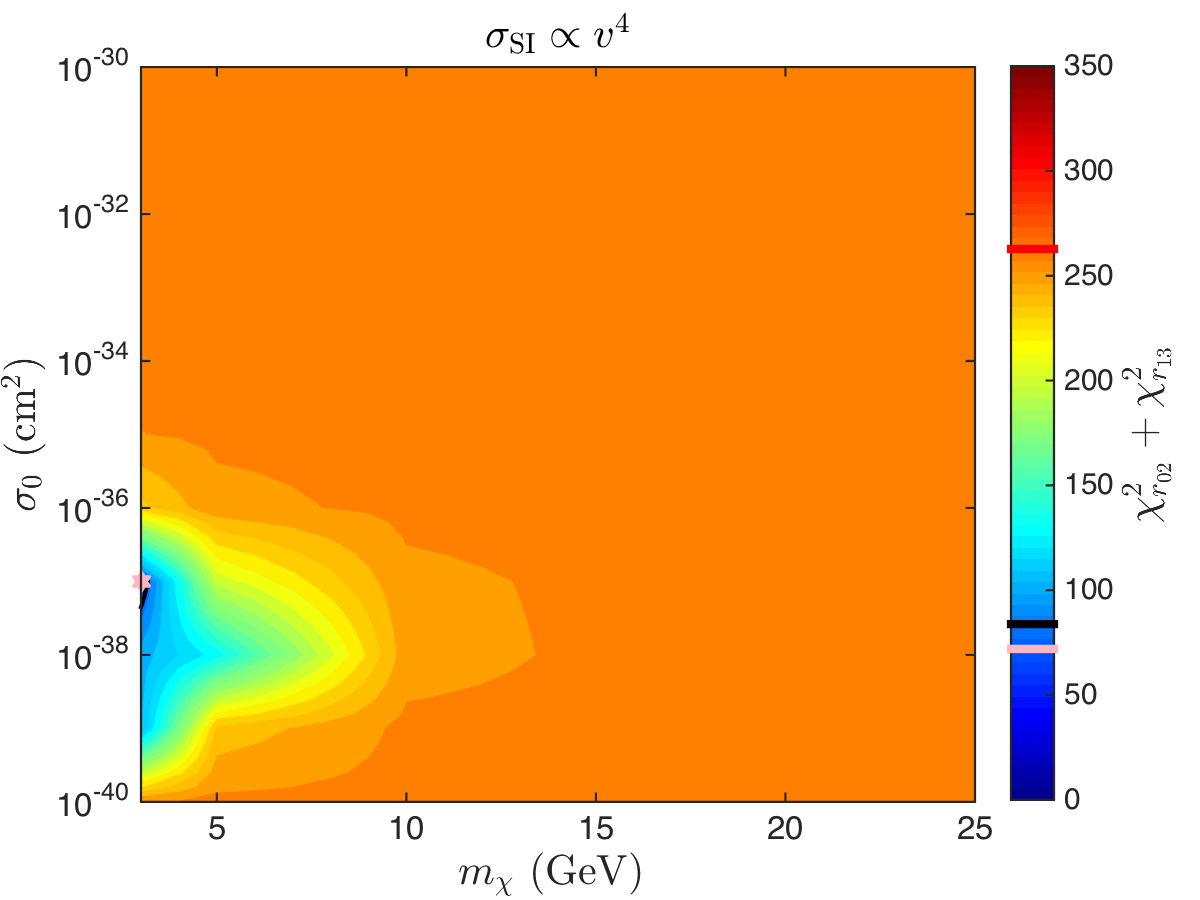} & \includegraphics[height = 0.32\textwidth]{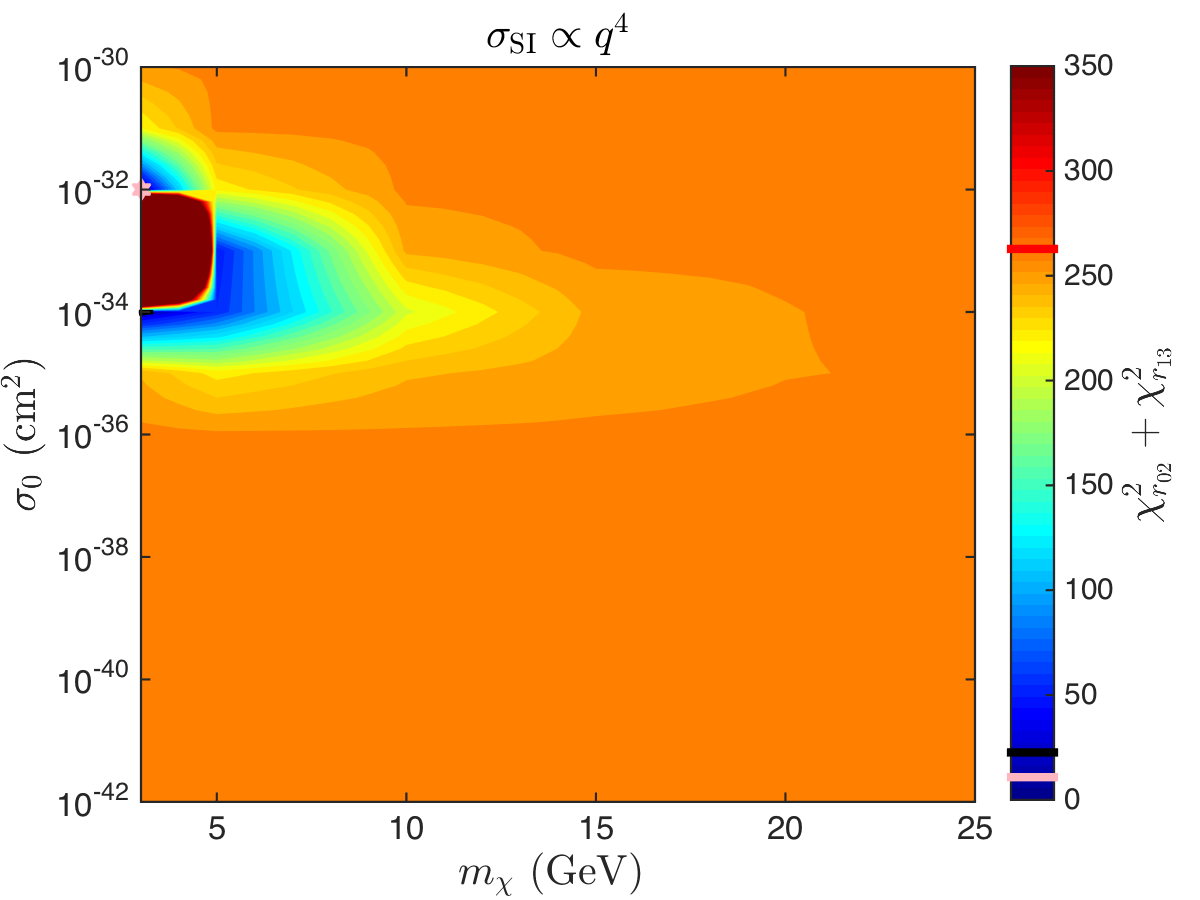} \\
\end{tabular}
\caption{Small frequency separation $\chi^2$. Blue regions represent a better fit to the observed value than the SSM.  Best fits are shown as pink stars and pink lines on colour bars. 3$\sigma$ deviations from the best fits are marked in black, and red lines show the $\chi^2$ value of the SSM. Regions masked in light blue correspond to parameter combinations where models did not converge.}
\label{SIrijchsq}
\end{figure}

\begin{figure}[!p]
\begin{tabular}{c@{\hspace{0.04\textwidth}}c}
\multicolumn{2}{c}{\includegraphics[height = 0.32\textwidth]{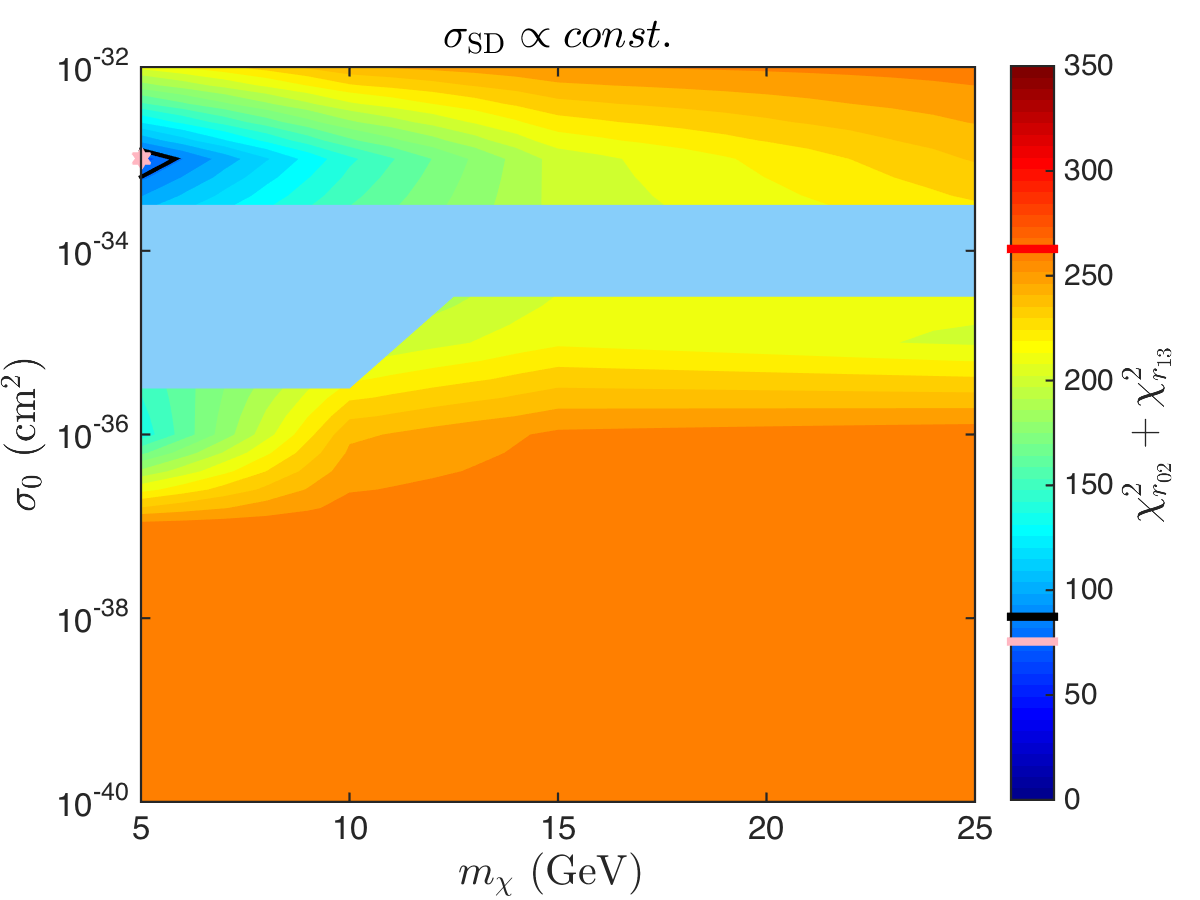}} \\
\includegraphics[height = 0.32\textwidth]{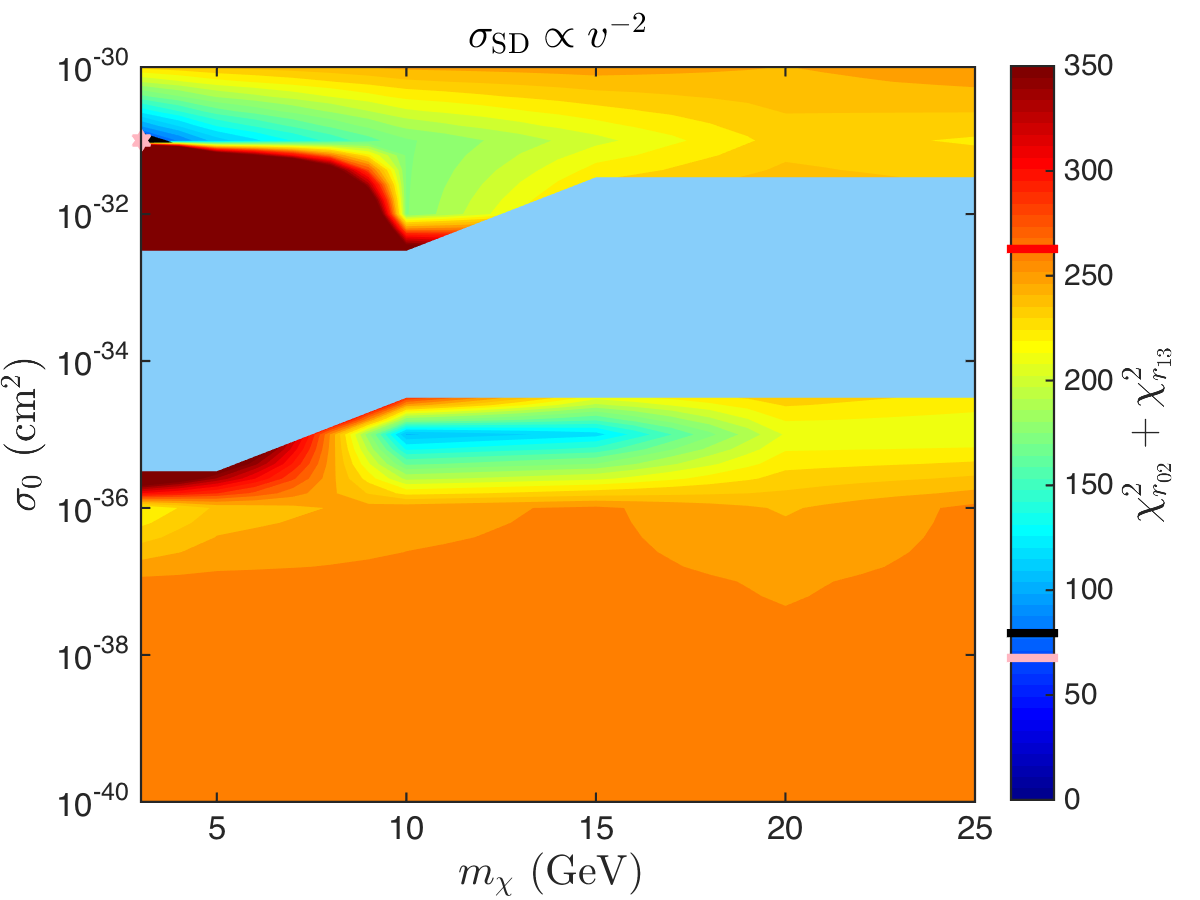} & \includegraphics[height = 0.32\textwidth]{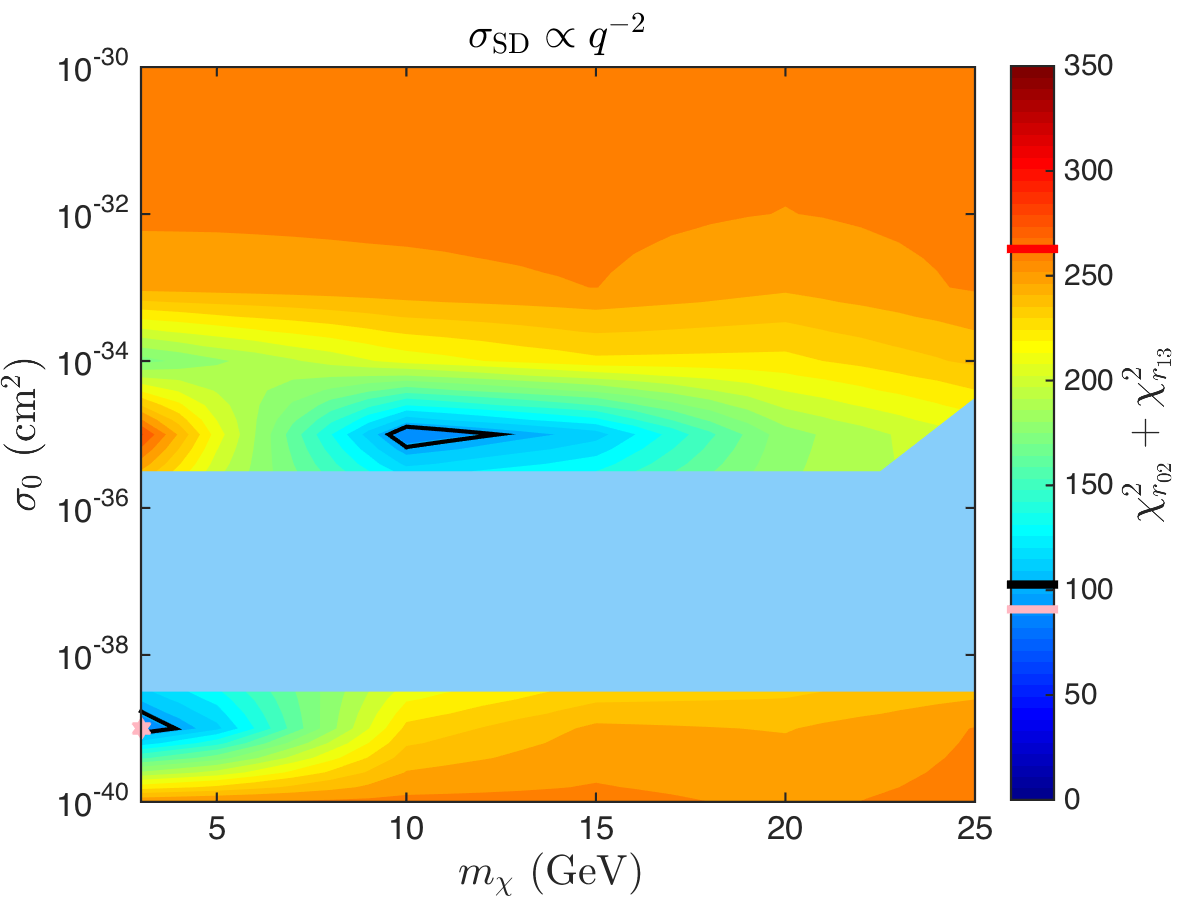} \\
\includegraphics[height = 0.32\textwidth]{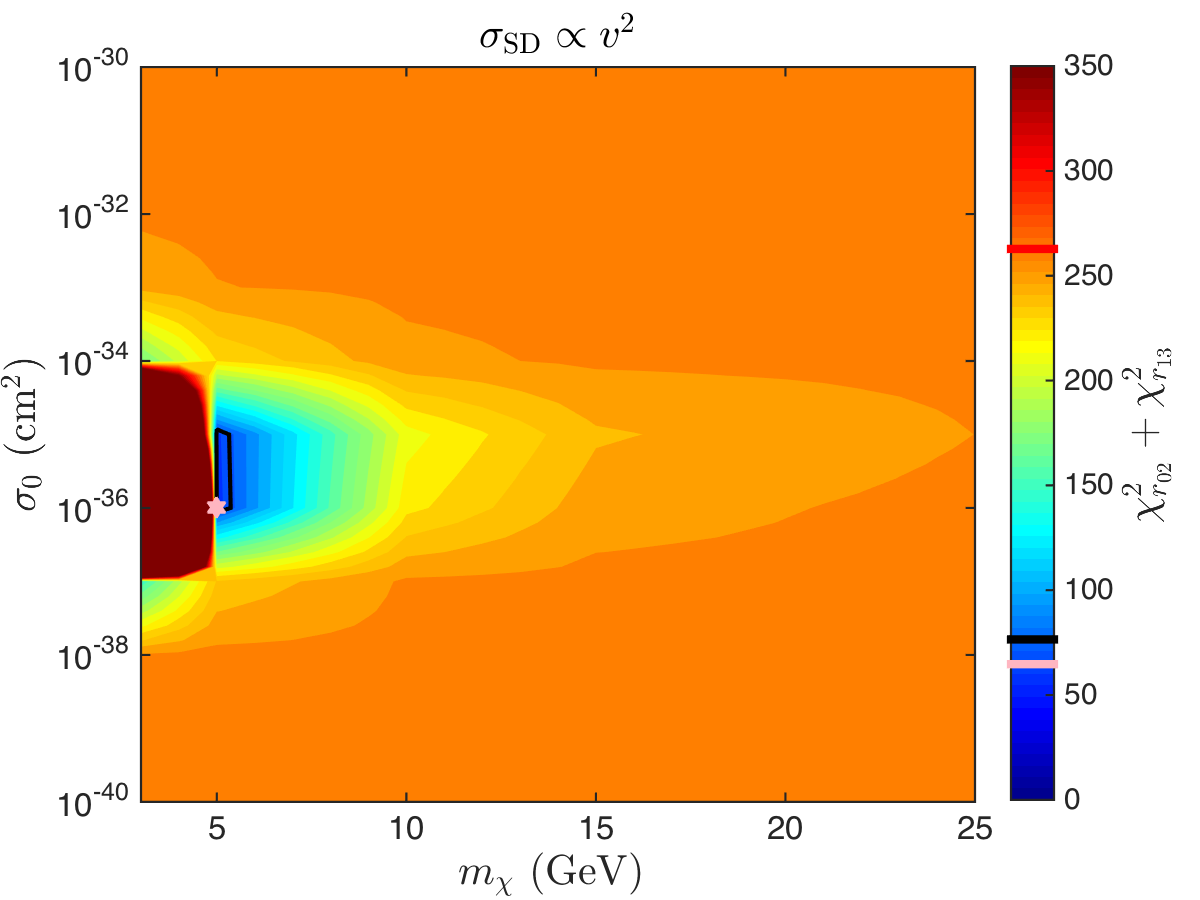} & \includegraphics[height = 0.32\textwidth]{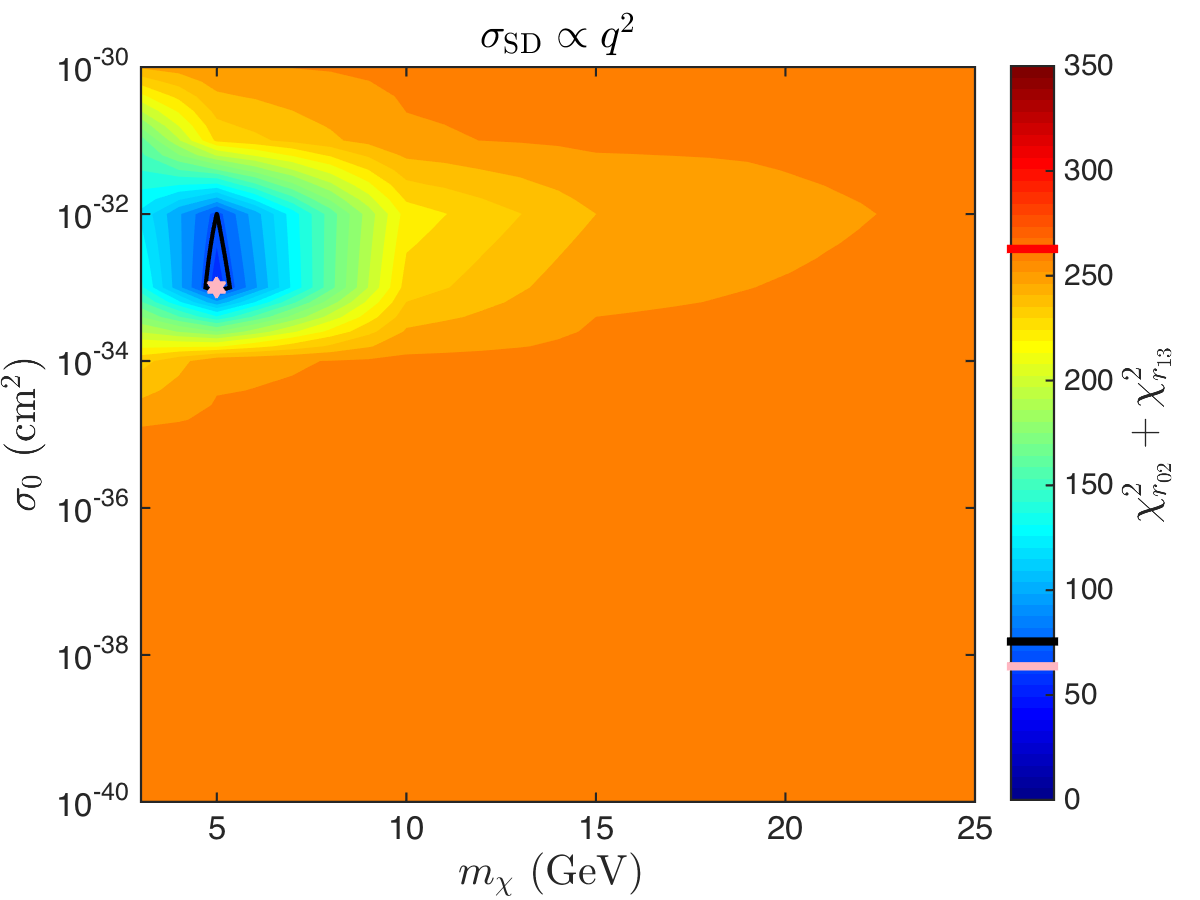} \\
\includegraphics[height = 0.32\textwidth]{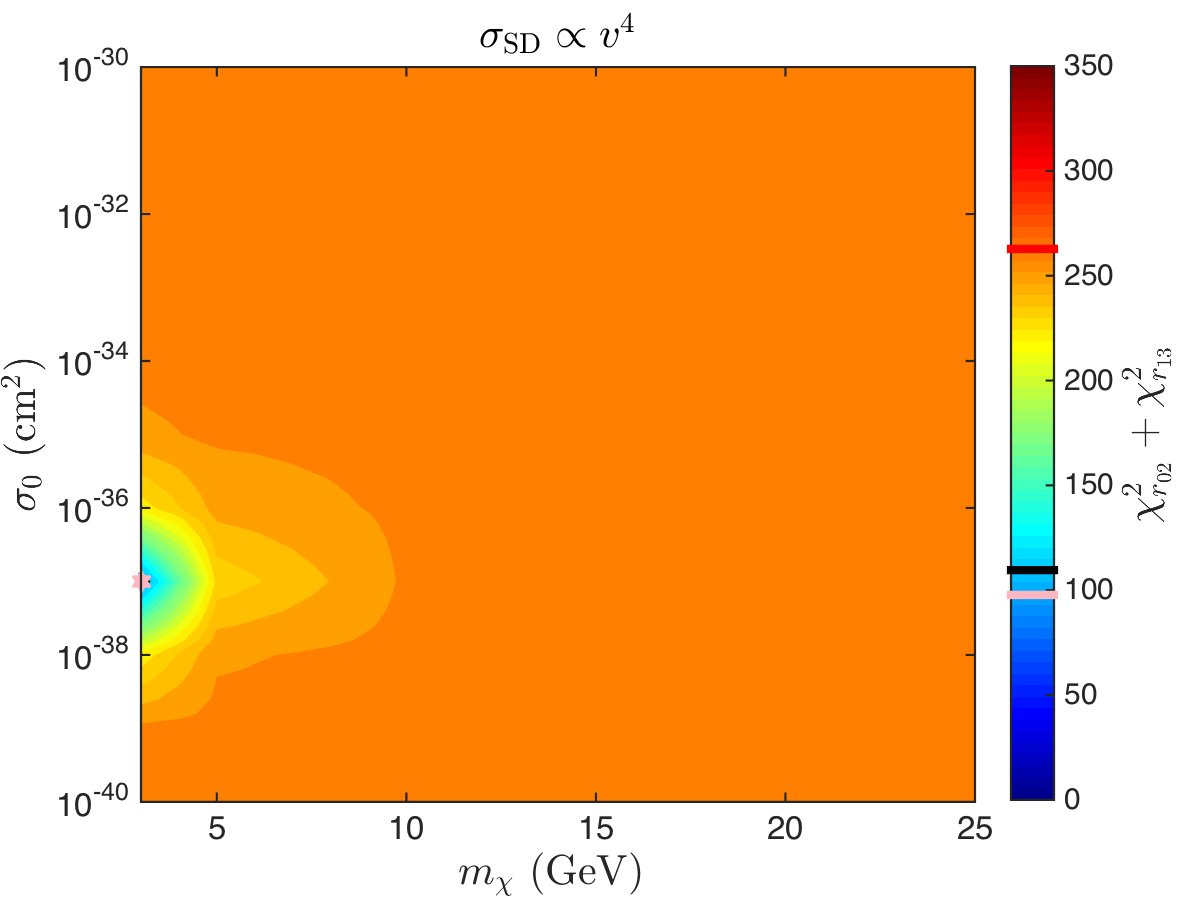} & \includegraphics[height = 0.32\textwidth]{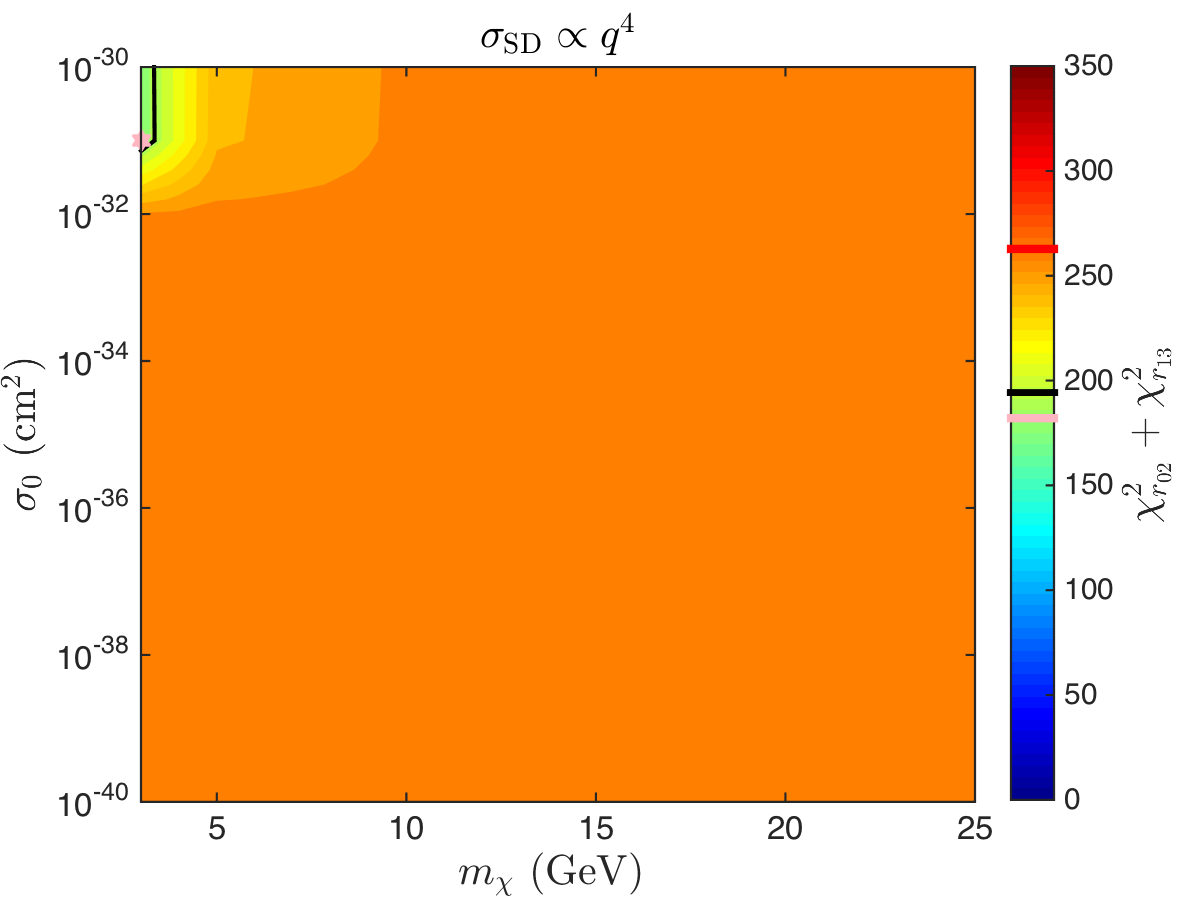} \\
\end{tabular}
\caption{As per Fig.\ \ref{SIrijchsq}, but for spin-dependent couplings.}
\label{SDrijchsq}
\end{figure}

\FloatBarrier

\bibliographystyle{JHEP_pat}
\bibliography{CandO,CObiblio,AbuGen,solarDM}

\end{document}